%% file: ms.tex
\documentclass[numberedappendix]{emulateapj}
\newcommand{\mode}{submit}
\usepackage{soul,color,natbib}
\input{defs.tex}
\shortauthors{Mahdavi et al.}
\shorttitle{JACO/CCCP Clusters of Galaxies}
\ifthenelse{\equal{\mode}{aastex}}
{

}{
\journalinfo{}
\submitted{Submitted October 13, 2012; Accepted February 25, 2013 for publication in \emph{The Astrophysical Journal}; Erraturm submitted July 27, 2014}

}

\definecolor{orange}{rgb}{0.8,0.8,0.8}
\definecolor{lb}{rgb}{0.7,0.7,1.0}
\newcommand{\hly}[1]{\sethlcolor{yellow}\hl{#1}}
\newcommand{\hlo}[1]{\sethlcolor{orange}\hl{#1}}
\newcommand{\hlb}[1]{\sethlcolor{lb}\hl{#1}}
\renewcommand{\hly}[1]{#1}
\renewcommand{\hlo}[1]{#1}
\renewcommand{\hlb}[1]{#1}

\begin{document}

\title{Joint Analysis of Cluster Observations: II.  Chandra/XMM-Newton X-ray
and Weak Lensing Scaling Relations for a Sample of 50 Rich Clusters of Galaxies}

\author{Andisheh Mahdavi}
\affil{Department of Physics
and Astronomy, San Francisco State University, San Francisco, CA 94131}
\author{Henk Hoekstra}
\affil{Leiden Observatory, Leiden University, Niels Bohrweg 2, NL-2333 CA Leiden, The Netherlands}
\author{Arif Babul and Chris Bildfell}
\affil{Department of Physics and Astronomy, University of Victoria, Victoria, BC V8W 3P6, Canada}
\author{Tesla Jeltema}
\affil{Santa Cruz Institute for Particle Physics, UC Santa Cruz 1156 High Street, Santa Cruz, CA 95064}
\author{J. Patrick Henry}
\affil{Institute for Astronomy, 2680 Woodlawn Drive, Honolulu HI 96822, U.~S.~A.}

\newcommand{\hh}[1]{\colorbox{yellow}{\parbox{#1}}}

\begin{abstract}
  We present a study of multiwavelength X-ray and weak lensing scaling
  relations for a sample of 50 clusters of galaxies. Our analysis
  combines \emph{Chandra} and \emph{XMM-Newton} data using an
  energy-dependent cross-calibration. After considering a number of
  scaling relations, we find that gas mass is the most robust
  estimator of weak lensing mass, yielding $15 \pm 6\%$ intrinsic
  scatter at $r_{500}^{WL}$ (the pseudo-pressure $Y_X$ yields a
  consistent scatter of $22\% \pm 5\%$).  The scatter does not change
  when measured within a fixed physical radius of $1$ Mpc.  Clusters
  with small BCG to X-ray peak offsets constitute a very regular
  population whose members have the same gas mass fractions and whose
  even smaller ($<10\%$) deviations from regularity can be ascribed to
  line of sight geometrical effects alone. Cool-core clusters, while a
  somewhat different population, also show the same ($<10\%$) scatter
  in the gas mass-lensing mass relation. There is a good correlation
  and a hint of bimodality in the plane defined by BCG offset and
  central entropy (or central cooling time).  The pseudo-pressure
  $Y_X$ does not discriminate between the more relaxed and less
  relaxed populations, making it perhaps the more even-handed mass
  proxy for surveys. Overall, hydrostatic masses underestimate weak
  lensing masses by $10\%$ on the average at $r_{500}^{WL}$; but
  cool-core clusters are consistent with no bias, while non-cool-core
  clusters have a large and constant $15-20\%$ bias between
  $r_{2500}^{WL}$ and $r_{500}^{WL}$, in agreement with N-body
  simulations incorporating unthermalized gas. For non-cool-core
  clusters, the bias correlates well with BCG ellipticity.  We also
  examine centroid shift variance and and power ratios to quantify
  substructure; these quantities do not correlate with residuals in
  the scaling relations. Individual clusters have for the most part
  forgotten the source of their departures from self-similarity.
\end{abstract}


\section{Introduction}
\label{sec:introduction}

Within the context of the currently favored hierarchical model for
structure formation, massive clusters of galaxies are, as a
population, the most recently formed gravitationally bound structures
in the cosmos.  Consequently, characteristics such as the shape and
evolutionary behavior of their mass function can, in principle,
be exploited as precision probes of cosmology. The resulting estimates
of parameters---such as the amplitude of the primordial fluctuations
and the density and equation of state of the mysterious dark
energy---can certainly complement and even compete with determinations
based on studies of the cosmic microwave background \citep[for a
review see ][]{Allen11}.

The efficacy of clusters as cosmological probes depends on three
factors: (1) the ability to compile a large well-understood catalog of
clusters; (2) the identification of an easily determined survey
observable (or combinations thereof) --- hereafter referred to as a
``mass proxy'' --- that can offer an accurate measure of cluster
masses; and (3) the existence of a well-calibrated relationship
between the mass proxy and the actual mass of the cluster.  Of these,
we shall focus our attention on the latter two since at present, the
effective use of clusters as cosmological probes is primarily limited
by systematic errors in the estimates of the true mass of the cluster
\citep{Henry09,Vikhlinin09,Mantz10}.

One of the first---and still among the most commonly used---mass
proxies is the "hydrostatic mass estimate", derived from X-ray
observations under the assumption that the clusters are spherically
symmetric and that the hot, diffuse, X-ray emitting gas in galaxy
clusters is in thermal pressure-supported hydrostatic equilibrium
(HSE).  Over the years, mismatch between hydrostatic mass estimates
and mass estimates derived by alternate means have led a number of
researchers to question the use of this proxy
\citep[e.g][]{MiraldaEscude95,Fischer97,Girardi97b,Ota04}.  Recent
studies suggest that the HSE masses of relaxed clusters are subject to
a systematic 10\%-20\% underestimate which grows to 30\% or more for
unrelaxed systems \citep{Arnaud07,Mahdavi08,Lau09}.  Numerical
simulation studies suggest that this bias is due to incomplete
thermalization of the hot diffuse intracluster medium (ICM)
\citep{Evrard90,Rasia06,Nagai07,Shaw10,Rasia12}.

Concerns with the HSE mass estimate have renewed interest in
identifying more well-behaved mass proxies that can give unbiased
estimates of the cluster mass.  One example of such an X-ray mass
proxy is $Y_X$, the product of the gas mass $M_g$ and ICM temperature
$T_X$ within a given aperture \citep{Kravtsov06}.  In numerical
simulation studies, this pressure-like quantity has been shown be a
much better mass proxy and has been successfully deployed in
measurements of cosmological parameters including the dark energy
equation of state \citep{Vikhlinin09a,Vikhlinin09}. More recently, the
gas mass $M_g$ has also emerged as a mass proxy with similar
predictive power to $Y_X$ \citep{Okabe10,Mantz11}. Success in tests
involving simulated clusters is necessary but far from sufficient.  At
present, numerically simulated clusters capture only a fraction of the
physical processes that affect the intracluster medium in real clusters.

An alternative way of independently testing the validity of the
individual mass proxies is via multiwavelength observations.
Specifically, comparisons of X-ray proxies and weak gravitational
lensing masses ($M_L$) are particularly interesting given the fact
that gravitational lensing provides a \emph{total} mass estimate that
neither depends on baryonic physics nor requires any strong
assumptions about the equilibrium state of the gas and dark matter,
and which can be determined over a wide range of spatial
scales. However, lensing measures the projected (2D) mass and
converting this to a unprojected (3D) mass has the effect of adding an
amount of scatter that is related to the geometry of the mass
distribution, its orientation along the line of sight, and projection
of extra-cluster mass along the line of sight \citep{Rasia12}.  In extreme cases,
these effects can result in an under- or over-estimate of the cluster
mass of as much as a factor of 2 \citep{Feroz12}, depending on the
specific technique used.
%(NEW REF: e.g. Feroz, F., Hobson,
%M.P., 2012, arXiv e-print 1108.2202, accepted for publication in
%MNRAS)

In this work, we employ a technique that achieves a low systematic
weak lensing mass bias of 3-4\%, thanks to the procedure described in
detail in \cite{Hoekstra12}. This bias level is lower than the 5-10\%
that is usual for numerical simulations, which also have a typical
scatter of $20\%-30\%$
{\protect\citep{Becker11,Bahe12,Rasia12,High12}}; \hlb{the actual
  amount of bias depends on the range of physical radii used in the
  weak lensing analysis.}

At any rate, weak lensing masses are, at present, the best measures of
cluster mass and very well suited for use in calibrating the different
mass proxies and identifying the best one of the lot.  Moreover, the
study of the relationship between the weak lensing mass estimate and
an observable mass proxy can potentially yield important insights into
the physics at play within cluster environments.  These are the goals
of the present paper.

To facilitate our study, we have assembled a sample of galaxy clusters
named the Canadian Cluster Comparison Project\footnote{Not to be
  confused with the Chandra Cluster Cosmology Project \protect\citep{Vikhlinin09}, which forms an
  identical acronym.}. We describe this sample in \S\ref{sec:data}. In
the present study, we restrict ourselves to studying the relationships
between weak lensing mass determinations and the mass proxies derived
jointly from \emph{Chandra} and \emph{XMM-Newton} observations. We use
the Joint Analysis of Observations (JACO) code base \citep{Mahdavi07}
to derive the mass proxies of interest from the X-ray data. JACO makes
maximal use of the available data while incorporating detailed
corrections for instrumental effects (for example, we model spatial
and energy variations of the PSF for both Chandra and XMM-Newton) to
yield self-consistent radial profiles for both the dark and the
baryonic components. Further details are given in \S\ref{sec:mass}.  In
\S\ref{sec:data} we summarize our data reduction procedure; in
\S{\ref{sec:mass}} we describe our mass modeling technique.  Our
quantitative measures of substructure, the luminosity-temperature
relation, the lensing mass-observable relations, and deviations from
hydrostatic equilibrium are discussed in \S\ref{sec:struct},
\S\ref{sec:lt}, \S\ref{sec:proxy}, and \S\ref{sec:hydro},
respectively. We conclude in \S\ref{sec:conclusion}. Throughout the
paper we take $H_0 = 70$ km/s/Mpc, $\Omega_M = 0.3$, and $\Omega_\Lambda = 0.7$.

\section{Sample and Data Reduction}
\label{sec:data}

\subsection{Sample Characterization}

The Canadian Cluster Comparison Project (CCCP) was established
primarily to study the different baryonic tracers of cluster mass and
to explore insights about the thermal properties of the hot diffuse
gas and the dynamical states of the clusters that can be gained from
cluster-to-cluster variations in these relationships.

For this purpose, we assembled a sample of 50 clusters of galaxies in
the redshift range $0.15 < z < 0.55$.  Since we wanted to carry out a
weak lensing analysis, we required that the clusters be observable
from the Canada-France-Hawaii Telescope (CFHT) so we could take
advantage of the excellent capabilities of this facility.  The latter
constraint restricts our cluster sample to systems at
$-15^{\circ}\;<\;{\rm declination}\;<65^{\circ}$.  We also required our
clusters to have an ASCA temperature $k_BT_X > 3$ keV.  To establish
cluster temperature, we primarily relied on a systematically reduced
cluster catalog of \cite{Horner01} based on ASCA archival data,
although in a few instances we used temperatures from other
(published) sources.

As a starting point, we scoured the CFHT archives for clusters with
high quality optical data suitable for weak lensing analysis,
including observations in two bands.  We identified \hlb{20} suitable
clusters observed with the CFH12k camera and with B and R band data
meeting our criteria.  Nearly half of these clusters were originally
observed as part of the Canadian Network for Observational Cosmology
(CNOC1) Survey \citep{Yee96,Carlberg96} and comprise the brightest
clusters in the {\it Einstein Observatory} Extended Medium Sensitivity
Survey (EMSS) \citep{Gioia90}.  Since the EMSS sample is known to have
a mild bias against X-ray luminous clusters with pronounced
substructure \citep{Pesce90,Donahue92,Ebeling00}, and we were
specifically interested in putting together a representative sample of
clusters that encompassed the spectrum of observed variations in
thermal and dynamical states, we randomly selected \hlb{30} additional
clusters from the Horner sample that met our temperature, declination
and redshift constraints and additionally, guaranteed that our final
sample fully sampled the scatter in the $L_X$ vs. $T_X$ plane.  Of
these systems, those without deep, high quality optical data were
observed with the CFHT MegaCam wide-field imager, using the $g^\prime$
and $r^\prime$ optical filter sets. \hlb{The resulting weak lensing masses for
  this sample are discussed in \protect\cite{Hoekstra12}}.

Our final sample comprises 50 clusters listed in Table 1.  All except
3 clusters have been observed by the {\it Chandra Observatory}.  These
three, plus 21 others, have also been observed by {\it XMM-Newton}.
Subsets of the CCCP cluster sample have been used in several prior
studies \citep{Hoekstra07,Mahdavi08,Bildfell08,Bildfell12}. \hly{The
  CCCP sample has served as the source for studies of individual
  clusters that are interesting in their own right, such as Abell 520
  and IRAS 09104+4109 \protect\citep{Mahdavi07,Jee12,OSullivan12}.}

In the left panel of Figure 1, we compare the distribution of the CCCP
clusters in the $L_X$---$T_X$ plane to those of two better
characterized samples of galaxies clusters: MACS \citep{Ebeling10} and
HIFLUGCS \citep{Reiprich02}, both of which employ well-defined
flux-based selection criteria based on the ROSAT All-Sky
Survey. HIFLUGS is on the average a lower redshift sample compared to
our CCCP sample, and MACS is on the average at a higher redshift.  The
samples have comparable scatter, suggesting that our CCCP sample is
not significantly more biased than HIFLUGCS or MACS, which have better
understood selection functions.  In the right panel of Figure 1, we plot
the distribution of the orthogonal scatter about the mean $L_X$--$T_X$
of the all three samples combined.  A KS test indicates that the three
distributions are statistically indistinguishable.  This confirms that
while the CCCP sample may not be a complete sample, it is a
representative sample in that it properly captures the scatter in the
$L_X$---$T_X$ and to the extent that these have physical origins, the
range of cluster thermal and dynamical states.

% Our set of 50 clusters is essentially an extension of
% \cite{Hoekstra07} and \cite{Mahdavi08}, who present a weak lensing and
% X-ray analysis of a sample of 20 clusters, respectively. For this work
% we assemble a non-complete yet representative sample incorporating a
% variety of dynamical states.  The parent sample was the systematically
% reduced cluster catalog of \cite{Horner01} based on ASCA archival
% data. The criteria for selection were visibility from Mauna Kea, ASCA
% temperature $k T_X > 4$ keV, and uniform scatter about the
% luminosity-temperature relation. 

% Because of reliance on archival ASCA data, the selection function for
% JACO/CCCP is not directly calculable. However, we set out to show that
% the JACO/CCCP sample is consistent with being a randomly selected
% subset of a more complete sample. in Figure \ref{fig:sample}, we
% compare the distribution of the clusters in this sample to two better
% characterized samples of galaxies clusters: MACS \citep{Ebeling10} and
% HIFLUGCS \citep{Reiprich02}, both of which employ well-defined
% flux-based selection criteria based on the ROSAT All-Sky
% Survey. HIFLUGS is on the average at a lower redshift, and MACS is on
% the average at a higher redshift, than our sample. We find that the
% samples have comparable scatter in the L-T plane, thus suggesting that
% our JACO/CCCP sample is not significantly more biased than HIFLUGCS or
% MACS, which have better understood selection functions.

\begin{figure*}
\resizebox{6.1in}{!}{\includegraphics{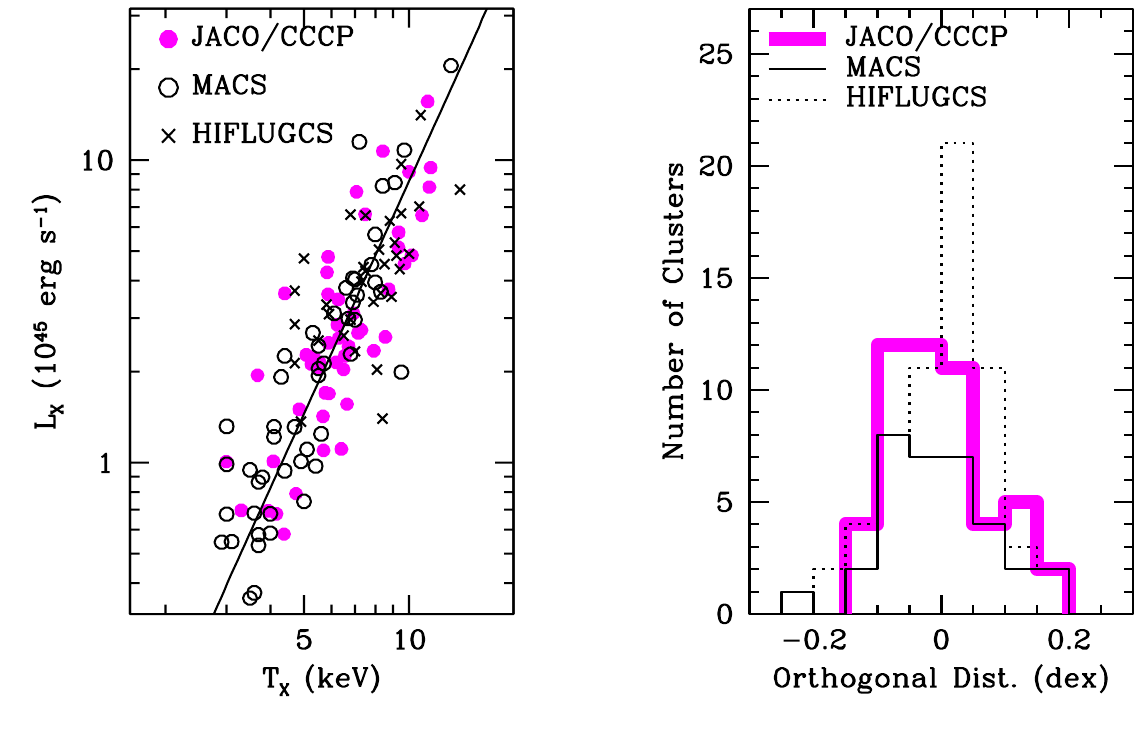}}
\caption{Comparison of the luminosity-temperature relationship for JACO/CCCP  sample (solid dots), HIFLUGS (open dots)
and MACS (stars) }.
\label{fig:sample}
\end{figure*}

\begin{figure*}
\begin{tabular}{cc}
\resizebox{3in}{!}{\includegraphics{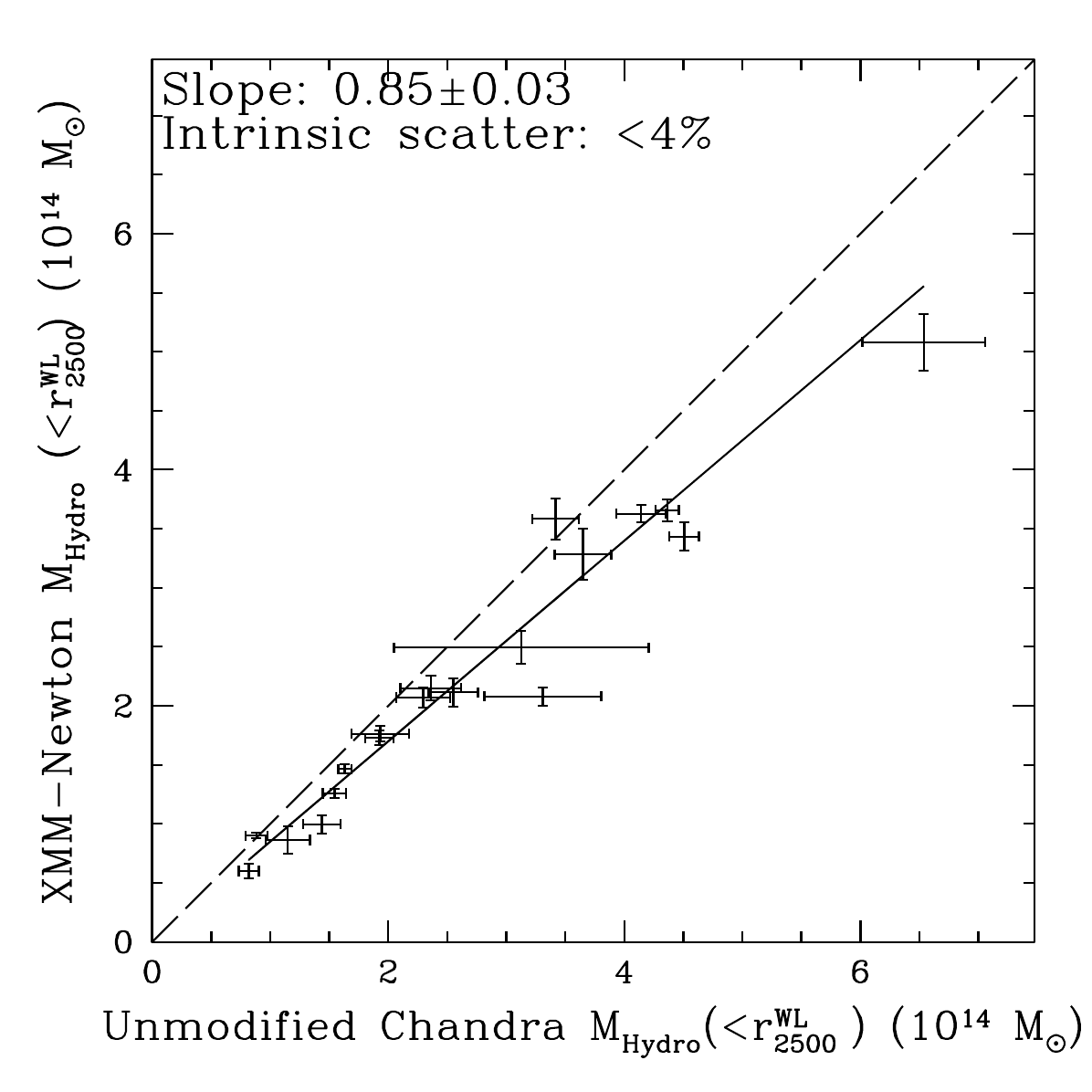}} &
\resizebox{3in}{!}{\includegraphics{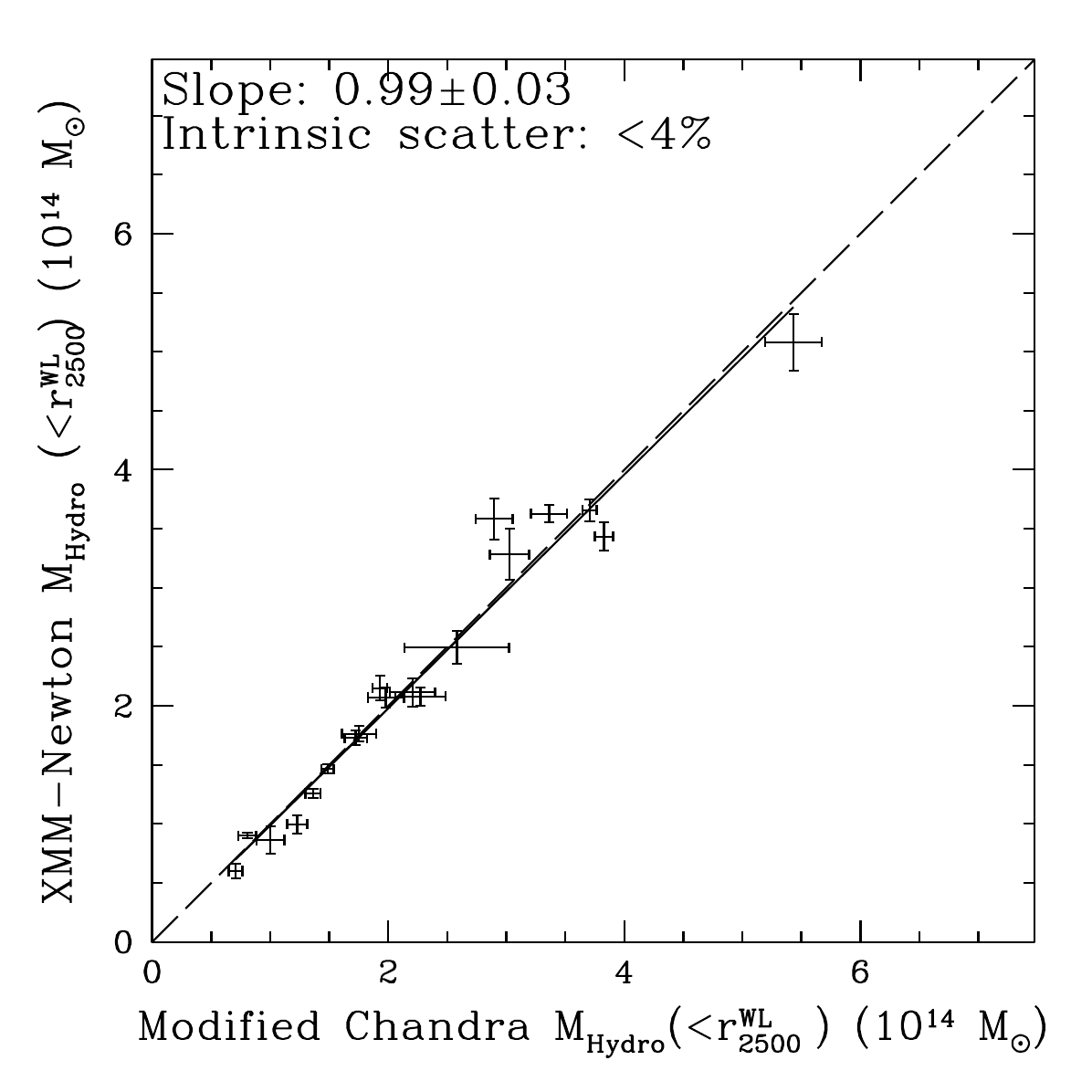}} \\
\resizebox{3in}{!}{\includegraphics{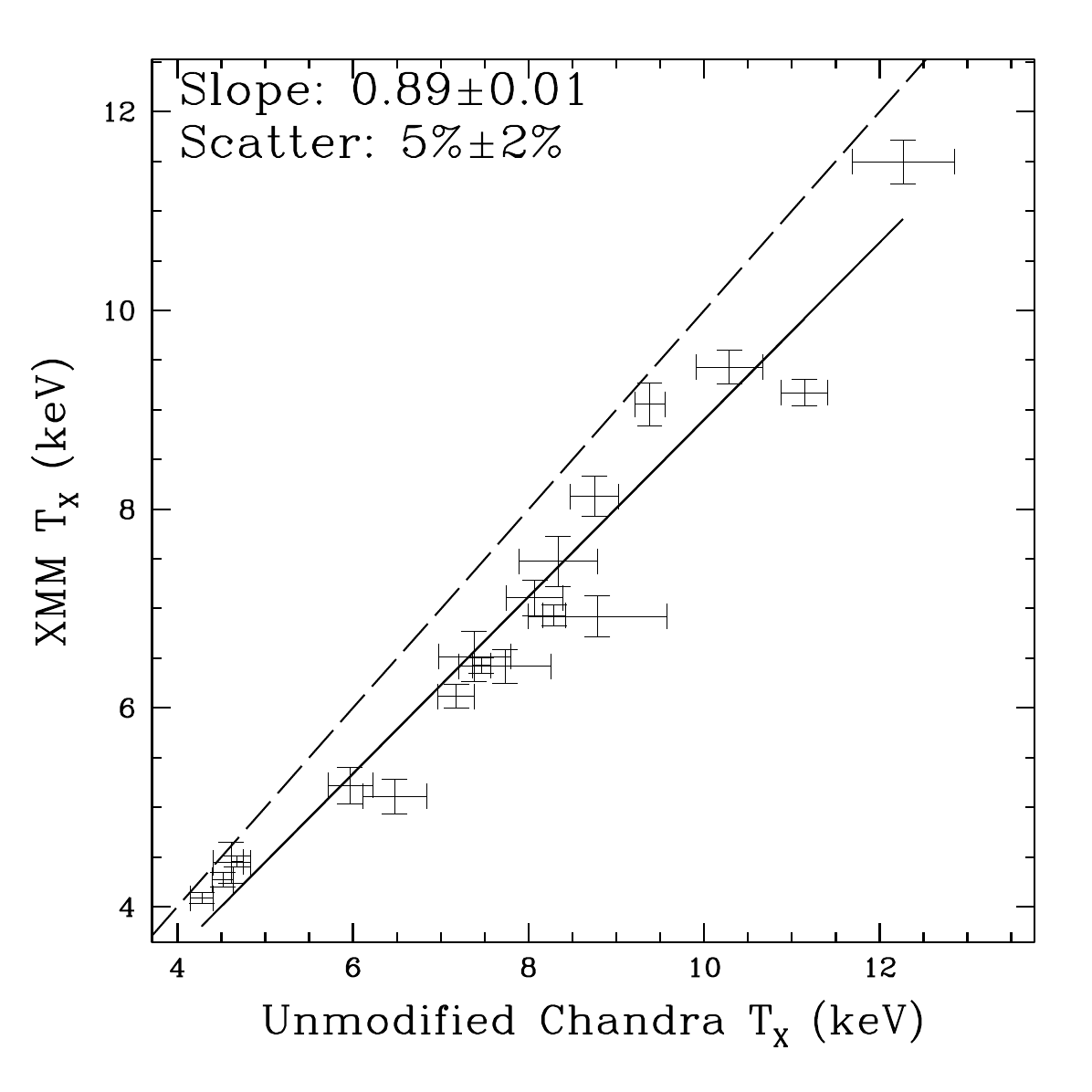}} & 
\resizebox{3in}{!}{\includegraphics{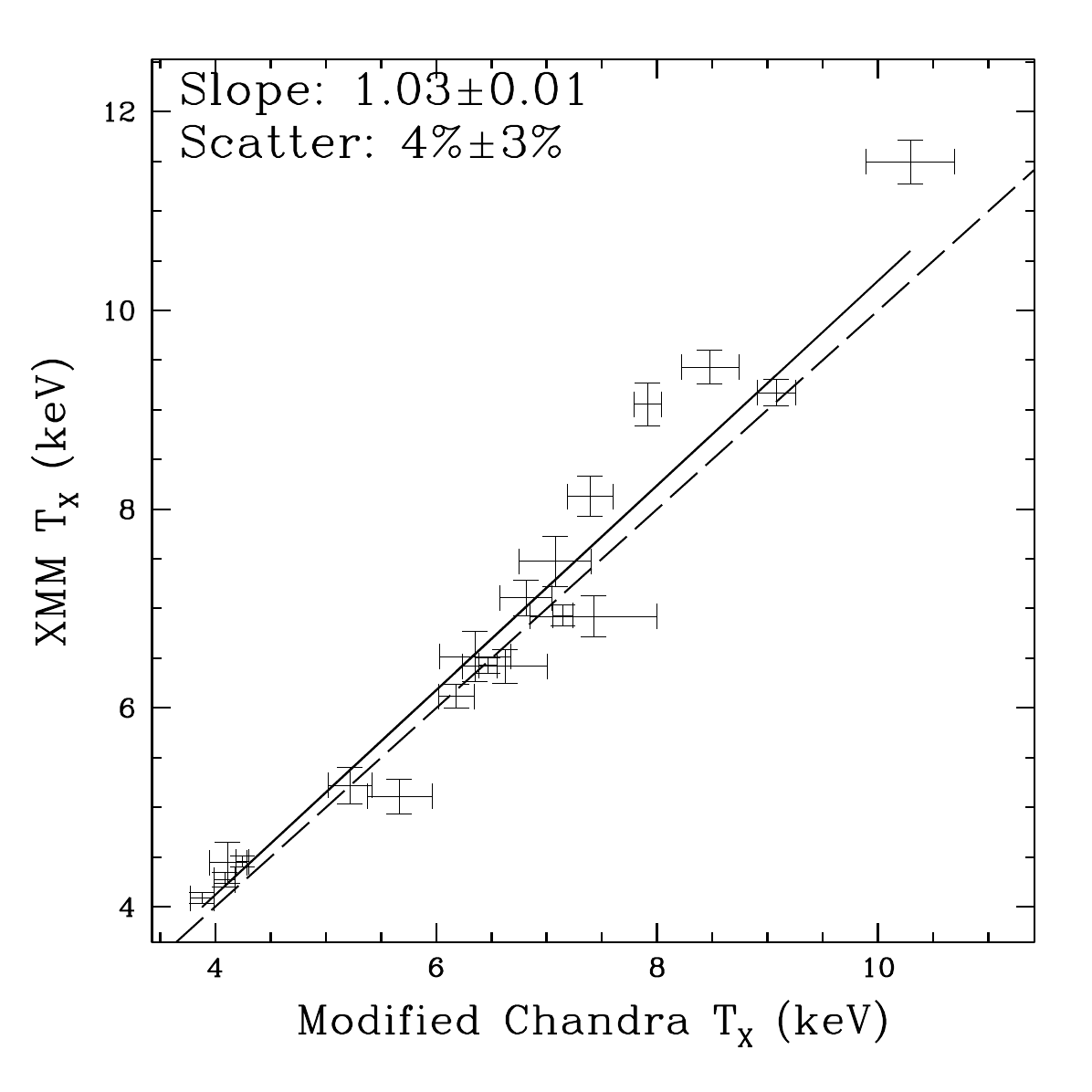}}\\
\resizebox{3in}{!}{\includegraphics{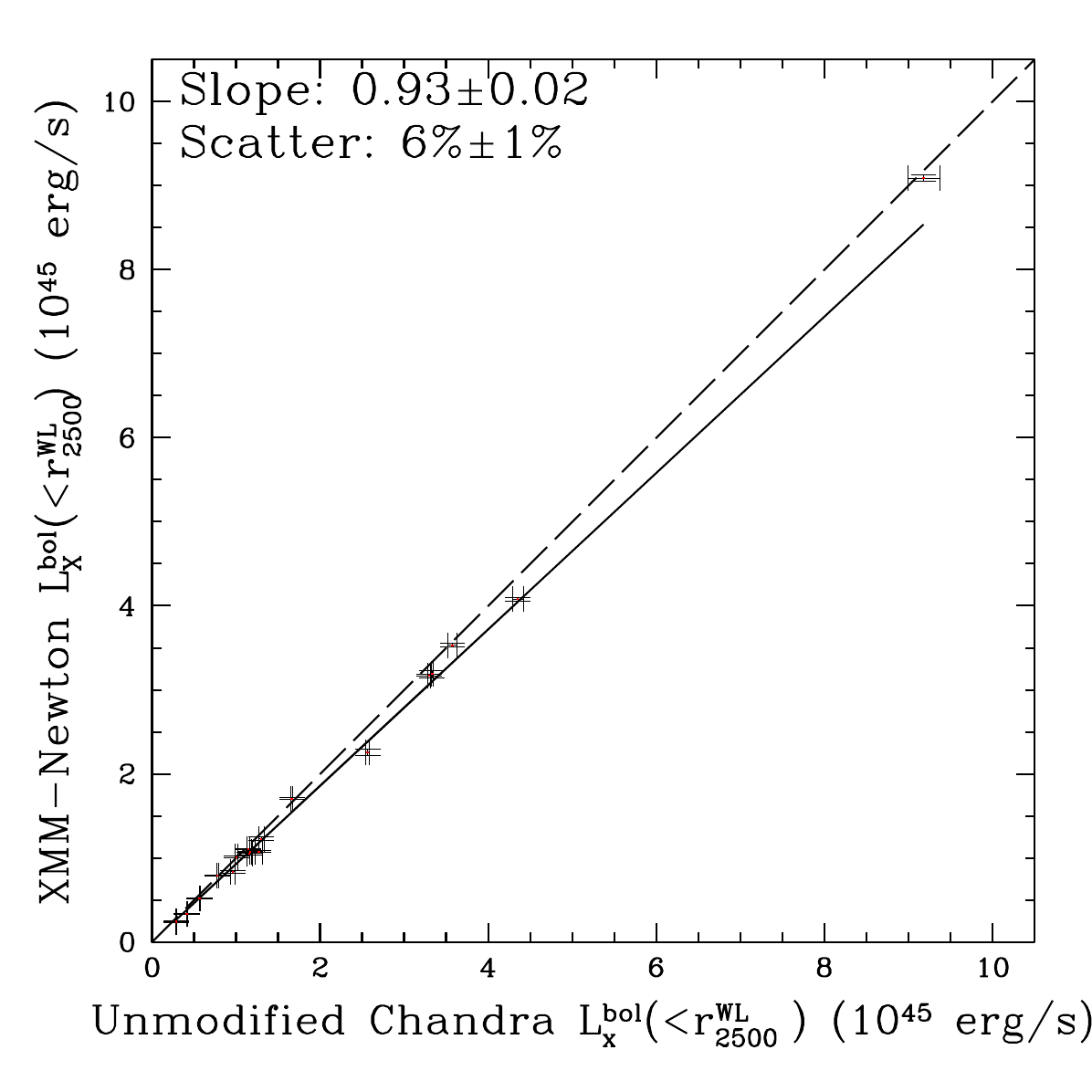}}& 
\resizebox{3in}{!}{\includegraphics{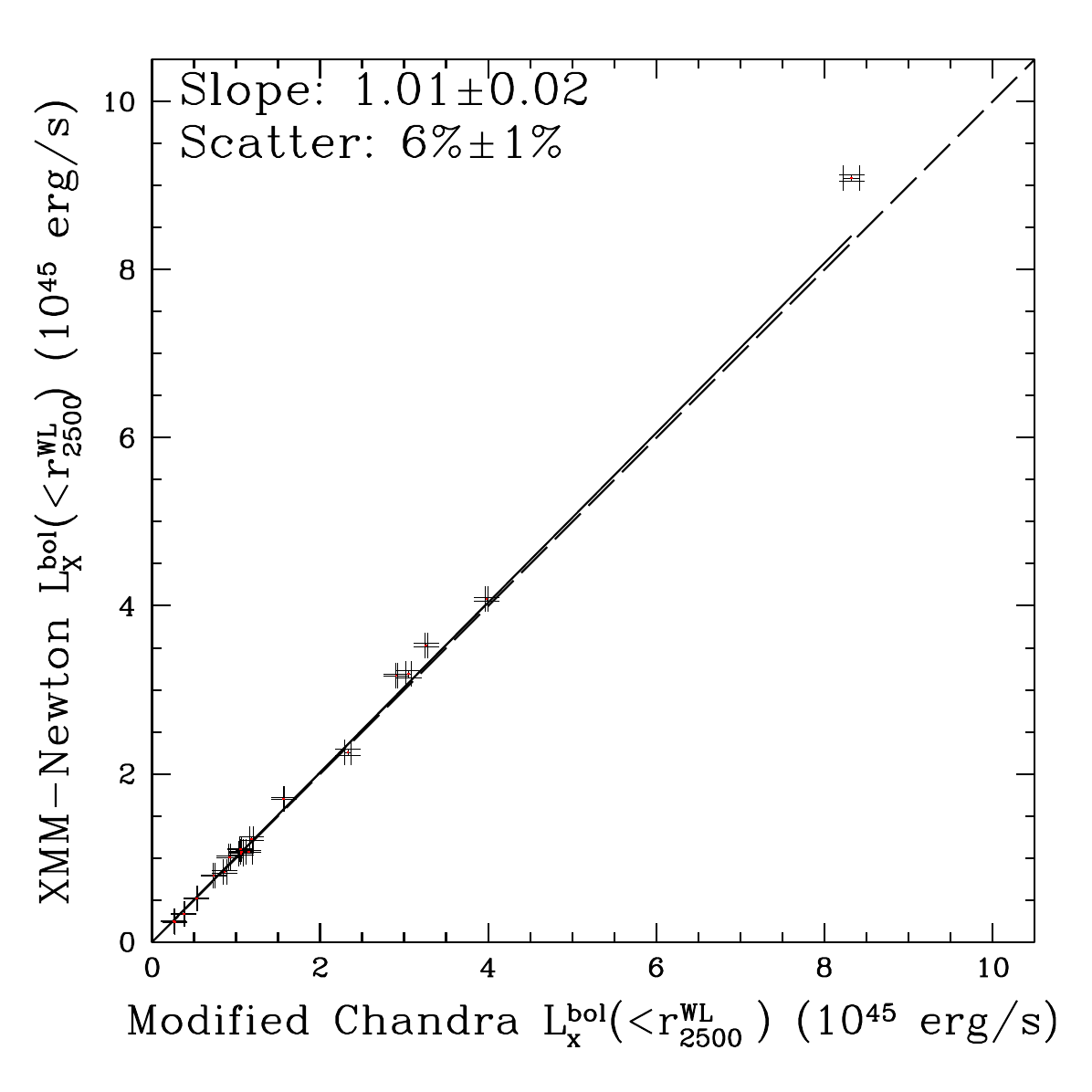}}
\end{tabular}
\caption{Comparison of XMM-Newton and Chandra X-ray masses
  (\emph{top}), temperatures (\emph{middle}), and bolometric X-ray
  luminosities (\emph{bottom}) within lensing $r_{2500}^{WL}$. The
  left-hand column shows the unmodified Chandra values, while the
  right-hand column shows the result of scaling the Chandra effective
  area by a power law in energy of slope $\zeta=0.07$, Chandra and
  XMM-Newton observables come into better agreement. The dashed
  line shows equality in all cases. \label{fig:crosscal}}
\end{figure*}

\subsection{Choice of density contrast}

For most of what follows, we study masses, temperatures, substructure
measures, and other thermodynamic quantities integrated within a
specific spherical radius. The choice of this radius is not obvious;
using fixed physical radii has the advantage of straighforwardness,
but the disadvantage that we would be probing characteristically
different regions of clusters as a function of masses. \hlb{Using fixed
overdensity radii $r_\Delta$ (defined such that $r_\Delta$ contains a
mean matter density of $\Delta$ times the critical density of the
universe at the redshift of the cluster)} is a better choice, but even here, the
value of $\Delta$ to use is not quite obvious. At the redshift of our
sample, X-ray data quality tends to be best around $r_{2500}$, but
most of the literature lists properties at $r_{500}$. Even after a choice of
$\Delta$, one must still decide whether to use the lensing or X-ray
value, since they are not guaranteed to agree.

We choose to standardize the \hlb{bulk of our discussion} on the
weak-lensing overdensity radius $r_{500}^{WL}$, because lensing masses
are likely to be \hly{more unbiased for non-relaxed clusters
  \protect\citep{Meneghetti10}}. \hly{For the most part, our results
  do not significantly change if we switch to X-ray $r_{500}$; one
  exception is the mass-temperature relation below, which tightens
  significantly with the switch. }  In \S\ref{sec:mpc}, we also
consider scaling relations with observables measured within fixed
physical radii, because these are more likely to be useful for
calibrating large data sets.

\subsection{Weak Lensing Overview}

The clusters in our sample were drawn from \cite{Hoekstra12}, which
contains a weak lensing analysis of CFH12k and Megacam data from the
Canada-France-Hawaii Telescope. We refer interested readers to
\cite{Hoekstra12} for details of the data reduction and weak lensing
analysis procedure. 

We base our lensing masses on the aperture mass estimates \citep[for
details see the discussion in \S3.5 in][]{Hoekstra07}. This approach
has the advantage that it is practically model
independent. Additionally, as the mass estimate relies only on shear
measurements at large radii, contamination by cluster members is
minimal. \cite{Hoekstra07} and \cite{Hoekstra12} removed galaxies that
lie on the cluster red-sequence and boosted the signal based on excess
number counts of galaxies. As an extreme scenario we omitted those
corrections and found that the lensing masses change by only a few
percent; for details see \citep{Hoekstra12}. Hence our masses are
robust against contamination by cluster members at the percent level.

The weak lensing signal, however, only provides a direct estimate of
the {\it projected} mass. To calculate 3D masses from the
model-independent 2D aperture masses we project and renormalize a
density profile of the form $\rho_\mr{tot}(r) \propto r^{-1}
(r_{200}+c r)^{-2}$ \citep{NFW}. The relationship between the
concentration $c$ and the virial mass is fixed at $c \propto
M_\mr{200}^{-0.14}/(1+z)$ from numerical simulations
\citep{Duffy08}. Hence, the deprojection itself, though well motivated
based on numerical simulations, is model dependent.  However, the
model dependence is weak---\hlb{20\% variations in the normalization
  of the mass-concentration relationship yield  $\approx 5\%$ variations
  in the measured masses} \citep[\S4.3]{Hoekstra12}. We
also note that the lensing analysis differs from the X-ray analysis in
that in the X-ray analysis, no mass-concentration relationship is
assumed (i.e., the concentrations and masses are allowed to vary
independently). We plan to address the effects of relaxing the lensing
mass-concentration relation in a future paper.

\subsection{X-ray Data Reduction}

We refer the reader to \cite{Mahdavi07} for details of the X-ray data
reduction procedure, which we briefly summarize and update here. We
use both Chandra CALDB 4.2.2 (April 2010) and CALDB 4.4.7 (December
2011). We also check our results against the latest CALDB (4.5.1) at
the time of writing. For XMM-Newton we use calibration files
up-to-date to January 2012; we also checked calibration files dating
as far back as April 2010. We detected no statistically significant
changes in the calibration files over this period for either Chandra
or XMM-Newton, except as detailed in \S\ref{sec:crosscal} below.

We follow a standard data reduction procedure. We use the software
packages CIAO (Chandra) and SAS (XMM-Newton) to process raw event
files using the recommended settings for each observation mode and
detector temperature. Where possible, we make event grade selections
that maximize the data quality for extended sources (including the
VFAINT mode optimizations for Chandra). We use the wavelet detection
algorithm WAVDETECT on exposure-corrected images to identify
contaminating sources; we masked out point and extended sources using
the detected wavelet radius. Each masking was checked by eye for
missing extended sources or underestimated masking radii.

The bulk of the X-ray background consists of a particle component
which bypasses the mirror assembly, plus an astrophysical component
that is folded through the mirror response. To remove the particle
background we match the 8-12 keV photon count rate from the outer
regions of each detector to the recommended blank sky observations for
each detector, and then subtract the renormalized blank-sky spectra.
What remains is the source plus \hly{an} over- or under-subtracted
astrophysical background, plus in some cases residual particle
background. All these residual backgrounds are modeled jointly with
the spatially resolved ICM model spectra, and their parameters
marginalized over for the final results.

To extract spatially resolved spectra, we find the surface brightness
peak in the Chandra image (if available) or XMM-Newton image (if
Chandra is not available). We then draw circular annuli that contain a
minimum of 1500 background-subtracted photon counts; where both
Chandra and XMM-Newton data are available, the annuli are taken to be
exactly the same for both sets of observations, with the minimum count
requirement being imposed on the Chandra data (for photons within
8$\arcmin$) or XMM-Newton data (for photons outside 8$\arcmin$). We
then compute appropriately weighted ancilliary response files (ARF)
and redistribution matrix files (RMF) for each spectrum, and subtract
appropriately scaled particle background spectra. We emphasize that
all spectra for each cluster undergo a simultaneous joint fit using a
forward-convolved spectral model of the entire cluster, so that the
choice of 1500 background-subtracted counts is not a
sensitivity-limiting factor. That is to say, in no case is a single
measurement derived from a single spectrum of 1500 counts, but rather
such spectra are fit together in large batches on a cluster-by-cluster
basis.

The detailed properties of the sample, including global X-ray
temperatures and bolometric X-ray luminosities, masses, and
substructure measures are listed in tables
\ref{tbl:sample} and \ref{tbl:data}.

\input{table1}

\label{sec:mass}
\subsection{X-ray Mass Modeling}

Here we summarize and update the modeling procedure of
\cite{Mahdavi07}, in which the cluster is spherically symmetric and
that the gas is in thermal pressure supported hydrostatic equilibrium
within the cluster potential. The essence of the technique is to
directly compare the observed spatially resolved spectra with model
predictions. For a spectrum observed in an annulus with inner and
outer radii $R_1$ and $R_2$, the model is
\begin{equation}
L_\nu = \int_{R_1}^{R_2} 2 \pi R dR \int_{R}^{r_\mr{max}} n_e n_H
\Lambda_\nu[T(r),Z(r)] \frac{2 r dr}{\sqrt{r^2-R^2}}
\end{equation}
 where $r$
denotes unprojected radius, $R$ denotes projected radius, $r_\mr{max}$
is the termination radius of the X-ray gas (taken to be $r_{100}$ in
this paper), $\Lambda_\nu$ is the frequency-dependent cooling function
which is a function of temperature $T$ and metallicity $Z$, and $n_e$
and $n_H$ are the electron and hydrogen number density, respectively.

One feature of the above method is that the unprojected temperature
profile is calculated self-consistently assuming hydrostatic
equilibrium of assumed gas and dark matter density profiles.  As a
result, we never have to specify or fit a temperature profile;
temperature is merely an intermediate ``dummy'' quantity connecting
the gas and dark matter mass distributions to the X-ray spectra. This
avoids subjective weighting involved in the fitting of 2D projected
temperature profiles \citep{Mazzotta04,Rasia05,Vikhlinin06b}, which
are more difficult to correct for the effects of PSF distortion.

% The model spectrum for each 3D spherical shell is generated in two
% separate ways:

% \begin{itemize}
% \item ``Hydrostatic'' fits: using highly flexible, parametric forms
%   for the gas, metallicity, and total matter distribution in the
%   cluster, the temperature in each shell is calculated using the
%   equation of hydrostatic equilibrium;

% \item ``Unconstrained'' fits: each spherical shell is assumed to have
%   its own constant temperature, density and metallicity. This method
%   is akin to the ``deprojection'' method in the literature, except
%   that all the thermodynamic variables of all the shells are fit and
%   forward-projected at the same time (i.e., for $N$ annuli, $3 N$
%   thermodynamic parameters are fit simulataneously).
% \end{itemize}

% Thus the ``hydrostatic'' fits yield a parametric estimate of the mass
% profile, while the ``unconstrained'' fits are free from the assumption
% of hydrostatic equilibrium but yield no information on the gravitating
% mass. In both cases, plasma pressure ($n_e k T$) and pseudo-entropy ($
% k T / n_e^{2/3}$) are also calculated as derived quantities for each
% 3D shell.  Any discrepancy in the ``hydrostatic'' fit vs. the
% ``unconstrained'' fit could be due to departures from equilibrium, to
% an incorrect gravitating mass distribution model, or to both.

\subsection{Parameters of the Hydrostatic Model}

The hydrostatic model assumes a flexible spherical electron
density distribution
\begin{eqnarray}
n(r) & = & n_{e_0} \left(\frac{r}{r_{x_0}}\right)^{-\alpha} B(r,r_{x_0},\beta_0)+\\
\nonumber & & n_{e_1} B(r,r_{x_1},\beta_1)+ n_{e_2} B(r,r_{x_2},\beta_2) 
\end{eqnarray}
where the familiar ``beta'' model is
\begin{equation}
B(r,r_{x_i},\beta_i) = \left(1+\frac{r}{r_{x_i}} \right)^{-\frac{3 \beta_i}{2}}
\nonumber
\end{equation}
In other words, the gas mass profile consists of a fully general
triple ``beta'' model profile, where the first beta model is further
allowed to be multiplied by a single power law. The metallicity distribution
is modeled as \citep[e.g.][]{Pizzolato03}
\begin{equation}
\frac{Z}{Z_\odot} = Z_0 \left(1 + \frac{r^2}{r_z^2}\right)^{-3 \beta_z} 
\end{equation}
with $r_Z$, $\beta_z$, and $Z_0$ free parameters. Finally, the total
mass distribution (baryons and dark matter) are modeled as a
\cite{NFW} profile:
\begin{equation}
 \rho_\mr{tot} = \frac{M_0}{r (c r + r_\Delta)^2}
\end{equation}
 where $M_0$ is
the normalization, $c$ is the halo concentration, and $r_\Delta$ is
the overdensity radius (see above). These are also free parameters,
except that rather than fitting $M_0$, we fit $M_\Delta$---the mass
within $r_\Delta$---as the normalization constant (because there is a one-to-one
relationship between $M_0$ and $M_\Delta$).

In general, some of the above parameters are better determined than
the others. For example, the inner slope of the gas density
distribution, $\alpha$, is always well measured (with a typical
uncertainty of $\pm 0.1$, and follows the well-known trend
\citep[e.g][]{Sanderson10} that low central entropy clusters have
steeper inner profiles, $\alpha \approx 0.5$, whereas high entropy
clusters have flatter profiles, $\alpha \approx 0$). The central
metallicities are similarly well-determined. On the other hand,
quantities such as the slopes and core radii of multiple $\beta$-model
profiles---such as $\beta_2$ and $\beta_3$ or $r_Z$ and
$\beta_Z$---frequently reveal significant degeneracies with each
other. In all cases, such degeneracies are properly marginalized over
using the Hrothgar Markov chain monte carlo procedure described in
\cite{Mahdavi07}, and the one-dimensional error bars in Table
\ref{tbl:data} always properly reflect any and all degeneracies among
the many parameters in this many-dimensional model.

\input{table2}

\label{sec:crosscal}

\subsection{Joint Calibration of Chandra and XMM-Newton Masses}

%% The Chandra and XMM-Newton X-ray Observatories have been active for
%% over a decade and have brought dramatic insights on the physical
%% processes governing clusters of galaxies. From cold fronts
%% \citep[e.g.][]{Vikhlinin01,Dupke03,Hallman04,Owers09} to shock fronts
%% \citep[e.g.][]{Markevitch02,Markevitch05,Russel10,Macario11,Gitti11},
%% Chandra's arcsecond angular resolution has revealed many
%% non-equilibrium processes with transformative effects on the
%% intracluster medium (ICM). Similarly, the large effective area and
%% spectroscopic sensitivity of XMM-Newton served as the strongest
%% motivator for the need for a central heating mechanism in cool-core
%% clusters \citep[e.g.][]{Peterson01,Sakelliou02,Tanaka06,Sanders11}.

Where available, we use both Chandra and XMM-Newton data for a
cluster.  This has several advantages: in the inner regions, Chandra
is able to resolve the cluster cores well; while XMM-Newton's wider
field of view yields better coverage of the outer regions of the
cluster. The simultaneous coverage of intermediate regions helps
constrain residual backgrounds following blank sky subtraction.

When combining Chandra and XMM-Newton data, cross-calibration is a
significant issue. In general, there are slight differences among the
responses of the Chandra ACIS and the XMM-Newton pn, MOS1, and MOS2
detectors. Even after over a decade in flight, the source of these
differences has not been conclusively identified. Typically,
comparisons show that Chandra temperatures are $5-15\%$ higher
\citep{Snowden08,Reese10}. The most recent calibration tests
\citep{Tsujimoto11} use the G21.5-0.9 pulsar (which is fainter than
the usual source, the Crab nebula, and hence not subject to detector
pileup). \cite{Tsujimoto11} find that the XMM-Newton pn has a 15\%
lower flux in the 2.0-8.0 keV energy band compared to the Chandra
ACIS-S.  This confirms an earlier finding by \cite{Nevalainen10} who
found similar results. Lower hard band flux naturally leads to lower
X-ray temperatures when 0.5-2.0 keV photons are also included. This
primarily affects masses for which spectral line emission is not
dominant (i.e., in hot, k T $ > 4$ keV clusters). \hly{It is at this
  point unknown where the source of the disagreement lies and which
  instrument is better calibrated}.

\begin{figure*}
\begin{center}
\resizebox{4.3in}{!}{\includegraphics{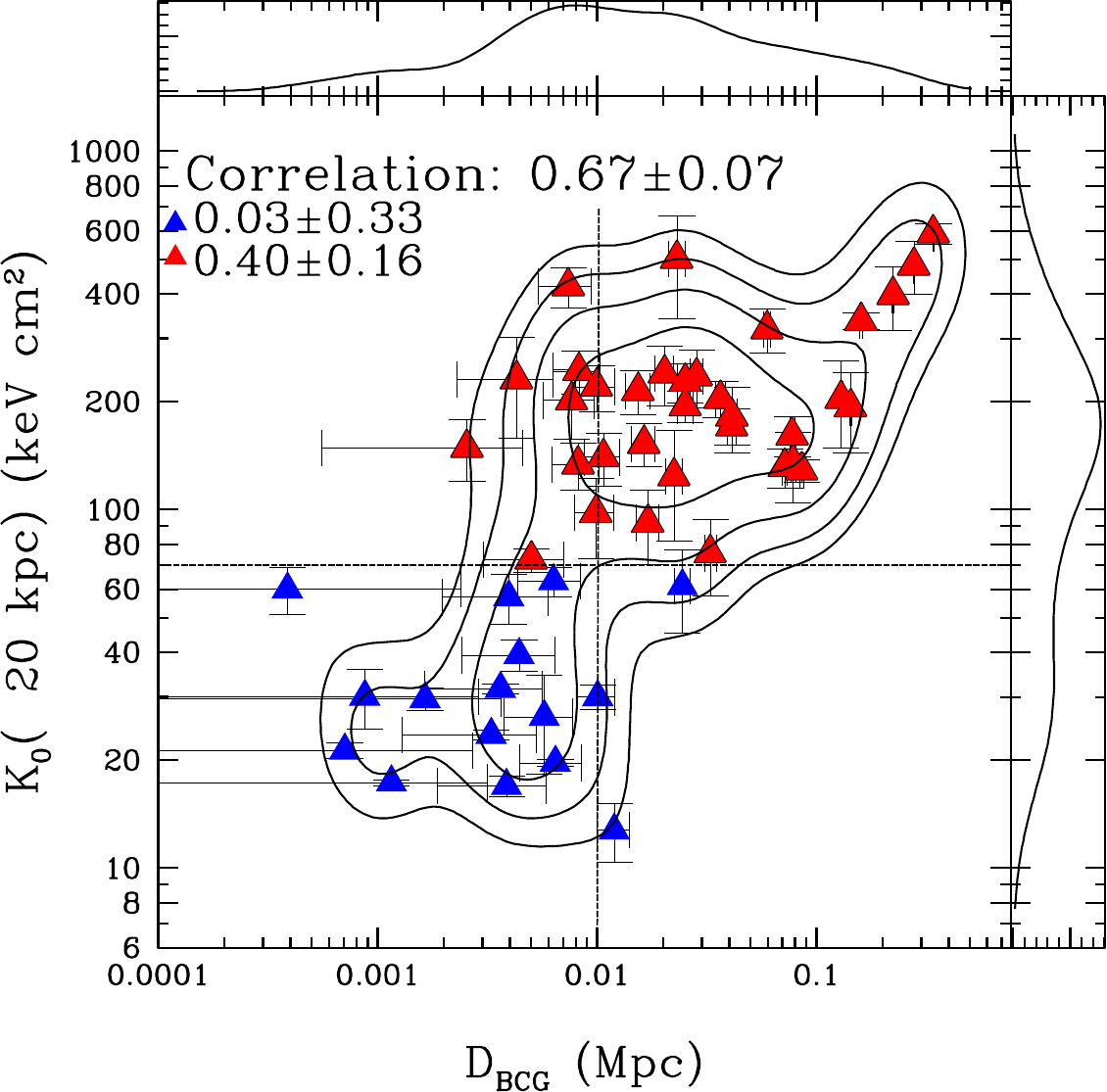}}
\end{center}
\caption{Bimodality in the joint distribution of BCG offset and
  central entropy; contours show lines of constant probability density
  after the points have been smoothed with a 0.25 dex Gaussian. The
  top and right axes show the 1D probability density for central
  entropy and BCG offset. Blue triangles show cool-core clusters and
  red triangles show non-cool-core clusters. The horizontal thin line
  shows our chosen division between cool-core and non-cool-core
  systems, while the vertical line shows our chosen division between
  low BCG offset and high BCG offset systems. \label{fig:k0bcg}}
\end{figure*}

\begin{figure*}
\begin{tabular}{cc}
\vspace*{-0.2in}\resizebox{3.5in}{!}{\includegraphics{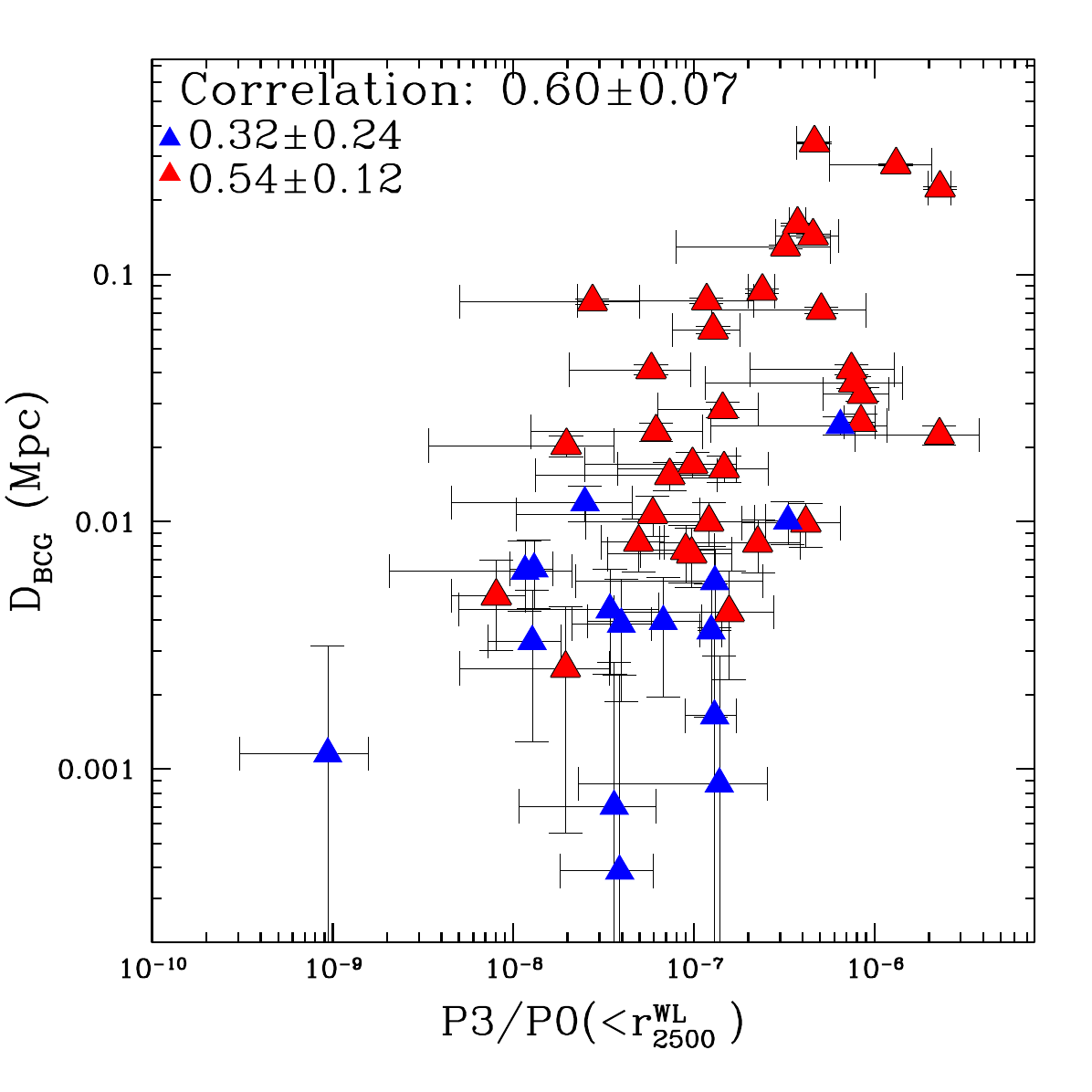}} &
\resizebox{3.5in}{!}{\includegraphics{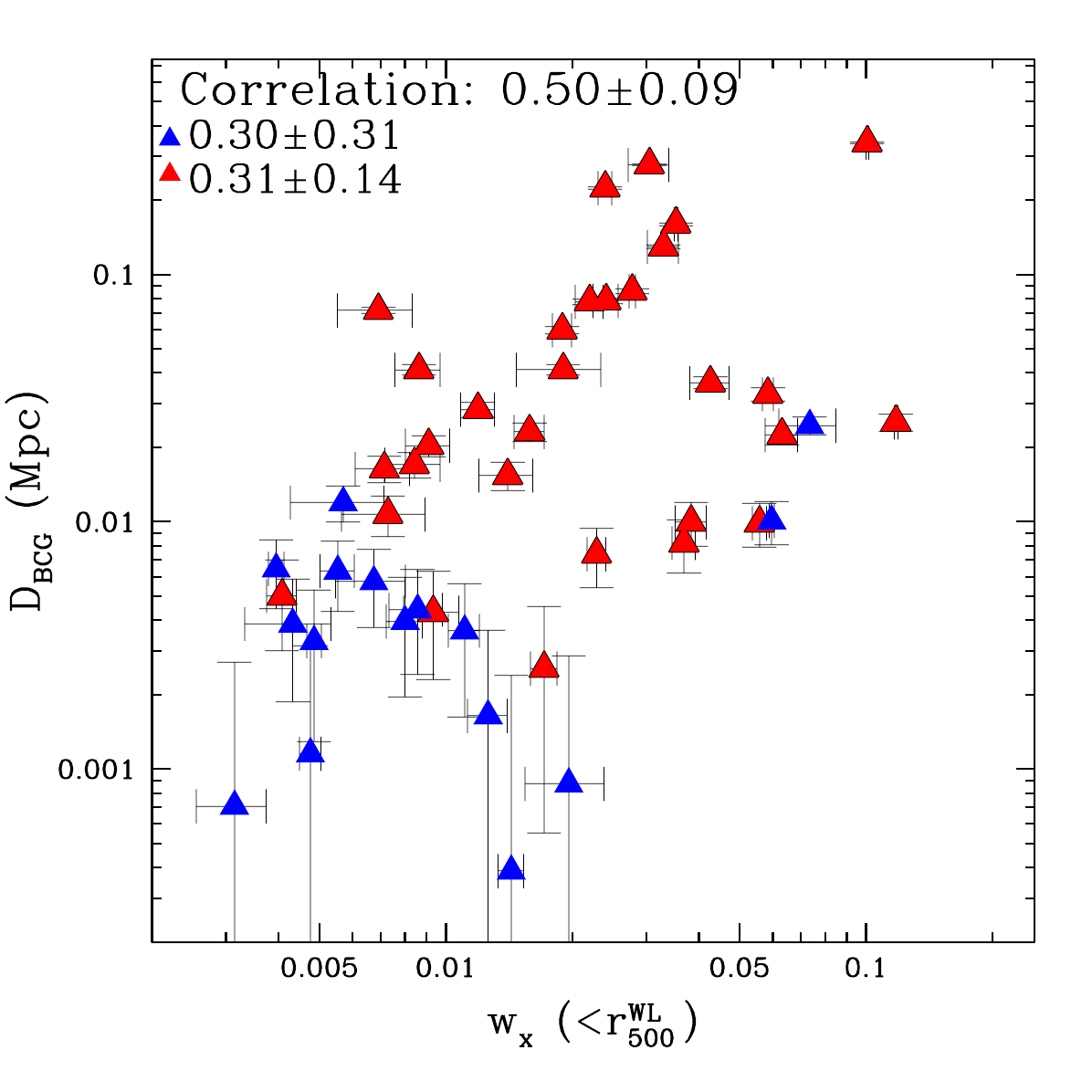}} \\
\vspace*{-0.2in}\resizebox{3.5in}{!}{\includegraphics{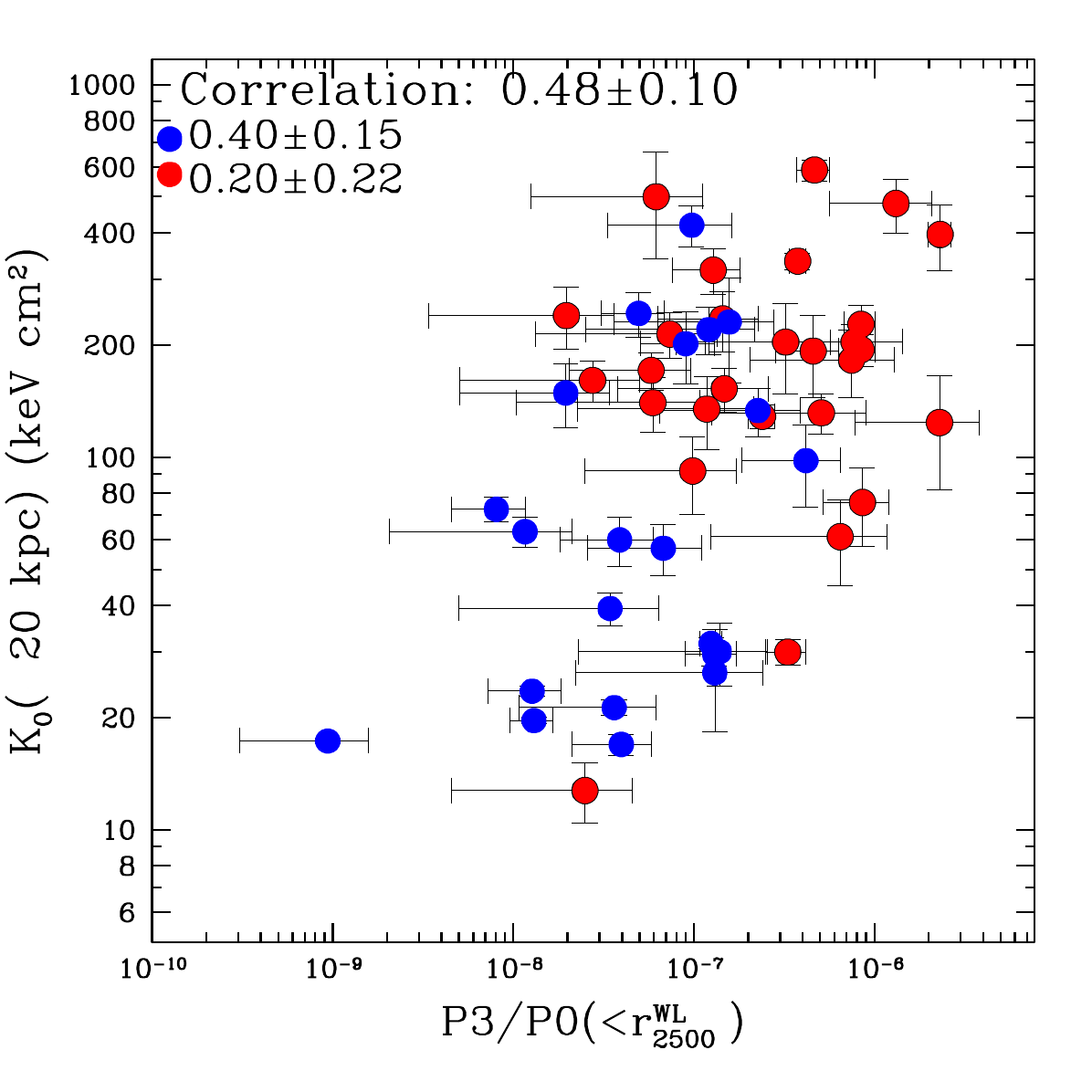}} &
\resizebox{3.5in}{!}{\includegraphics{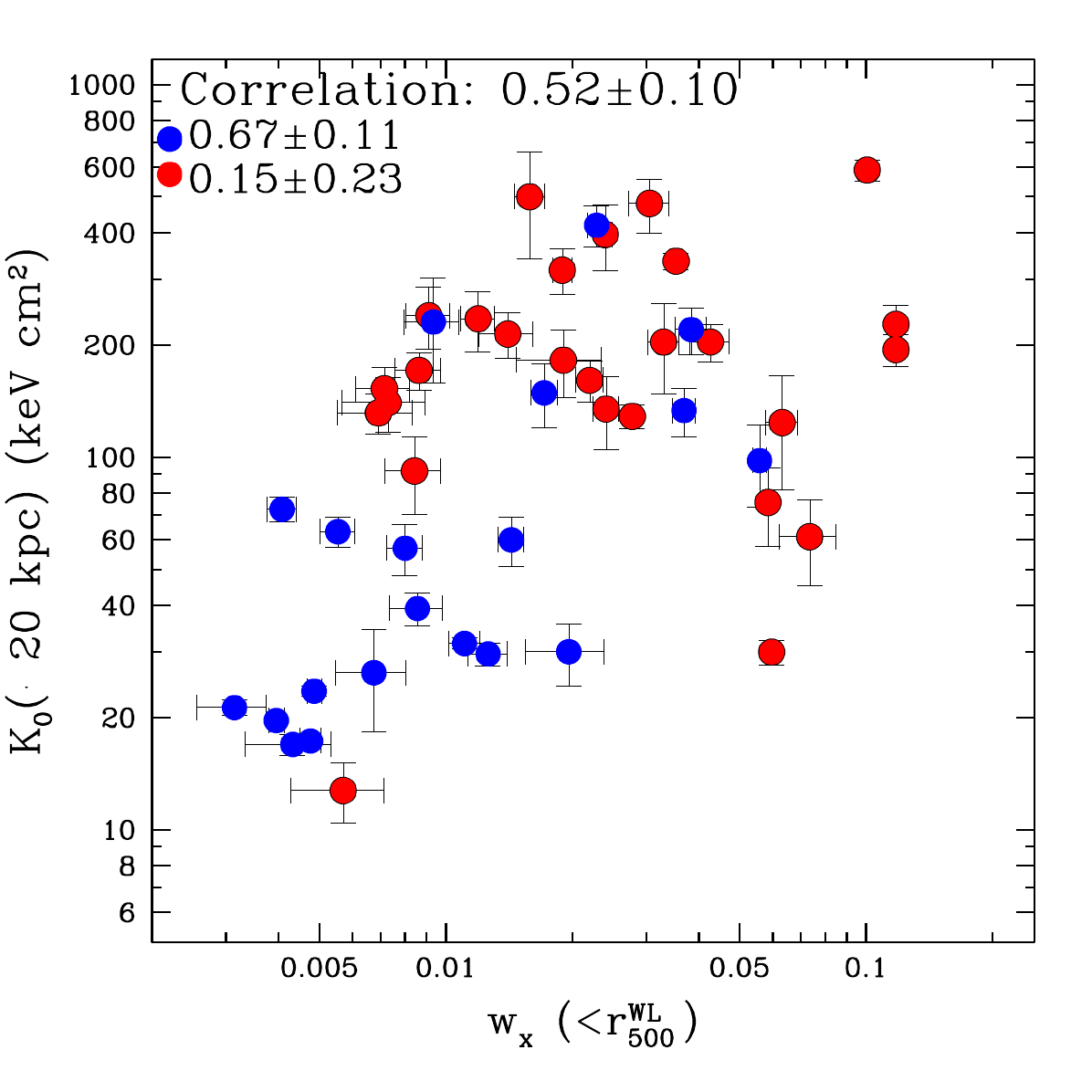}} \\
%\vspace*{-0.2in}\resizebox{3.1in}{!}{\includegraphics{p3p0-wx.pdf}} &
%\resizebox{3.1in}{!}{\includegraphics{p3p0-bcg.pdf}} \\
\end{tabular}
\caption{Correlation of four different substructure measures (central
  entropy $K_0$, BCG offset $D_\mr{BCG}$, X-ray centroid variance
  $w_X$, and $P3/P0$ power ratio) against each other. Blue triangles
  show cool-core clusters and red triangles show non-cool-core
  clusters;  blue circles show
  low-BCG-offset systems and red circles show high-BCG-offset
  systems. \label{fig:substruct}}

\end{figure*}

Figure \ref{fig:crosscal} shows the X-ray mass measured within lensing
$r_{2500}^{WL}$ for the 19 clusters in our sample which contain both
Chandra and XMM-Newton data.  Shown are the results for CALDB 4.2.2
(April 2010). We also checked CALDB 4.4.7 (December 2011) and CALDB
4.5.1 (June 2012). The calibration for our sample changed little
during this period, and in all three cases, we find that Chandra
masses are higher than XMM-Newton masses by roughly 15\%.  All
observations were recorded prior to 2010, and taken as a whole, the
change in the Chandra masses of these systems is not statistically
significant between CALDB 4.2.2 and 4.4.7. % We do find that
% the intrinsic scatter in the Chandra vs.  XMM-Newton masses is
% slightly higher in the 2010 CALDB than for the 2011 CALDB.
% We find that the masses are slightly more discrepant for the 2011
% CALDB.
We adopt the 2010 CALDB for the remainder of this paper, stressing that
any changes to our results would be well within the statistical errors
presented were we to switch to a different calibration release.

To be able to combine Chandra and XMM-Newton data, one must
first ensure that they are consistent. We find that the following
simple cross-calibration prescription is able to bring the data
into self-consistency:
\begin{equation}
 A^\mr{corrected}_\mr{CXO}(E) =  A_\mr{CXO}(E) \left(\frac{E}{\mr{keV}} \right)^\zeta
\end{equation}
where $\zeta =0$ gives the unmodified CALDB area, and $\zeta > 0$ has
the effect of down-weighting the high energy effective area of
Chandra.  We find that setting $\zeta = 0.07$ brings Chandra and
XMM-Newton masses into agreement as shown in Figure
\ref{fig:crosscal}. In either case, the intrinsic scatter between
Chandra and XMM-Newton mass measurements at these fixed radii is
certainly less than 10\%, though inconsistent with zero at the
2$\sigma$ level.

Figure \ref{fig:crosscal} also shows that the integrated X-ray
temperatures and luminosities within lensing $r_{2500}$ are also
improved by our suggested calibration. The discrepancy between 
unmodified Chandra X-ray temperatures and XMM temperatures is
roughly the same as the discrepancy in hydrostatic masses. The
bolometric X-ray luminosities are also in better agreement as a
result of the effective area re-calibration, though in this case
the original discrepancy is less severe than in the realm of the
spectroscopic temperature.

We chose to modify the Chandra effective area, and not the XMM-Newton
effective area, based on the fact that the XMM-Newton has exhibited
the least variation over the years, whereas Chandra has enacted larger
10-15\% changes in its effective area calibration historically. 
\hly{ We note that had we modified the XMM-Newton effective
  area to match that of Chandra, then we would have found in what
  follows that clusters no longer exhibit self-similar behavior and
  that (a) those with obvious substructure would be the ones whose masses
  calculated assuming hydrostatic equilibrium would agree with their weak
  lensing masses,} \hlb{and (b) that clusters with cool cores would have
  hydrostatic masses greater than their weak lensing masses.} This
uncertainty in the telescopes' effective areas must be viewed as
a fundamental systematic limitation of X-ray astronomy at least as
related to cluster science.

\subsection{Online Data  and Regression Tool}

All data and analysis software used for this paper are available
online at \url{http://sfstar.sfsu.edu/cccp}.  Fits of scaling
relations (i.e., the modeling of linear or power law relationships
among measured quantities) are complicated by the fact that error in
both coordinates makes ordinary $\chi^2$ analysis invalid. A detailed
treatise of recent developments in the theory behind modeling 2D data
with errors in both coordinates appears in \cite{Hogg10}. These
techniques allow the simultaneous estimation of slope, intercept, and
intrinsic scatter in such relations. We implement the methods of \cite{Hogg10}
at the data website for this article.

\section{Measures of non-relaxed status}
\label{sec:struct}

The gas in all clusters of galaxies exhibits some degree of deviation
from an idealized smooth, triaxial distribution. Such deviation could
come in terms of subclumping, asymmetry, or both. Its presence gives
some clue as to the nature of its evolutionary history; for example,
asymmetry could indicate either the beginning or the end of a merger
event; subclumps could either be recently accreted small groups of
galaxies, or surviving cold cores from recent mergers.

Despite this ambiguity, objective measures of substructure are helpful
in arriving at quantitative estimates of departure from
equilibrium. To begin, we employ two common and well-tested measures of
substructure: power ratios and centroid shift variance. Power ratios
are Fourier-space estimators of fluctuations in the overall cluster
surface brightness distribution, while the centroid shift is a measure
of the variance of the distance between the X-ray surface brightness
peak (which is always well defined) and the centroid (which in a
non-relaxed cluster often varies significantly as a function isophote
used for its estimation). We refer the reader to
\cite{Buote95,Poole06,Jeltema05,Jeltema08} and \cite{Boehringer10} for details
on the calculation of these estimators.

As further tracers of the relaxed or nonrelaxed state of a system, we
also consider the somewhat more straightforward measures, central
entropy and the X-ray to optical center offset. Low central entropies
indicate the presence of a cool core, which tend to be associated
(non-exclusively) with relaxed clusters. 
\hlo{We define the central entropy as:}
\begin{equation}
 K_0 \equiv K(20 \mr{kpc})
\end{equation}
\hlb{In other words, the central entropy is defined as the 
deprojected entropy profile evaluated at a radius of 20kpc from the
cluster center.}
% The inner radius, $0.02 r_{500}^{WL}$ is typically $\approx 20$ kpc for
% clusters, small enough to be well inside the cool central regions for
% the clusters that possess them. We find that if we use larger radii
% (such as $0.1$ or $0.15 r_{500}^{WL}$) the correlations we discuss in
% \S\ref{sec:corr} are not as significant.

Similarly, the distance between the brightest cluster galaxy (BCG) and
the X-ray surface brightness peak can be a good predictor of relaxed
state, with large shifts indicating ongoing or residual merger
activity \citep{Poole07}. \hlo{We measure this distance via simple
  astrometry on X-ray and optical images, and call it $D_\mr{BCG}$.}

% However, the extent of the merger-induced
% displacement should depend on the total mass of the merging subsystems;
% therefore, to correctly compare clusters of different masses, we 
% define the BCG offset as
% \begin{equation}
%  w_\mr{BCG} = \frac{D_{XO}}{r_{500}^{WL}}
% \end{equation}

One would expect relaxed halos to be more representative of idealized
halo growth models. Hence we expect scaling relations among the
various thermodynamic and dark matter parameters to be tighter for
clusters selected on the basis of the more well-behaved substructure
indicators. We also expect the most powerful substructure measures to
be correlated with each other.

\subsection{Correlations among measures of substructure}
\label{sec:corr}

We explore the possibility of whether our substructure measures show
inherent correlation. The presence of such correlations, particularly
when involving both X-ray and optical data, can serve as road maps
towards our goal of quantifying departures from equilibrium as
economically as possible. We use the Spearman's rank correlation
coefficient, with bootstrap resampling for determining $1\sigma$ 
uncertainties.

The relationship between central entropy and BCG offset is the most
significant correlation in our sample. This also happens to be the
most interesting correlation due to the relative ease of deriving
central entropy and BCG offset from observables. Figure
\ref{fig:k0bcg} shows that the two substructures measures appear to
form a two-peaked joint distribution, with low central entropy, low
BCG offsets in one corner, and high central entropy, high BCG offset
clusters in another. The dividing line is best seen as a curve with
equation
\begin{equation}
K_0 = 7 \mr{\,keV\,cm}^2 \left(\frac{D_\mr{BCG}}{\mr{Mpc}}\right)^{-1/2}
\end{equation}

The high correlation coefficient between $K_0$ and $D_\mr{BCG}$
appears to be due to bimodality: \hly{when we calculate the correlation
  coefficient separately for either cloud, we find that the
clouds individually do not contain significant internal correlation.}
Though the above formula offers the most clean separation between
the two clouds, most of the separation can be captured by imposing
cuts in entropy, or, somewhat less cleanly, \hly{in} BCG offset.

For this reason, throughout the rest of the paper, we introduce a
labeling system that represents cuts in these two most easily measured
substructure estimators. We use blue triangles to indicate $K_0< 70$
\hly{keV} cm$^2$ (``cool core systems'' or CC), and red triangles to
indicate $K_0 > 70$ \hly{keV} cm$^2$ (``non-cool-core
systems'' or NCC). \hlo{This nomenclature is based on the fact that of 70 keV cm$^{2}$ corresponds to a cooling time of $\approx 1.5$ Gyr; most cool core clusters
have central cooling times below this value.}

Similarly, we use blue circles to indicate systems with $D_\mr{BCG} <
0.01$ \hlo{Mpc} (``low BCG offset systems'') and red circles to
indicate $D_\mr{BCG} > 0.01$ \hlo{Mpc} (``high BCG offset systems'').

In Figure \ref{fig:substruct}, we look for inherent correlations among
the other various indicators of substructure. Strong correlations
exist between the BCG offset $D_\mr{BCG}$, the central entropy $K_0$,
the X-ray centroid shift $w_X$ at $r_{500}^{WL}$, and the P3/P0 ratio at
$r_{2500}^{WL}$ (in measuring the latter two, we cut out the central 0.15
$r_{500}^{WL}$ to avoid dilution of the signal by the cool
core). Interestingly, the P3/P0 ratio measured at $r_{500}^{WL}$ (instead
of $r_{2500}^{WL}$\hly{)} showed much larger scatter (presumably due to noise) and
proved much less tightly correlated with the other substructure
measures than the P3/P0 ratio at $r_{2500}^{WL}$.

In particular, it should be noted that \hlb{$P3/P0$ exhibits almost as
  strong a correlation with BCG offset as does central entropy},
though there is no evidence for bimodality. For non-cool-core
clusters, the \hlb{$P3/P0$} is \emph{\protect\hlo{significantly} more}
correlated with BCG offset than is the central entropy. This is quite
a surprising result, since \hlb{$P3/P0$} traces cluster dynamics
outside the cool core, whereas the central entropy is more sensitive
to the inner parts.

The BCG correlation trends are consistent with the well-known tendency
of cool cores to occur in smoother (i.e.  more relaxed, hence lower
$w_X$, low power ratio) clusters where a BCG sits close to the bottom
of the potential well \citep{Bildfell08}. This demonstrates the tight
quantitative link between these completely independent X-ray and
optical indicators of substructure.

\section{The $L_X$-$T_X$ Relation}
\label{sec:lt}

Similarly to previous studies
\citep[e.g.][]{Morandi07,Pratt10,Mittal11}, we find that the
\hlb{luminosity-temperature ($L_X-T_X$)} relationship exhibits a significant scatter of
$\approx 50\%$ when the core of the cluster is included---a scatter
which is diminished considerably, to 36\%, when the core is
excised. This effect is due to the overall non-self-similarity of
cluster cool cores in comparison to the regions outside the cool core
\citep[e.g][]{Vikhlinin06}. When the core is not excised, the
cool-core clusters lie significantly above the non-cool-core clusters,
an effect first noted by \cite{Fabian94b} and subsequently
studied in detail by \cite{McCarthy04} and \cite{Maughan12}.

\begin{figure*}
\begin{tabular}{cc}
\vspace*{-0.2in}\resizebox{3.1in}{!}{\includegraphics{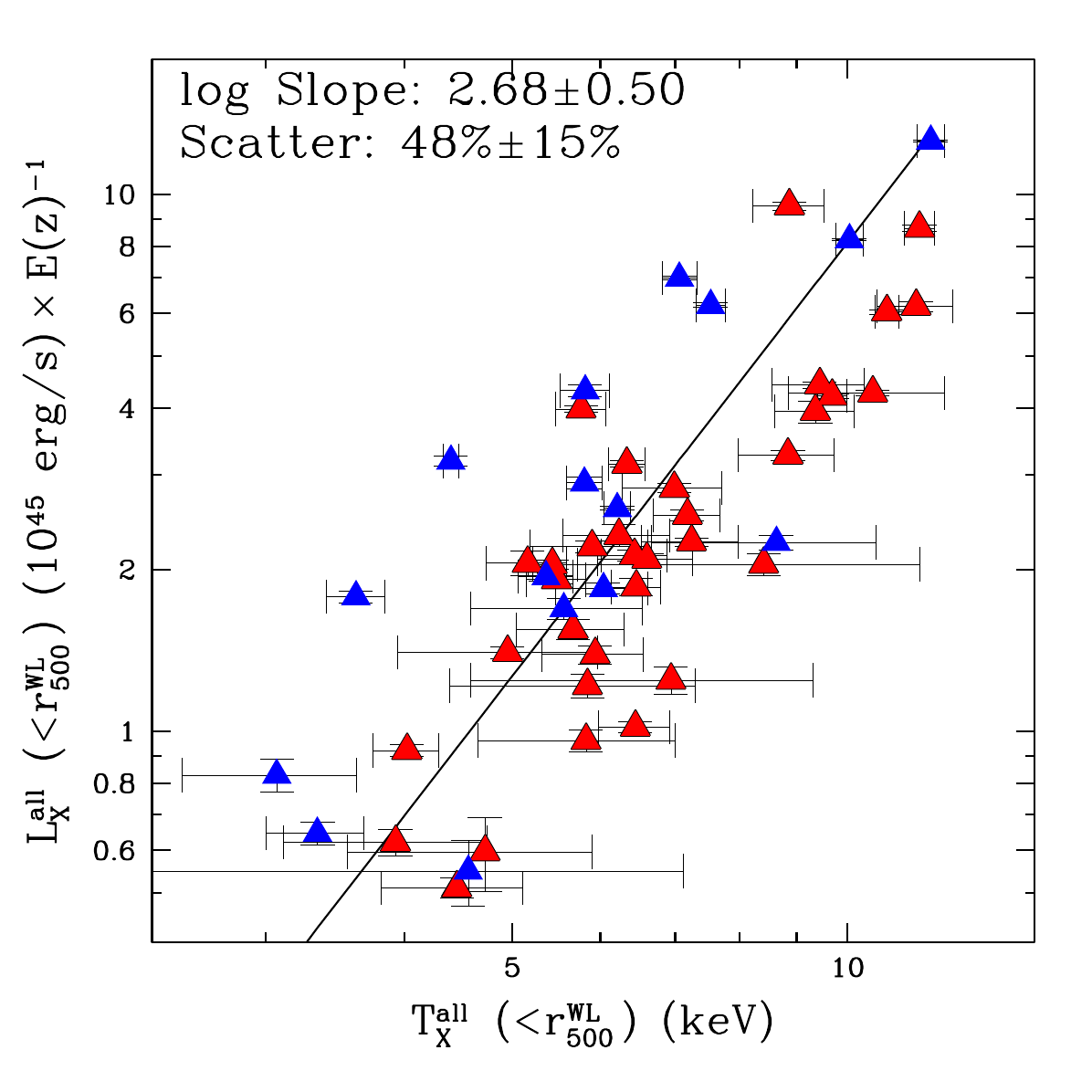}} &
\resizebox{3.1in}{!}{\includegraphics{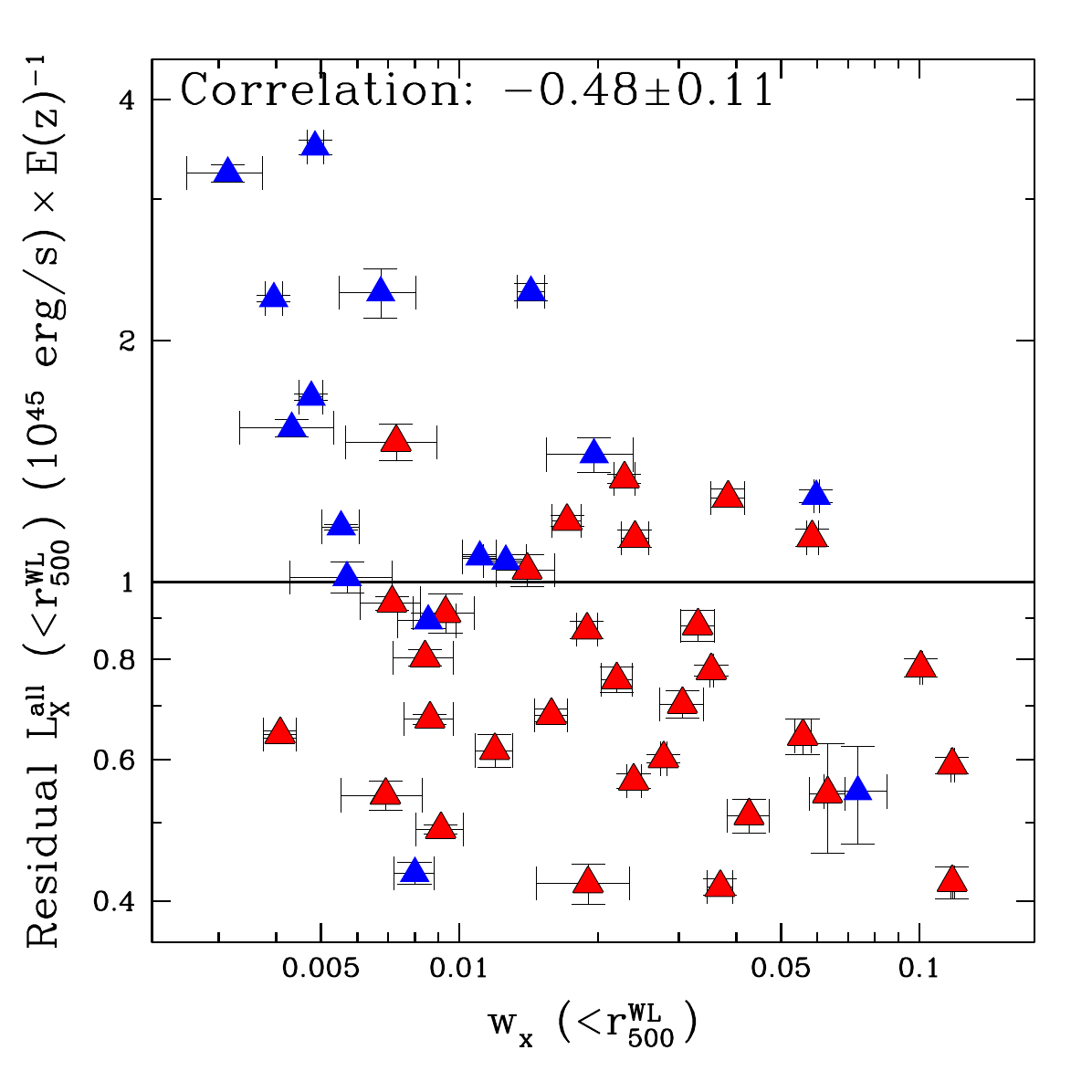}} \\
\vspace*{-0.2in}\resizebox{3.1in}{!}{\includegraphics{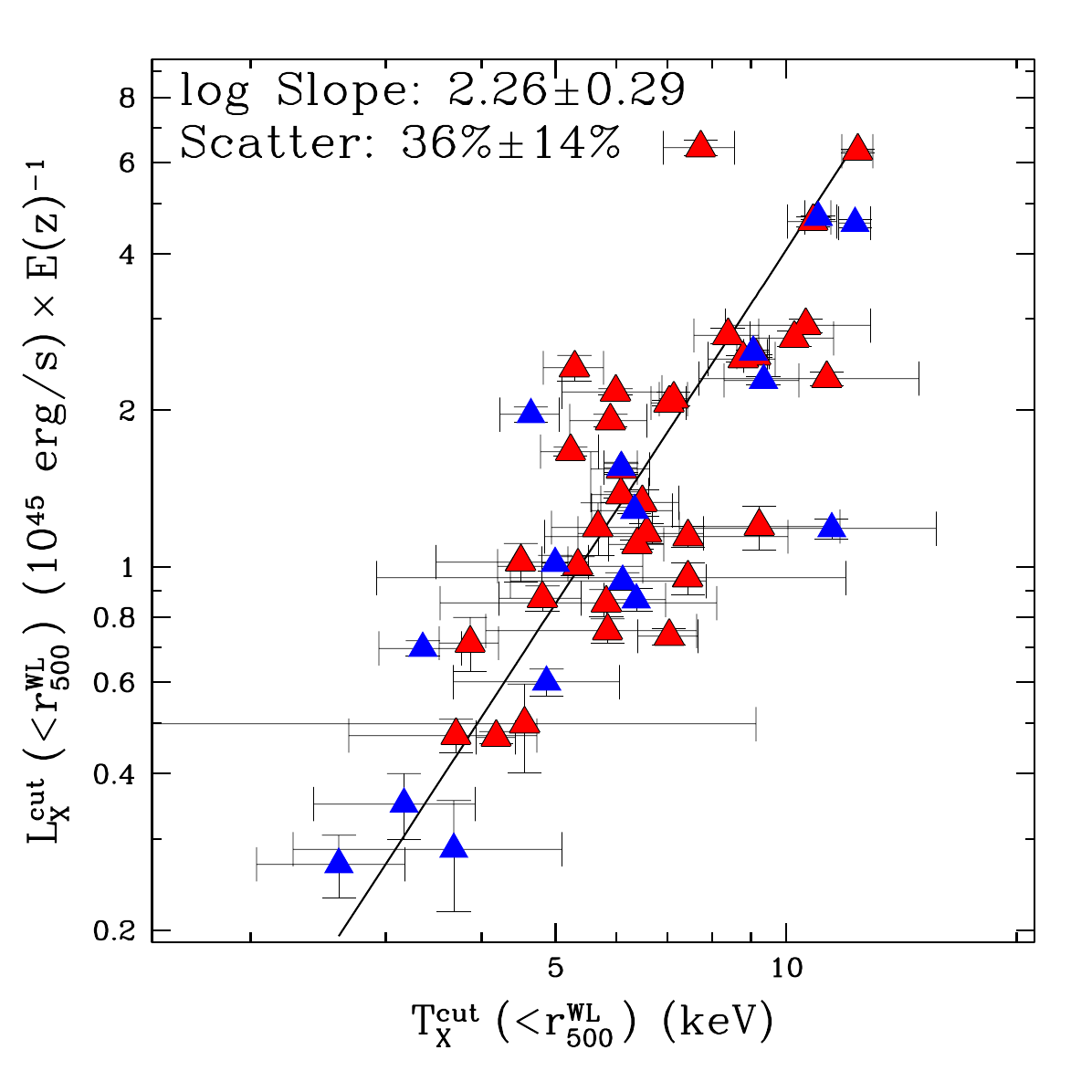}} &
\resizebox{3.1in}{!}{\includegraphics{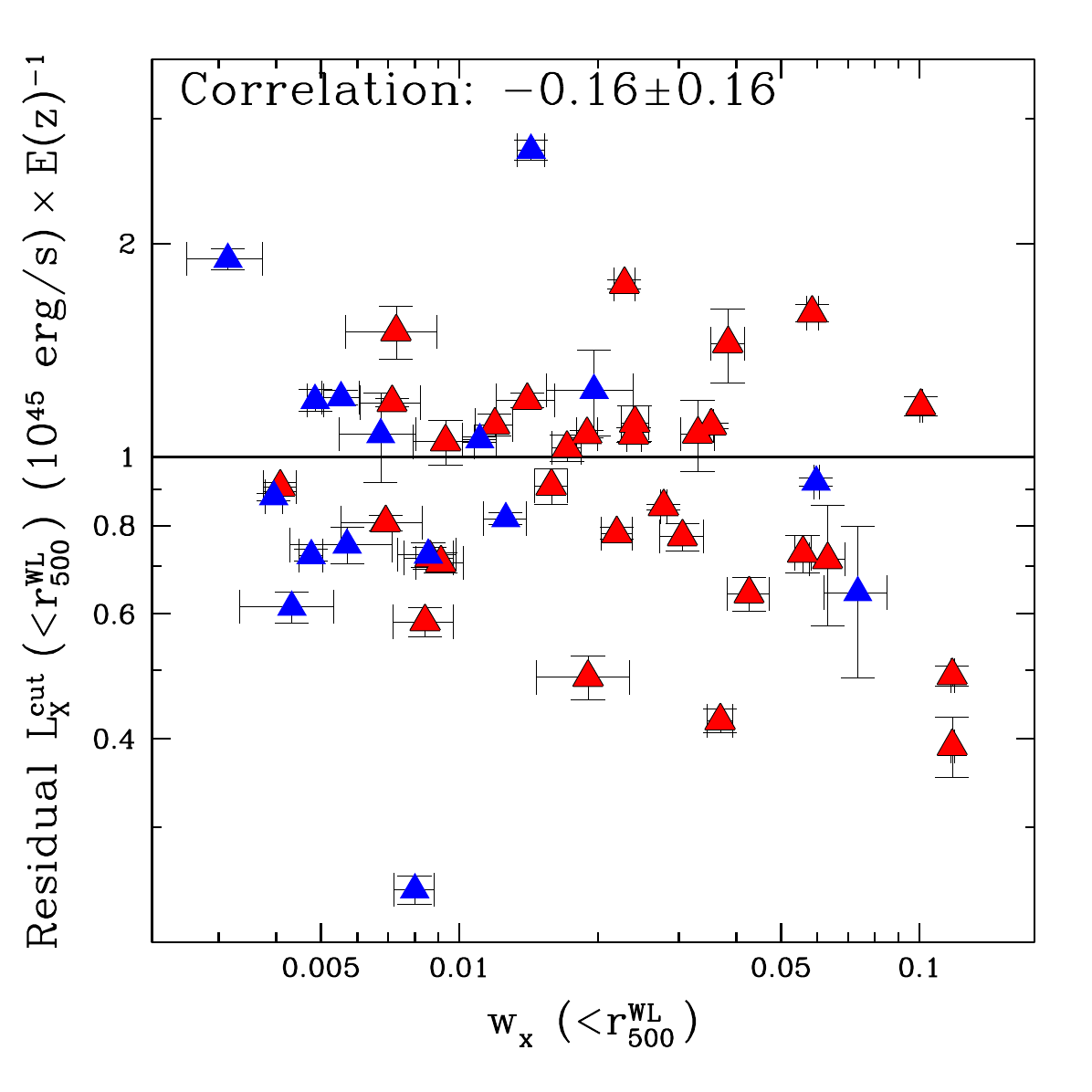}} \\
\end{tabular}
\caption{(top panels) The luminosity-temperature relationship at
  lensing $r_{500}^{WL}$ and its residuals compared to centroid shift
  variance $w_X$.  (bottom panels) same as top, except that the
  inner 0.15 $r_{500}^{WL}$ has been removed. The residuals are
  uncorrelated with all four substructure measures.  Blue
  triangles show cool-core clusters and red triangles show
  non-cool-core clusters. \label{fig:lt}}
\end{figure*}

In Figure \ref{fig:lt} and Table \ref{tbl:scaling}, we show that when
we include all cluster emission, the residuals of the $L_X-T_X$
relation show a strong and significant correlation with both the
central entropy of the cluster and the centroid shift $w_X$ (we choose
$w_X$ because of the four measures discussed in \S\ref{sec:corr} it
offers the strongest correlation). However, when we cut out the
central 0.15 $r_{500}^{WL}$, the distinction disappears, and the
cool-core and non-cool-core clusters become indistinguishable in terms
of entropy as well as $w_X$. This is consistent with the findings of
\cite{Maughan12} in the sense that once the cool core is taken out of
consideration, residuals in the L-T relation no longer carry information
regarding the dynamical state of the cluster.

% However, intriguingly, even after cutting out the cool core, a
% significant correlation of the $L_X$ residuals for the
% \emph{non-cool-core} clusters with $w_X$ persists. It would appear
% that $w_X$ continues to carry some information regarding the deviation
% of unrelaxed clusters from self-similarity (whereas for the cool core
% clusters, it does not, regardless of whether we cut out the core or not).

% This brings up the question of whether, despite the lack of
% correlation in the core-cut residuals, it is still possible to
% ``correct'' the $L_X-T_X$ relation using the centroid shift variance
% $w_X$.  Can we lower the scatter in the $L_X-T_X$ relation even more
% by using substructure estimators such as $w_X$ as a calibrator and
% applying corrections to $L_X$?  The simple answer is, as expected, no:
% when we apply the correction for $w_X$ ($\Delta L_X \propto
% w_X^{-0.29\pm 0.1})$ the intrinsic
% scatter in the non-cool-core $L_X-T_X$ relation decreases from $36\%
% \pm 13\%$ to $34\% \pm 17\%$, i.e., not significantly. A similar
% argument applies to all the other substructure measures, where the
%  correlation is even weaker.

% The $L$-$T$ relationship is the most straightforward X-ray scaling
% relation (in that very little modeling is required to determine the
% observables). The residuals in this relation, even after excising the
% core, are correlated with substructure estimates such as
% $w_X$. However, the residual correlation itself has scatter, which
% means that it has little power to reduce the scatter in the original
% $L_X-T_X$ relation. 

This is an example of ``irreversible scatter''---in other words,
outside their cores, the clusters of galaxies in our sample have
``forgotten'' the cause of the intrinsic scatter in the $L_X$-$T_X$
relation.  This has implications for scaling relation correction
procedures such as described in \cite[e.g.]{Jeltema08}, where
relationship between the residuals and the substructure measures for
simulated clusters are used to produce corrected observables which sit
more tightly on the scaling relations. The lack of correlation in our
case implies that such procedures will not reduce the scatter in the
measured scaling relations (at least for the JACO/CCCP sample).

% Finally, we note that the slopes of the $L$-$T$ relation are
% consistent whether derived within a projected aperture corresponding
% to lensing $r_{500}$, X-ray $r_{500}$, or 1 Mpc.  Interestingly,
% however, the scatter is smallest when using a fixed aperture. This may
% be due to inaccuracies in the estimate of $r_{500}$. Errors inherent in
% the mass measurement process translate directly into errors in
% $r_{500}$. It would so far appear that at least for the $L$-$T$
% relation, such errors are large enough that they overwhelm any
% benefits derived from applying self-similar scaling.

\section{Lensing Mass-Observable Relation}
\label{sec:proxy}
%% However, reconciling Chandra and XMM-Newton results has not been free
%% of challenges. One problematic area for Chandra and XMM-Newton has
%% been the discrepancy at the $15\%$ level between the spectral
%% properties of diffuse sources as determined by either telescope. This
%% has particular relevance for X-ray derived masses, given that the

The mass-observable relationship is an important ingredient in the
determination of the cosmological parameters with clusters of
galaxies.  Because the mass function is the ultimate connector between
the cosmological parameters and the data, finding accurate mass
proxies \hly{using multiple methods and wavelength regimes is
  important}. Comparison of X-ray derived observables with weak
gravitational lensing masses, which do not require the assumption of
hydrostatic equilibrium, has proved a fruitful path towards this end
\citep[e.g.][]{Mahdavi08,Okabe10,Jee11}. We list our results for
several different mass-observable relations in Table
\ref{tbl:scaling}. 

\subsection{Temperature, Gas Mass, and Pseuo-Pressure}

\begin{figure*}
\begin{tabular}{cc}
\resizebox{3.2in}{!}{\includegraphics{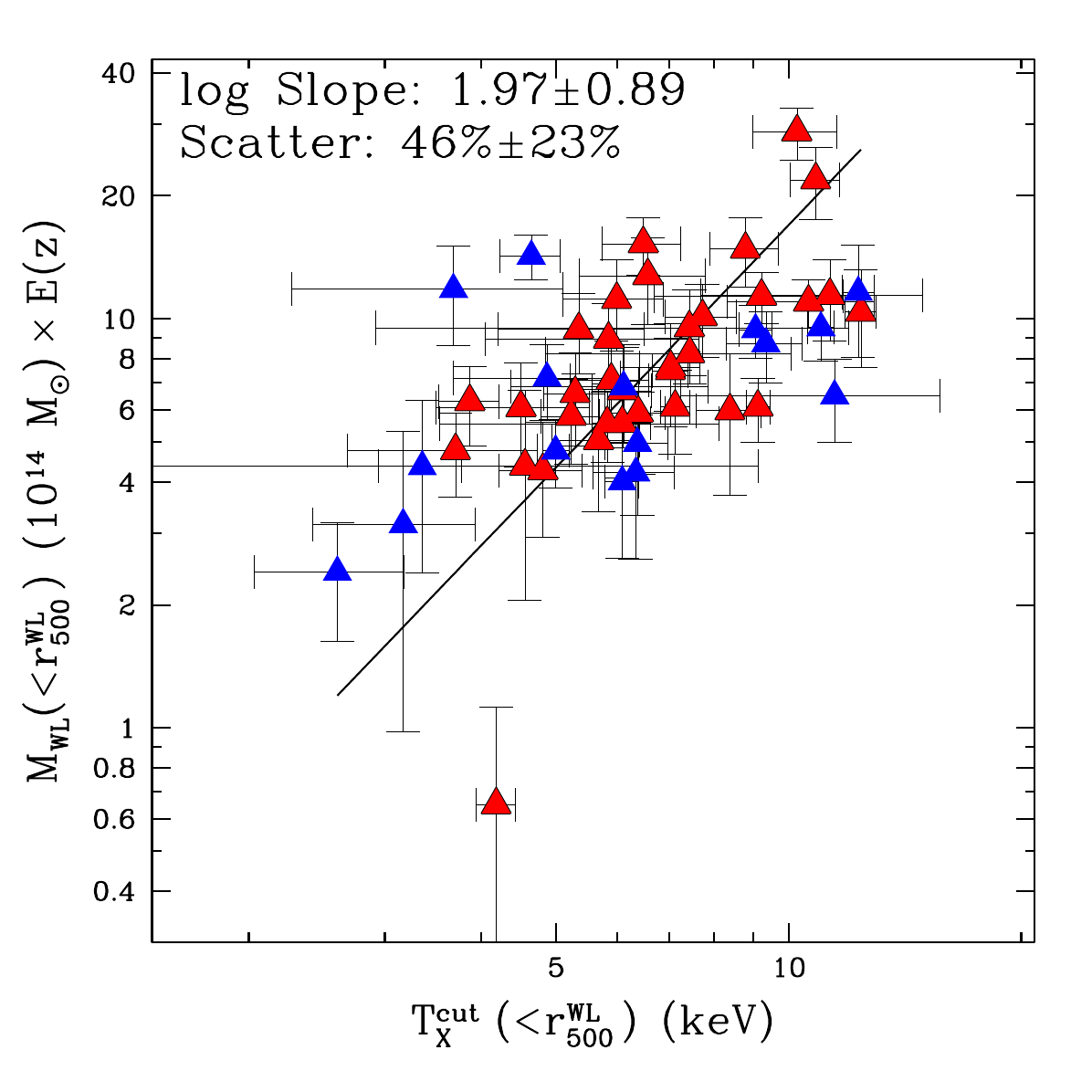}} &
%\resizebox{3.1in}{!}{\includegraphics{m-t-residual.pdf}} \\
\resizebox{3.2in}{!}{\includegraphics{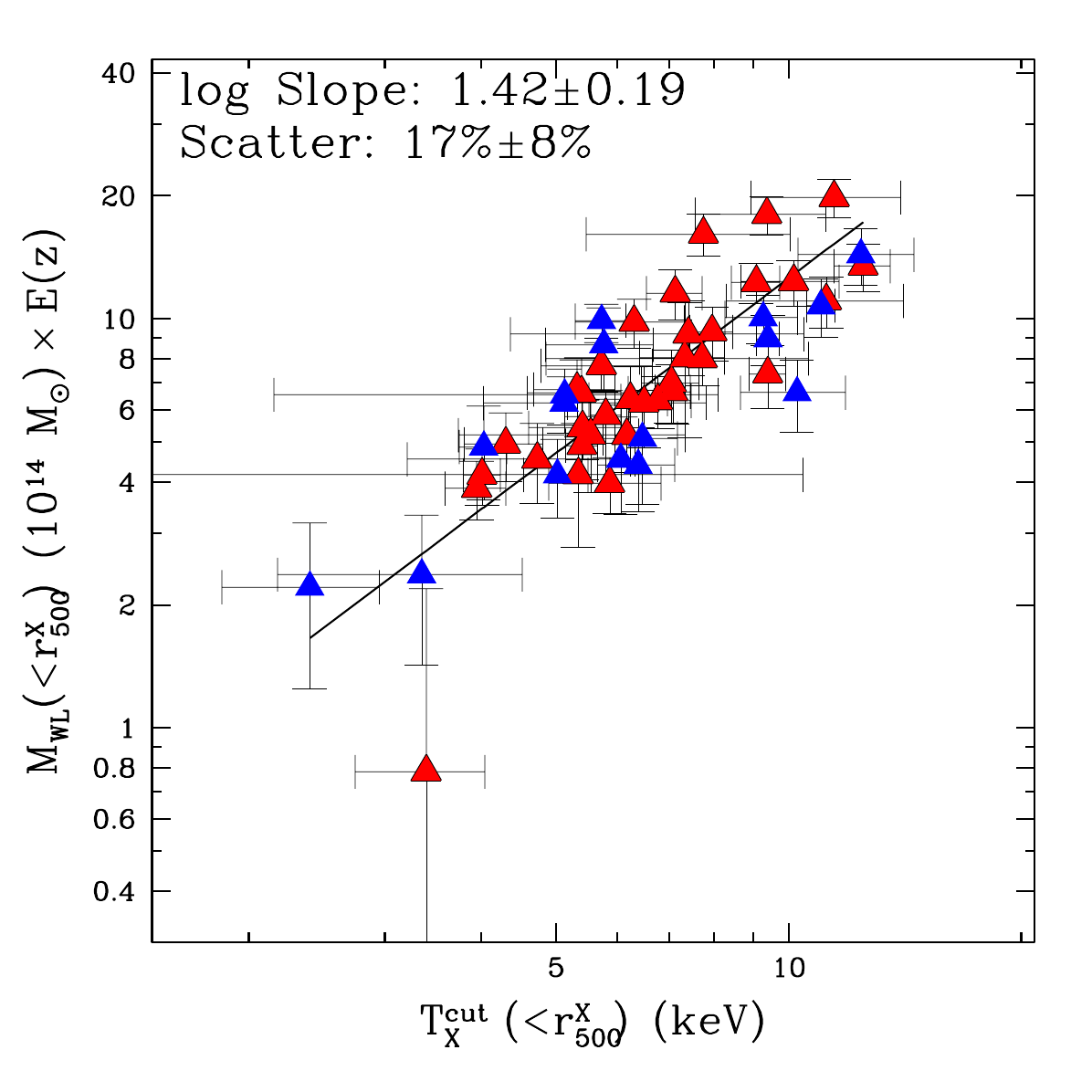}} \\
%\resizebox{3.1in}{!}{\includegraphics{m-t-mpc.pdf}} \\
\end{tabular}
\caption{The mass-temperature relationship at
  lensing $r_{500}$ (left) and at X-ray  $r_{500}$ (right). 
 The latter shows less scatter due to the intrinsic correlation of X-ray $r_{500}$ with temperature.
Blue
  triangles show cool-core clusters and red triangles show
  non-cool-core clusters. \label{fig:mt}}
\end{figure*}

We begin by examining the lensing mass-gas temperature relationship in
Figure \ref{fig:mt}; while exhibiting significant intrinsic scatter
\citep{Ventimiglia08,Zhang08,Mantz10}, the $M$-$T$ relation is still a
worthwhile keystone for comparison with previous work.  We find that
the relationship is consistent with being self-similar, with a larger
scatter and uncertainty at lensing $r_{500}$ than at X-ray
$r_{500}$. Regardless of whether we consider the cool-core or the
non-cool-core subsamples, the scatter is roughly $46\%$. The scatter
drops dramatically to $17\% \pm 8\%$ when we use X-ray $r_{500}$
because of the inherent correlation between the gas temperature and
X-ray $r_{500}$ itself, which do not attempt to model. The phenomenon
of inherent correlation is discussed in greater detail by
\cite{Kravtsov12}, and arises because the \emph{aperture} used to
measure the mass is highly correlated with the observable on the other
axis (in this case, X-ray $r_{500}$ and X-ray temperature are highly
correlated).

The normalization derived for the mass-temperature relation is
consistent with previous work, for example \cite{Pedersen07},
\cite{Henry09} and \cite{Okabe10}.

\hlb{Table {\protect\ref{tbl:scaling}} also shows similar results for
  the core-excised X-ray luminosity-lensing mass ($L_X-M_{WL}$) relation. The
  instrinsic scatter ($35\% \pm 13\%$) is consistent with that of the
  mass-temperature relation, and as before, the scatter is
  dramatically lower at $r_{500}^X$ than at $r_{500}^{WL}$, again
  likely due to internal correlation between $r_{500}^X$ and $L_X$
  which we do not model.}

Far more impressive is the gas mass-lensing mass relationship.  The
gas mass has been shown in previous work to be a useful mass proxy
\citep{Mantz10,Okabe10}---essentially, the assumption that rich
clusters of galaxies have the same gas fraction is turning out to be a
remarkably robust one. \hlb{We improve the significance of the }
\protect\cite{Okabe10} \hlb{finding with our sample of $50$ clusters}:
at $r_{500}^{WL}$, the gas mass is consistent with being proportional
to the lensing mass, with a log slope of $1.04 \pm 0.1$, and a
normalization implying a gas fraction $f_\mr{gas} = 0.12 \pm 0.01$.

\begin{figure*}
%\begin{tabular}{cc}
%\%vspace*{-0.2in}\resizebox{3.2in}{!}{\includegraphics{K0-mg-mlens}} &
\begin{center}
\resizebox{4in}{!}{\includegraphics{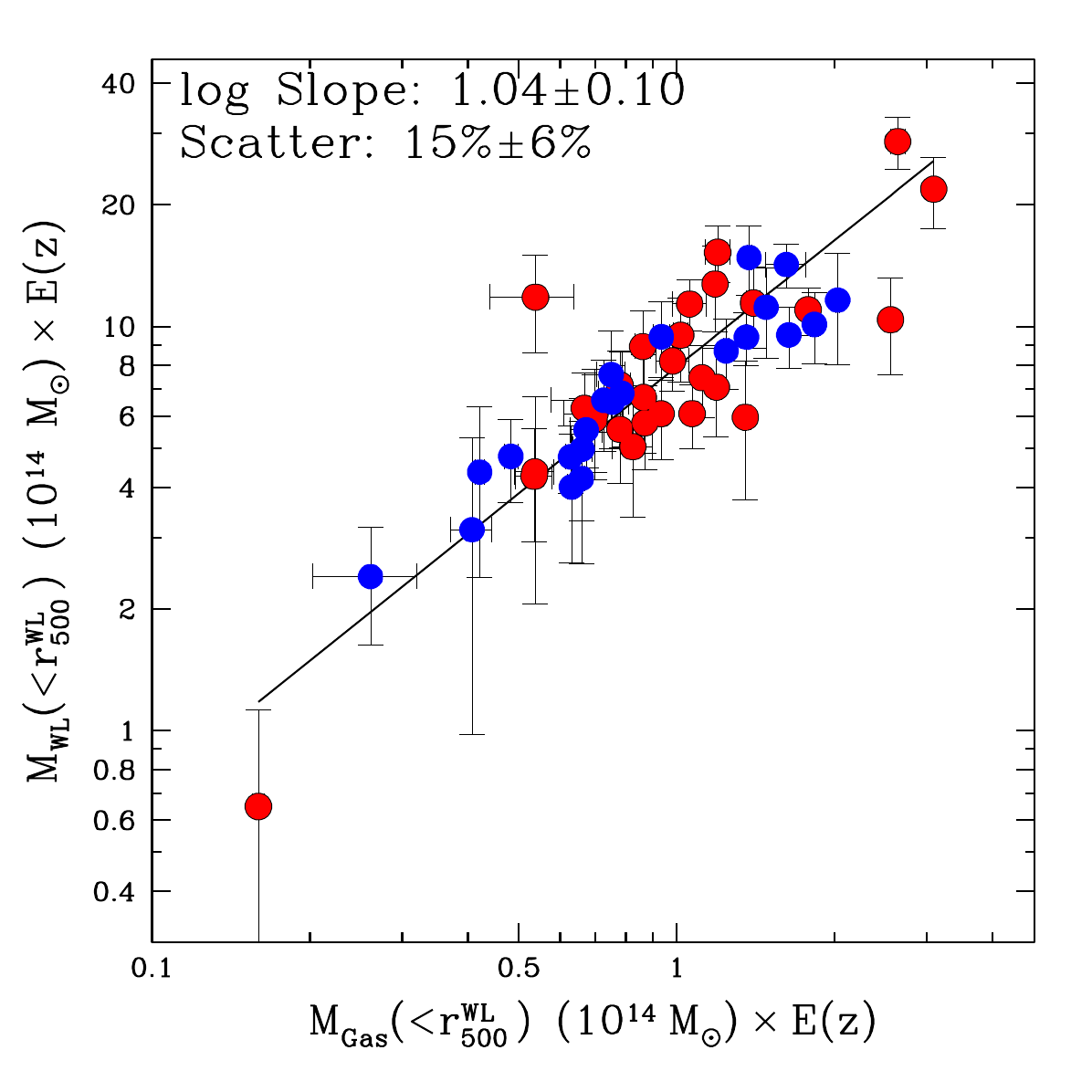}} 
\end{center}
%\\
%\end{tabular}
\caption{The gas mass-lensing mass relationship at lensing $r_{500}$.
\protect\hlb{Blue
circles show low-BCG-offset systems and red circles
show high-BCG-offset systems. Most of the low BCG offset systems are
also low central entropy clusters.} \label{fig:mg}}
\end{figure*}

We find a low scatter of $15 \pm 8\%$ for the $M_\mr{gas}-M_L$
relation (Figure \ref{fig:mg}) \hly{for all clusters, regardless of
dynamical state}. Interestingly, the same scatter holds
regardless of whether we use lensing $r_{500}^{WL}$ or a fixed aperture of
1 Mpc.

This low scatter at fixed radius is important. Recently, sophisticated
treatments of the covariance between the axes in the mass-observable
relation have become possible \citep{Hogg10}. Specifically, in the case
of gas mass and lensing mass measured at $r_{500}^{WL}$, there is a subtle
correlation between the two axes, even though one quantity (lensing
mass) is measured using optical data and the other quantity (gas mass) is
measured using X-ray data. The issue is that the aperture
itself, $r_{500}^{WL}$, depends on the lensing mass, and therefore, by
choosing the same aperture for the gas mass, we might introduce a
correlation that produces artificially low scatter. \hlb{This effect was
described in detail by} \protect\cite{Becker11} \hlb{who find that such correlations
can result in the measured scatter being $\approx 50\%$ smaller than
the true scatter.}


\begin{figure}
\resizebox{3.2in}{!}{\includegraphics{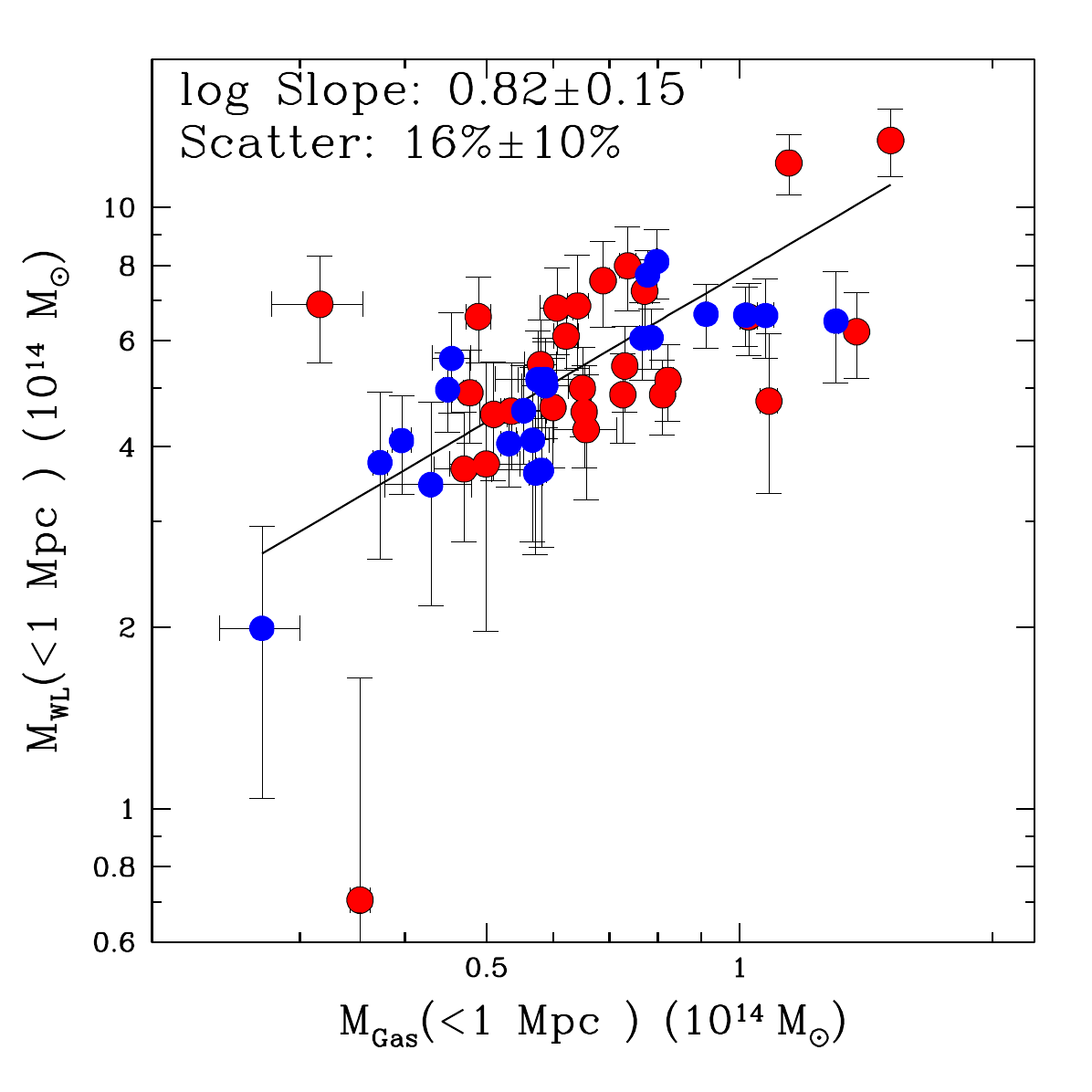}}
\caption{Lensing mass vs. gas mass at a fixed physical radius of 1 Mpc.
Blue circles show low-BCG-offset clusters and red
  \protect\hly{circles} show high-BCG-offset clusters. \protect\hlo{The relation retains the low scatter of the relations at
fixed density contrast.} \label{fig:mgmpc} }
\end{figure}

\begin{figure}
\resizebox{3.2in}{!}{\includegraphics{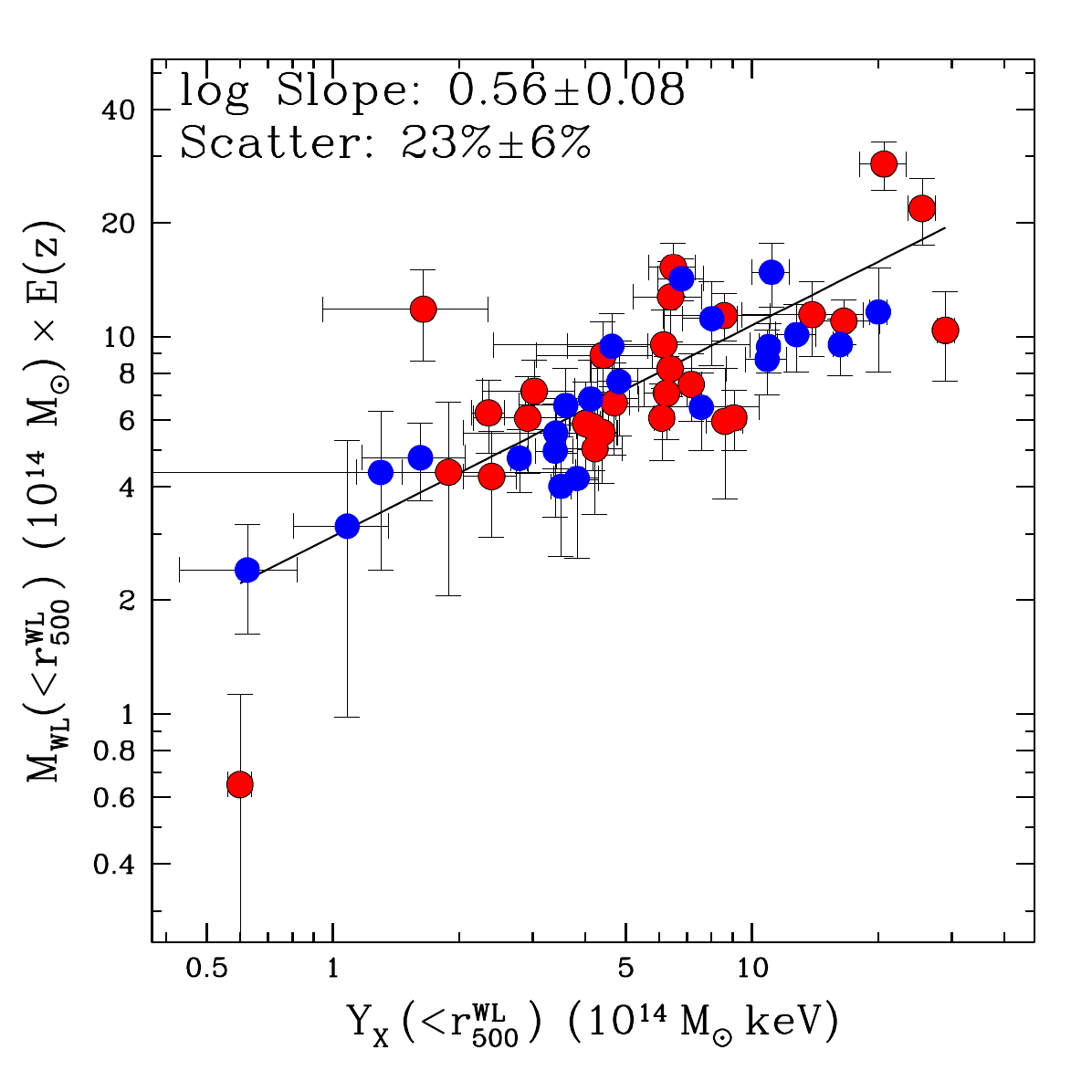}}
\caption{Lensing mass vs. pseudo-pressure $Y_X$.  Blue
  circles show low-BCG-offset clusters and red circles show
  high-BCG-offset clusters.\label{fig:yx}}t
\end{figure}

However, using a physical aperture of 1 Mpc completely takes away any
possibility of covariance between the two axes. In \hlb{Figure
  \protect\ref{fig:mgmpc}}, we truly have two statistically
independent observations, and yet the intrinsic scatter remains remarkably low,
$16\% \pm 7\%$. The fact that the scatter does not change when
switching to a fixed physical aperture is reassuring. \hlb{The
  $1\sigma$ scatter uncertainties are just large enough to accommodate
  the scatter underestimate predicted by} {\protect\cite{Becker11}}
\hlb{(e.g.  if the ``true'' scatter at both $r_{500}^{WL}$ and 1 Mpc
  is 20\%, our $1\sigma$ errors would be consistent with a 50\%
  scatter underestimate at $r_{500}^{WL}$ and no scatter underestimate
  at 1 Mpc).} \hlb{In Table {\protect\ref{tbl:scaling}} we also list
  the performance of $Y_X$, $L_X$, and $T_X$, measured at fixed
  physical radius of 1 Mpc, as predictors of $M_{WL}(<1
  $Mpc$)$. Overall, we find little difference between the intrinsic scatter
  at 1 Mpc compared to $r_{500}^{WL}$.}

\subsection{Regularity of cool core and low BCG offset clusters}

  Another point of particular importance is the fact that for the
  cool-core clusters, the $1\sigma$ scatter is $<10\%$ (the scatter is
  $<6\%$ if we cut on BCG offset instead)---these numbers are low
  enough to be consistent with zero. Simulations and analytical work
  \citep[e.g][]{Becker11} show that \hlb{$\approx 15\%$} is roughly
  the amount of intrinsic scatter we can expect due to geometric
  errors from the assumption of spherical geometry. Thus deviations
  from spherical symmetry can produce scatter \hlb{we observe} in the
  cool-core $M_\mr{gas}-M_L$ relation, and as a result, we can begin
  to claim that we are approaching a full accounting of all sources of
  systematic error in the mass-observable scaling relation.

We note that the BCG offset works as well as central entropy in
identifying the low-scatter subsample. This is an interesting result,
because of our four substructure measures, BCG offset is by far the
least expensive to calculate, in that it does not require X-ray
temperature (spectral) information---\hlb{a set of X-ray and optical images}
is sufficient to calculate $D_\mr{BCG}$. \hlo{However, it is worth
  noting that while the low BCG offset and cool-core subsamples have
  significant overlap, they are not precisely the same, and the two
  cuts trace two different types of equilibrium (dynamical and
  thermal, respectively)}.

Another frequently used mass proxy is $Y_X$, the pseudo-integrated
pressure first pioneered by \cite{Kravtsov06} and examined by
\cite{Vikhlinin06}; being the product of the gas mass and the core-cut
temperature at $r_{500}^{WL}$, $Y_X$ is directly comparable to the
integrated Sunyaev-Zel'dovich compton $Y$ parameter
\citep{Plagge10,Andersson11}.

We show the $Y_X$-$M_L$ relation at $r_{500}^{WL}$ for our sample in
Figure \ref{fig:yx}; we find consistency with the expected
self-similar slope of 0.6, but slightly higher intrinsic scatter to
the gas mass when used as a mass proxy: the overall intrinsic
deviation is \hlb{$\approx 23\% \pm 6\%$} regardless of whether we
use the entire sample or the cool-core subsample.

\hlb{One might be tempted to argue that} gas mass is a superior mass proxy to
$Y_X$, not simply because of its ease of calculation and comparable
overall intrisic scatter, but also because of the systematically lower
intrinsic scatter that comes about when only cool-core clusters are
considered. \hlb{However, this discrimination between relaxed and
  non-relaxed clusters is perhaps not optimal in a cosmological
  context, where uniformity of scatter across the entire sample is
  important.  Where uniformity is most important, $Y_X$ is a superior
  choice to gas mass, because as we show in Table
  {\protect\ref{tbl:scaling}} it has uniform scatter regardless of
  cluster central entropy or BCG offset.}

\hlb{Finally, it is instructive to compare $Y_X$ with its radio
  counterpart, the cylindrically integrated Sunyaev-Zel'dovich (SZ)
  pressure $Y_{SZ}$}. {\protect\cite{Hoekstra12}} \hlb{consider direct
  correlations between $Y_\mr{SZ}$ from the \emph{Planck} mission and
  projected weak lensing masses; they find an intrinsic scatter of
  $12\pm5\%$ at projected $r_{2500}$. As a point of comparison, when
  we conduct a similar exercise on spherically determined $Y_X$ and
  $M_{WL}$ (both measured at spherical $r_{2500}^{WL}$), we find an
  intrinsic scatter of $18\% \pm 6\%$, consistent with the}
{\protect\cite{Hoekstra12}} \hlb{SZ comparison.}

\subsection{Predicting $M_{500}^{WL}$ with fixed aperture mass proxies for surveys}
\label{sec:mpc}

\hlb{In a blind X-ray survey, the aperture $r_{500}^{WL}$ or even
  $r_{500}^X$ may not be easily available.  For cosmology, we still
  need to know $M_{500}$.  It is therefore useful to investigate whether one
  can directly predict $M_{500}^{WL}$ without the need to calculate
  overdensity radii $r_\Delta$ for the various X-ray observables. }
For example, a wide-field all-sky X-ray survey may be able to measure
hundreds of thousands of gas masses within fixed physical apertures,
but lack the photon statistics to allow for the calculation of X-ray
overdensity radii.

\begin{figure*}
\begin{tabular}{cc}
\resizebox{3.5in}{!}{\includegraphics{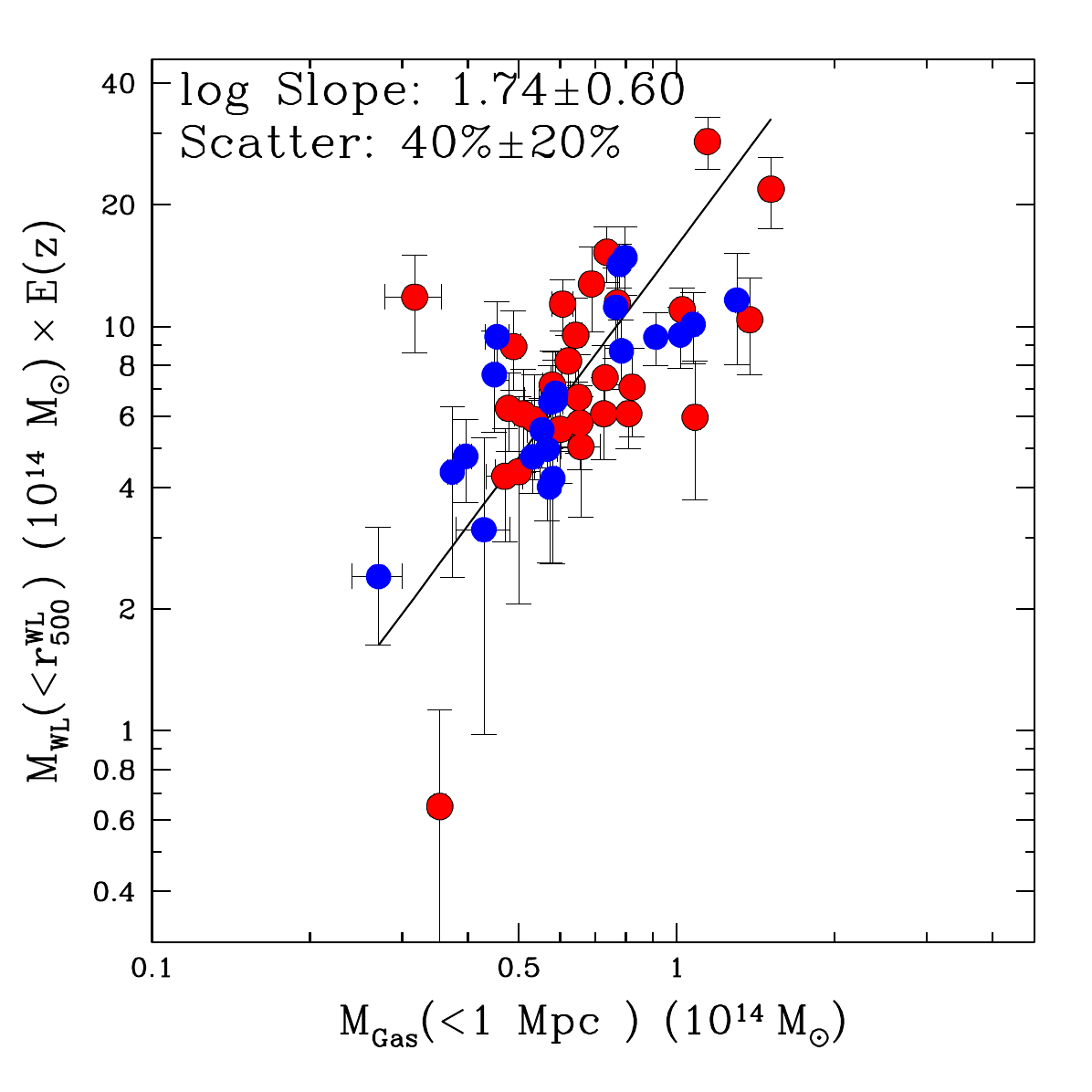}} &
%\resizebox{3.3in}{!}{\includegraphics{bcg-mhydro-mlens-r2500.pdf}} \\
\resizebox{3.5in}{!}{\includegraphics{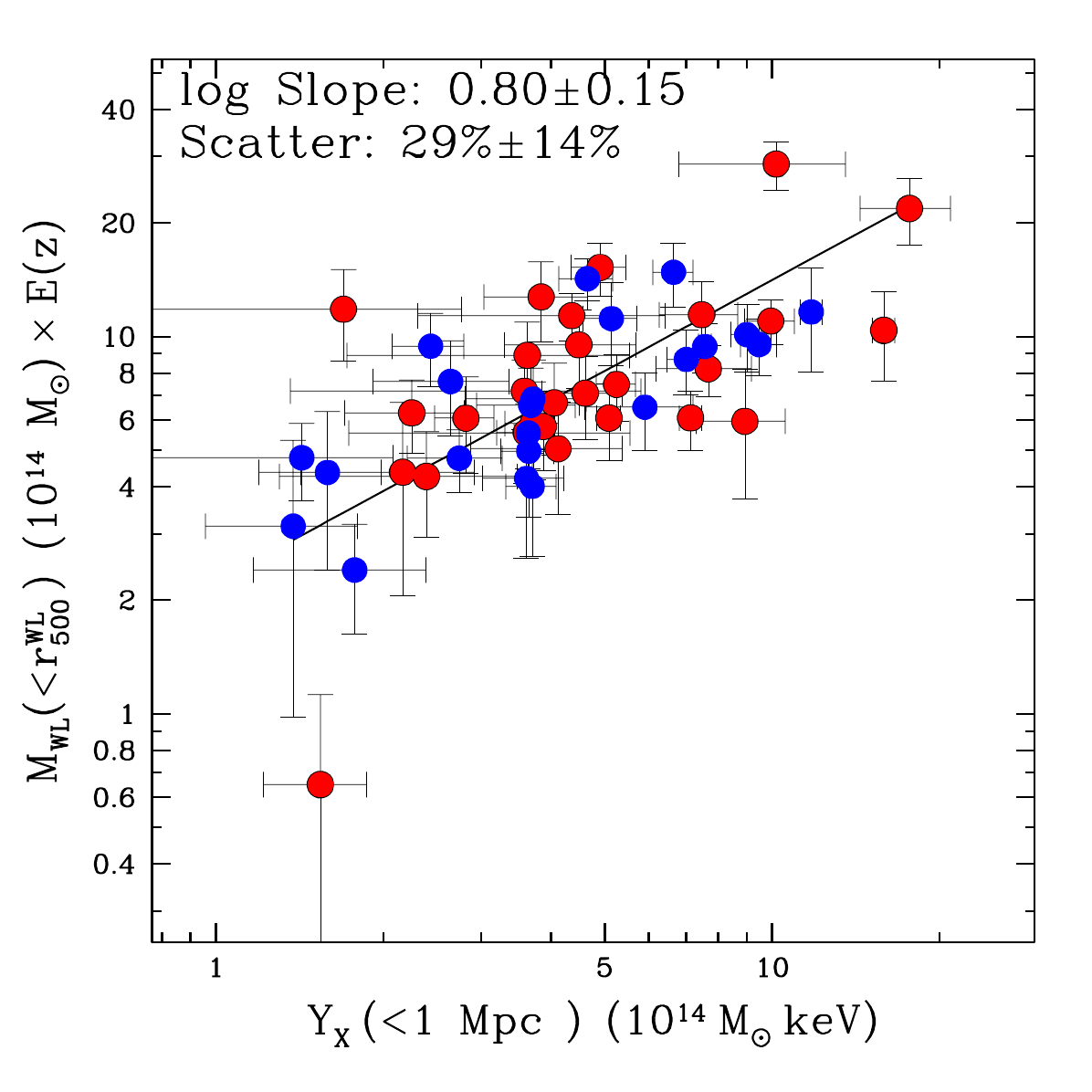}} 
%\resizebox{3.3in}{!}{\includegraphics{bcg-mhydro-mlens-r500.pdf}} \\
\end{tabular}
\caption{The lensing mass at $r_{500}$ vs. gas mass and $Y_X$ measured
  at \emph{fixed} radius of 1 Mpc. Blue circles show low BCG offset
  clusters, while red circles show high BCG offset clusters.
 \label{fig:mixed}}
\end{figure*}

In Figures \ref{fig:mixed} we consider this situation, plotting
$M_\mr{WL}(<r_{500}^{WL})$ against gas mass and $Y_X$ measured within
a fixed radius of $1$ Mpc. As expected, the slopes now deviate from
self-similar, and the intrinsic scatter is considerably higher than in
Figures \ref{fig:mg} and \ref{fig:yx}. However, interestingly, $Y_X$
exhibits somewhat less scatter (29\%) in this ``mixed'' scaling
relation than does gas mass (40\%). In surveys with poor photon
statistics where no X-ray or weak lensing estimates of $r_{500}$ are
readily available, $Y_X$ \emph{measured within a fixed physical
  aperture} may constitute a better mass proxy, because no separate
estimate of X-ray $r_{500}$ is required to use the relations shown in
Figure \ref{fig:mixed}. The results are summarized in Table
\ref{tbl:scaling}.

These data leave us with the perhaps dispiriting result that low
($<10\%$) scatter X-ray mass proxies may be derived either at fixed
physical radii, yielding total mass estimates within fixed physical
radii; or they may be derived at fixed overdensity radii, yielding
total mass estimates within fixed overdensities. But it seems
difficult to achieve very low scatter without either committing to fixed
physical radii (straightforward to measure, but more difficult to use
for cosmology); or to fixed overdensity radii (difficult to measure,
but more useful for cosmology) in both axes.

% \hly{A final useful survey scaling relation we consider is the $L_X-M$
% relation, shown in Figure \protect\ref{fig:lm}. We find that the 
% core-cut X-ray 
% luminosity measured within 1 Mpc is an adequate predictor of 
% weak lensing mass within $r_{500}^{WL}$; \hlo{cool-core clusters} are
% consistent with a low scatter of $28\% \pm 18\%$ in this relation.}

% \begin{figure}
% \resizebox{3.2in}{!}{\includegraphics{lxm}}
% \caption{\protect\hly{Lensing mass at $r_{500}^{WL}$ vs. core-cut X-ray luminosity within a
%   fixed aperture of 1 Mpc.  Blue circles show low-BCG-offset clusters
%   and red circles show high-BCG-offset clusters.\protect\label{fig:lm}}}
% \end{figure}

\begin{figure*}
\begin{tabular}{cc}
\resizebox{3.5in}{!}{\includegraphics{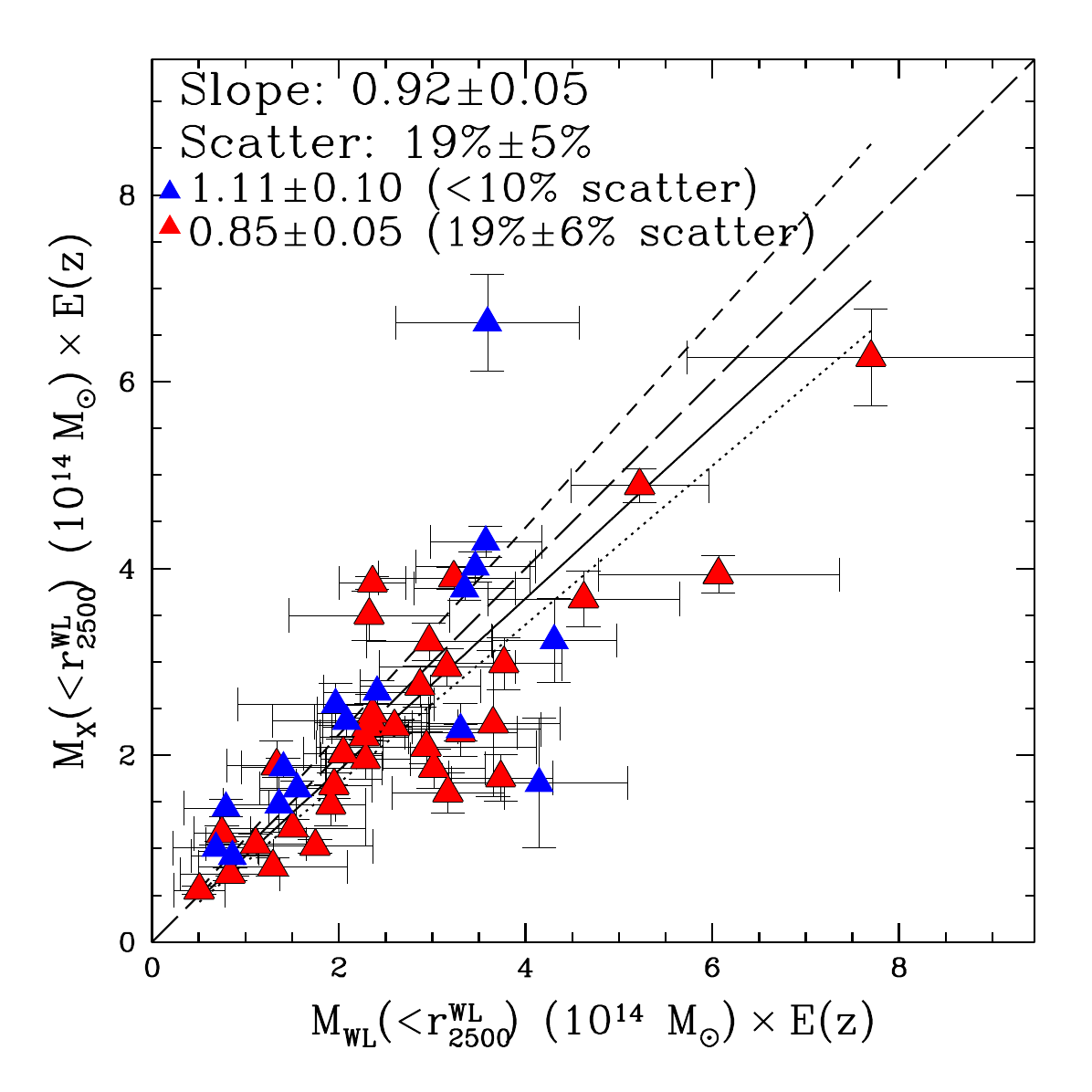}} &
%\resizebox{3.3in}{!}{\includegraphics{bcg-mhydro-mlens-r2500.pdf}} \\
\resizebox{3.5in}{!}{\includegraphics{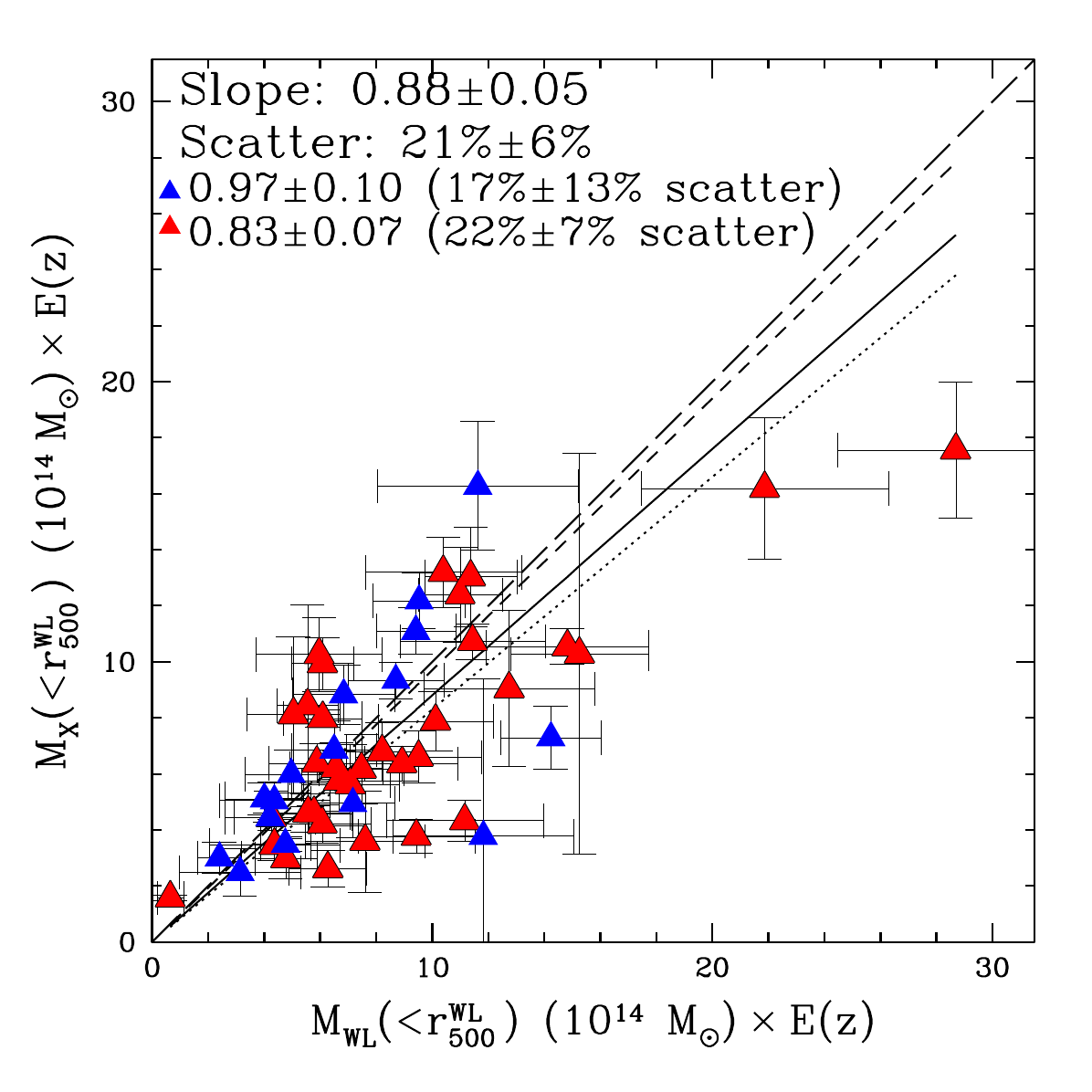}} 
%\resizebox{3.3in}{!}{\includegraphics{bcg-mhydro-mlens-r500.pdf}} \\
\end{tabular}
\caption{The relationship between hydrostatic mass and lensing mass at
  $r_{2500}^{WL}$ (\emph{left}) and $r_{500}^{WL}$ (\emph{right}).  Blue
  triangles show cool-core clusters and red triangles
  show non-cool-core clusters. Cool core clusters
  tend to have hydrostatic masses that agree with lensing masses;
  non-cool-core clusters tend to exhibit the
  hydrostatic mass underestimate.  \protect\hlb{The solid line indicates the best fit;
the long-dashed line indicates the line of equality; the short-dashed
line corresponds to the cool-core clusters, and the dotted line
corresponds to the non-cool-core clusters.}\label{fig:mhml}}
\end{figure*}

\begin{figure*}
\begin{center}
\resizebox{4in}{!}{\includegraphics{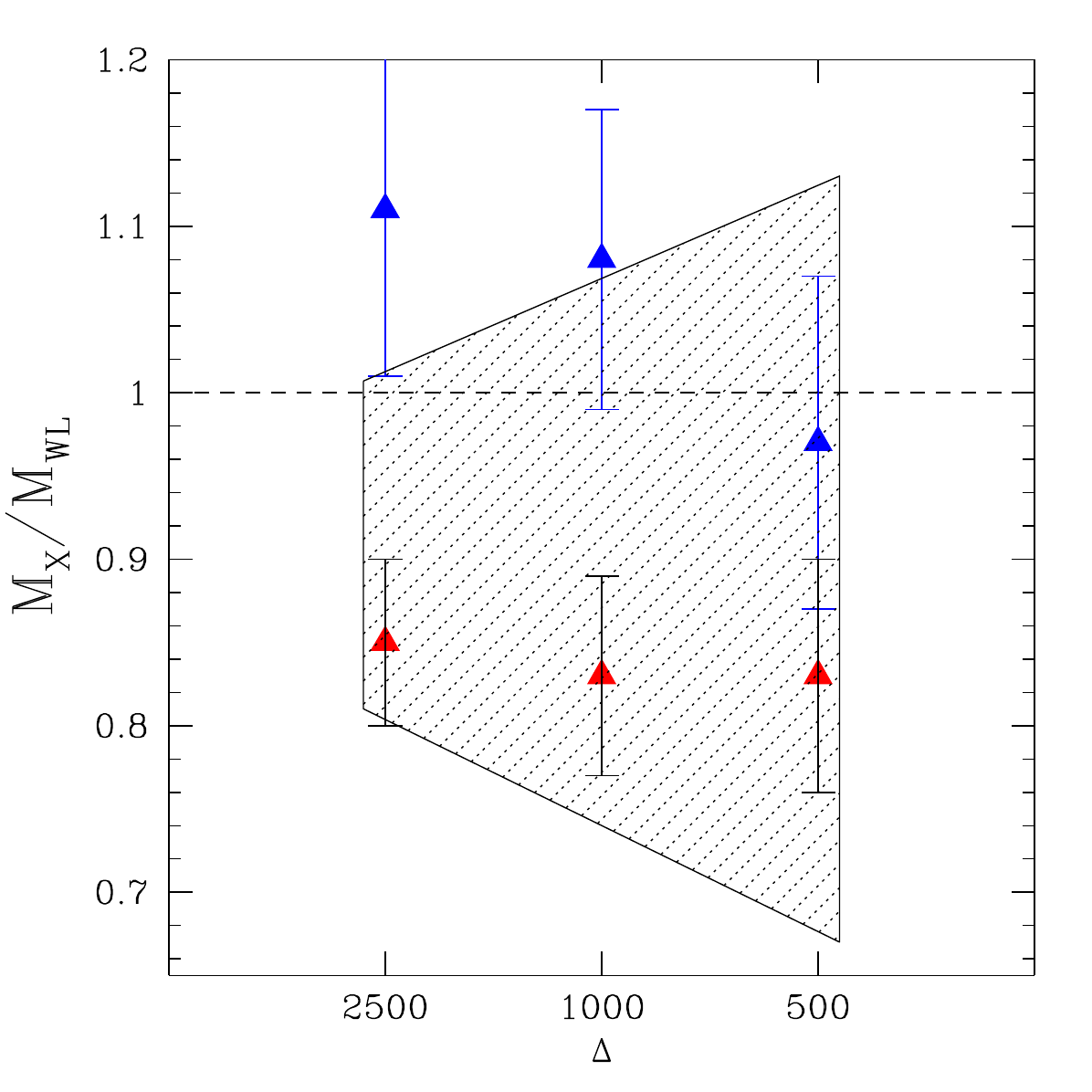}}
\end{center}
\caption{The X-ray to Weak-Lensing mass ratio as a function
of density contrast for cool-core systems (blue triangles)
and non-cool-core systems (red triangles). \protect\hlb{The error bars are not independent
  because the data within $r_{2500}$ also contributes to the measurement
  at $r_{500}$.} The shaded
region shows the range of X-ray cluster mass underestimate as 
determined by \protect\cite{Lau09}. \label{fig:mhmlsummary}}
\end{figure*}

\begin{figure*}
\begin{center}
\begin{tabular}{cc}
\resizebox{3.5in}{!}{\includegraphics{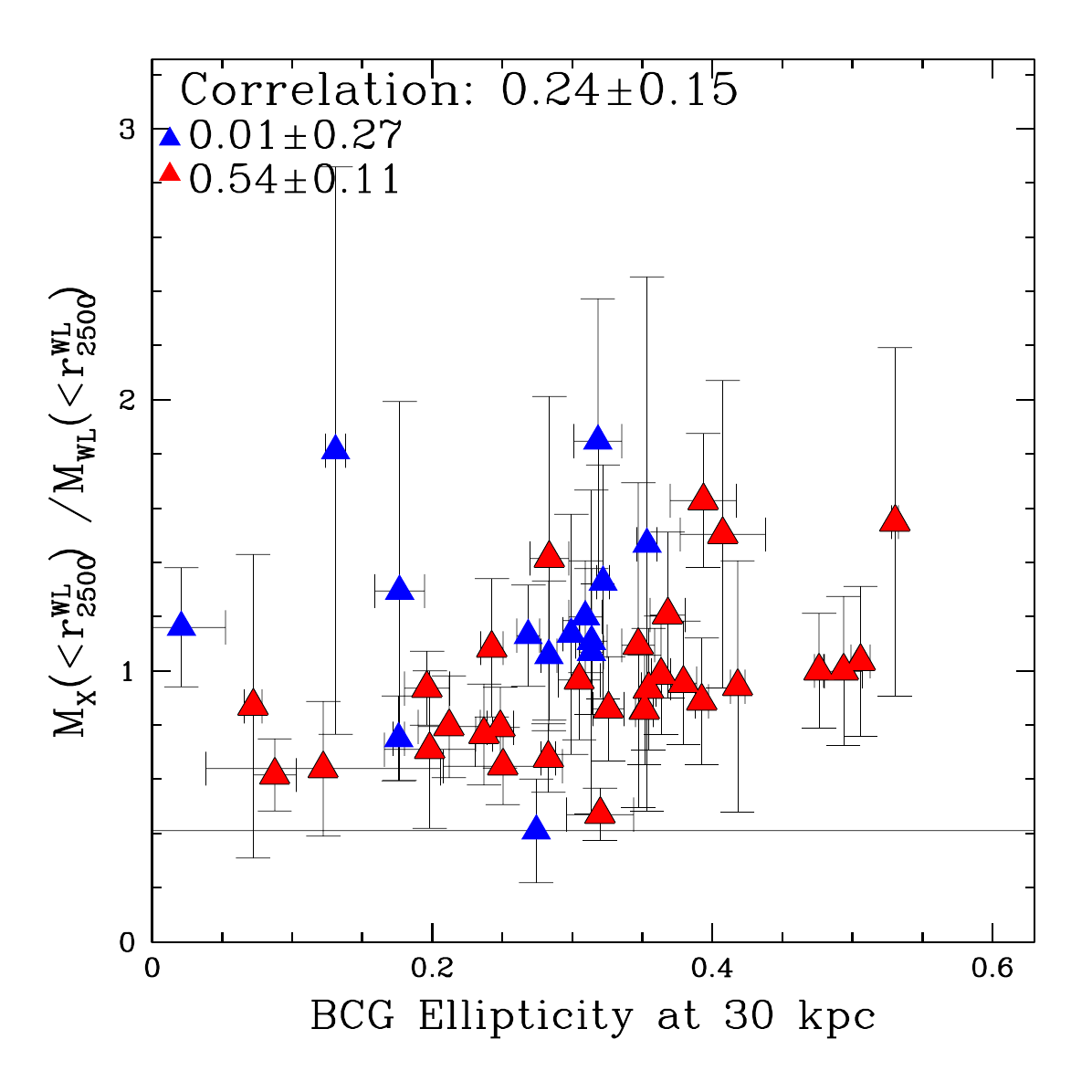}} &
\resizebox{3.5in}{!}{\includegraphics{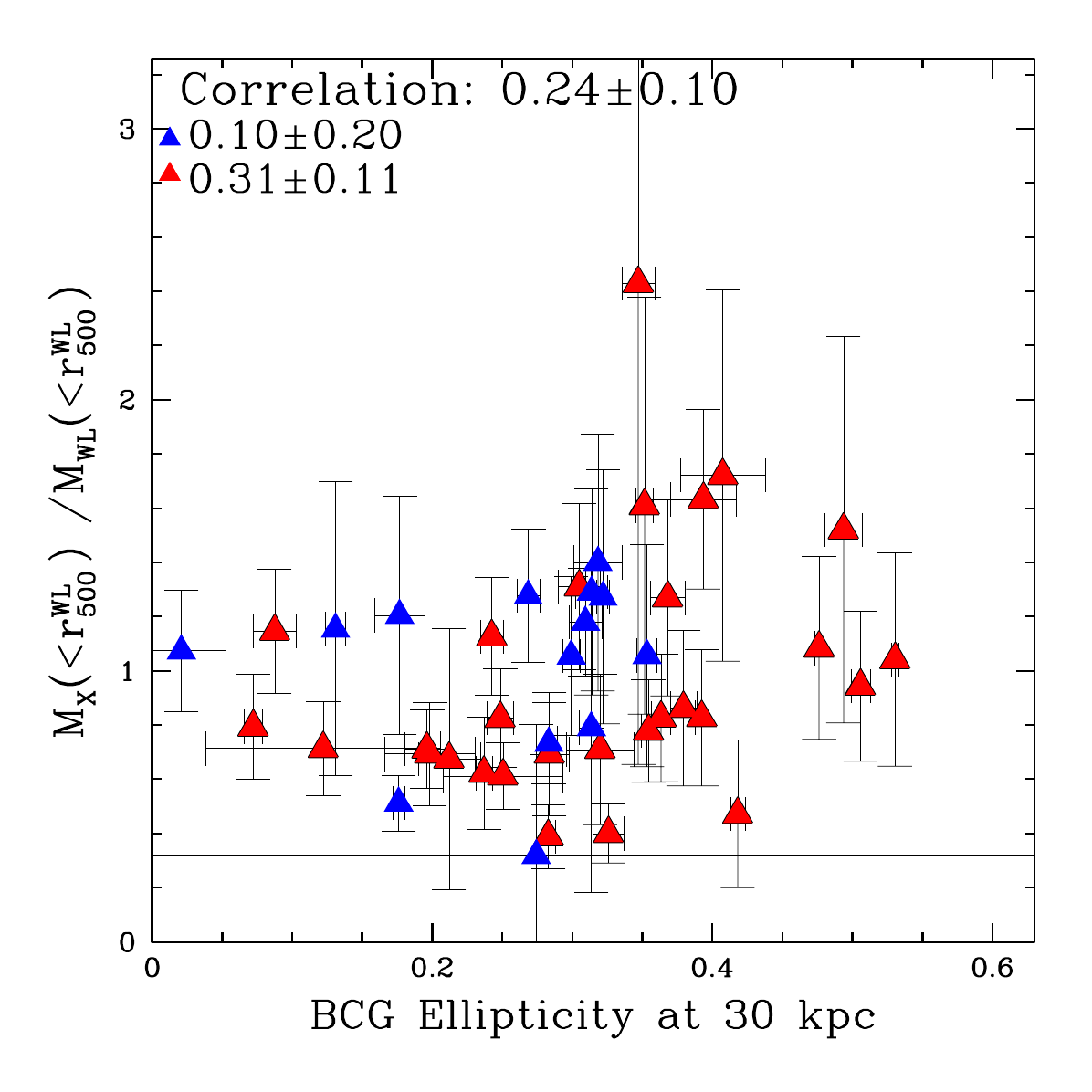}} \\
\end{tabular}
\end{center}
\caption{The X-ray to Weak-Lensing mass ratio as a function 
of BCG ellipticity for 43 BCGs with measureable ellipticities at
30 kpc. The largest error bar in ellipticity belongs to CL0024. Shown are non-cool-core systems (red triangles)
and cool-core systems (blue triangles). 
\protect\hlb{The correlation
is  significant only for the non-cool-core systems at $r_{2500}$; there is a marginal
correlation at $r_{500}$. Cool-core systems do not participate in the trend}.  \label{fig:ell}}
\end{figure*}

\subsection{Lack of Correlation with Substructure Measures}

We have already argued that the intrinsic scatter in the lensing mass
to X-ray observable relations is potentially fully accounted for by
the triaxiality of the clusters; nevertheless, it is still useful to
consider whether the scatter in such relation may be further minimized
via correlation with measures of substructure, at least as an
empirical means to gauge the effect of this triaxiality. However, we
find that none of the substructure measures---BCG offset, central
entropy, centroid shift variance, or power ratio---have any
significant correlation with residuals in the mass-observable
relation. We note that \cite{Marrone12} did find a residual correlation with BCG
\emph{ellipticity} in the relationship between weak lensing mass and
the integrated Sunyaev-Zel'dovich effect signal $Y_\mr{SZ}$.  We examine
a similar relation in \S\ref{sec:ell}.

It follows from this lack of correlation that the $M_x/M_{WL}$ ratio
itself is not correlated with morphological measures such as centroid
shift or power ratio, either. \cite{Rasia12} consider a similar question;
they examine whether the ratio of the X-ray mass to the true mass is 
correlated with centroid shift or power ratio. They find a weak correlation
between this ratio and the substructure measures, with a Pearson rank
coefficient of -0.2 to -0.3, significant at $2\sigma$. We do not observe
such a correlation, most likely because we do not measure
the X-ray to true mass ratio, but rather the X-ray to weak lensing mass
ratio, the latter of which has its own intrinsic scatter.

In our present sample, then, it is possible to minimize the scatter in
the mass-observable relation by conducting a cut on central entropy,
but it is not possible to ``correct'' this scatter for the
non-cool-core clusters by utilizing any of the four substructure
quantifiers we consider or even BCG ellipticity.

\section{Deviations from Hydrostatic Equilibrium}
\label{sec:hydro}
\subsection{Hydrostatic Mass Underestimate}

\cite{Mahdavi08} argued that a subsample of the clusters discussed
here have X-ray masses at $r_{500}^{WL}$ that are on the average
$15\%$ lower than their lensing masses at $r_{500}^{WL}$. This
discrepancy may be attributed to deviations from hydrostatic
equilibrium due to residual gas motions and incomplete thermalization
of the ICM; the fact that hydrostatic masses tend to underestimate the
true mass by 10-20\% was first discussed by \cite{Evrard90} and
continues to be important in grid-based simulations
\citep[e.g.][]{Lau09,Nelson12}, SPH simulations
\citep[e.g.][]{Rasia06,Battaglia12,Rasia12}, and observations of
distant clusters \citep{Andersson11,Jee11}. Biases in gravitational
lensing masses could in principle also affect the X-ray to
weak-lensing mass ratio, such systematic biases are only $\approx
5-10\%$, but would have the effect of increasing the X-ray to weak
lensing mass ratio \citep[e.g.][]{Becker11}. Note that in recent
N-body hydrodynamical simulations, even though the hydrostatic bias
(10-15\%) is roughly twice the level of the weak lensing bias, the
scatter about this bias is larger by a factor of two for weak lensing
masses than for the hydrostatic masses
\citep{Meneghetti10,Rasia12,Nelson12}.

It is worth pointing out, however, that the technique we use for our
lensing mass measurements should yield lower bias than suggested by
these simulations. The technique achieves this lower bias of 3-4\%
(rather than the expected bias of 5-10\%) by omitting the regions of
the shear map that are most susceptible, at the cost of increased statistical
uncertainty. We refer the reader to \cite{Hoekstra12} for details.

In Figure \ref{fig:mhml} we extend our results to the full sample of
50 clusters. The larger size of the sample allows us to resolve
differences between cool-core and non-cool core clusters.  We find that
cool-core clusters and non-cool-core clusters do not exhibit the
same level of departure from hydrostatic equilibrium.

Cool core clusters have hydrostatic masses that are proportional to
their weak lensing masses at all radii. The $M_X-M_L$ relation for
this subsample has a small scatter ($<20\%$), about the right level for
all the scatter to be accounted for by triaxiality. Overall, we find
that cool core clusters are consistent with having no difference
between their X-ray and weak lensing masses.

The picture is dramatically different for non-cool-core clusters. In
these systems, we find a roughly constant hydrostatic mass to lensing
mass ratio of $80\%$, regardless of whether we look at $r_{500}^{WL}$ or
$r_{2500}^{WL}$. Our results are consistent with N-body gas dynamical
simulations
%  However, our results are in slight tension with the
% simulations at $r_{2500}$. At $r_{2500}$ \cite{Lau09} suggest that
% even unrelaxed clusters should have only a $\approx 10\%$
% underestimate on the average, whereas we find an underestimate of
% $20\%$. 
as shown in Figure \ref{fig:mhmlsummary} and Table
\ref{tbl:mhmlsummary}. Broadly, these results are consistent the
hydrostatic mass underestimates predicted by gasdynamical simulations
that account for unthermalized gas, such as \cite{Nagai07},
\cite{Jeltema08} and \cite{Lau09}. We find that the non-cool-core
clusters populate the lower end of the region allowed by these
simulations, whereas the cool-core clusters populate the region
where X-ray and true mass agree within 10\%. Of these simulations
\cite{Jeltema08} is the most consistent with our measured $20\%$
average mass underestimate for disturbed systems.

% \subsection{Correlation with other substructure measures}
% \label{sec:lack}

% \hlb{We now consider whether the discrepancy between lensing and
%   hydrostatic masses can be related to other quantitative measures of
%   substructure, such as $w_X$ or $P3/P0$ power ratio.}

% While a simple cut in entropy (\hly{or BCG offset $D_\mr{BCG}$}) would
% appear to divide the sample into low-underestimate (relaxed) or
% high-underestimate (non-relaxed) subsamples, we find that no further
% reliable corrections are possible. In particular, the ratio of the
% hydrostatic mass to the lensing mass itself does not correlate well
% with any of the four substructure measures (central entropy, BCG
% offset, centroid shift variance, or power ratio). The reason for this
% effect is that, while the \emph{mean} X-ray to lensing mass ratio is
% sensitive to a cut in entropy \hly{or $D_\mr{BCG}$}, the direction of the
% scatter for individual \emph{individual} clusters is too noisy to make
% this correction possible.

\subsection{Correlation with BCG Ellipticity}
\label{sec:ell}

Finally, we consider the question of whether BCG ellipticity is
correlated with differences between hydrostatic and weak lensing
masses. Such a correlation is suggested by \cite{Marrone12}, who use
the integrated Compton parameter $Y_\mr{sph}$ as a mass proxy.  In figure
\ref{fig:ell}, we \hlb{show $M_X / M_\mr{WL}$ at $r_{2500}$ and
  $r_{500}$, plotted against CFHT ellipticities measured at 30kpc}.
\hlo{We find that cool-core systems are consistent with $M_X/M_L=1$
  $(\chi^2/\nu = 18/14)$; whereas non-cool-core systems are
  definitively not consistent with $M_x/M_L=1$ ($\chi^2/\nu=70/29$).}

Therefore, for non-cool-core systems, we find a good correlation
between BCG ellipticity and the X-ray to weak lensing ratio ratio at
$r_{2500}^{WL}$, and a weak correlation at $r_{500}$. \hlo{While this
  is similar to the trend found by \protect\cite{Marrone12} for
  $Y_\mr{sph}$, there is a difference in that our cool-core systems do
  not appear to participate in the correlation}. \hlo{Furthermore,
  also in apparent contrast with \protect\cite{Marrone12}}, \hly{our
  correlation becomes } \hlb{ less significant at $r_{500}^{WL}$}. We
interpret this result as suggesting that while cluster orientation
plays some role in low X-ray to weak lensing mass ratios, it is not
the only agent at work in this complex relationship \hlb{(indeed, the
  hydrostatic mass underestimate must also play a role).}
% \hly{The cool-core cluster have
%   lower than average ellipticities in our sample (cool-core
%   clusters have mean $\epsilon= 0.27$, where as non-cool-core clusters
%   have mean $\epsilon=).}.

We note that it is not altogether surprising that the trend of
ellipticity with $M_X/M_L$ for cool core clusters is insignificant. We
have shown in \S\ref{sec:hydro} that our X-ray and weak-lensing masses
are consistent for this sub-population (in contrast, \cite{Marrone12}
contained several undisturbed clusters with significant $Y_\mr{sph}$ to weak
lensing mass discrepancies).
% \hly{Formally, the low intrinsic scatter reported in
%   \S\protect\ref{sec:hydro} (see also Figure \protect\ref{fig:mhml} and Table
%   \protect\ref{tbl:mhmlsummary}) indicates that cool core clusters, regardless
%   of observed BCG ellipticity, have consistent X-ray and weak lensing
%   masses}.

% \hly{For non-cool-core clusters, our results point to a clear and
%   significant trend of increasing $M_X/M_L$ with ellipticity.}  \hlb{
%  We note that the trend is much more gradual than in} \cite{Marrone12}.
% % \hlb{We find that
% % the phenomenological model
% % }
% % \begin{equation}
% % \frac{M_X(<r_{2500}^{WL})}{M_L(<r_{2500}^{WL})} = (0.5 \pm 0.05) + \epsilon
% % \label{eq:ell}
% % \end{equation}
% % \hlb{adequately fits the non-cool-core clusters, with the relation
% %   having $15\% \pm 7\%$ intrinsic scatter}.  
\hlb{Furthermore,} 
it is difficult to
  untangle the effects of elongation along the line of sight (which
  would chiefly bias weak lensing masses high) and non-hydrostatic gas
  (which would chiefly bias the X-ray masses low). \hlb{We also stress
that the trend is altogether absent at $r_{500}$} However,
  empirically, we can point out that the non-cool-core clusters with
  the highest ellipticities have consistent X-ray and weak lensing
  masses, something corroborated by \protect\cite{Marrone12}.

% \begin{figure*}
% \begin{tabular}{cc}
% \vspace*{-0.2in}\resizebox{3.3in}{!}{\includegraphics{cc-residuals.pdf}} &
% \resizebox{3.3in}{!}{\includegraphics{ncc-residuals.pdf}} \\
% \end{tabular}
% \caption{Residuals in the  hydrostatic mass to lensing mass relation
%   $r_{500}$ compared to the P3/P0 power ratio. Blue squares (left) show
%   cool-core clusters and red squares (right) show non-cool-core
%   clusters. The P3/P0 ratio appears uncorrelated with cool core
%   residuals, but significantly correlated with non-cool-core residuals.
% \label{fig:nccp3}}
% \end{figure*}

\begin{deluxetable*}{lccrccc}
\tablecaption{Mass Proxy Fits with Lognormal Intrinsic Scatter \label{tbl:scaling}}
\tablehead{ & \colhead{Proxy} & \colhead{$M_{WL}$} & & \colhead{Log} & \colhead{Log} & \colhead{Fractional Scatter} \\
Proxy&Aperture&Aperture& Sample & \colhead{Slope} & \colhead{Intercept} & \colhead{in $M_{WL}$ at fixed proxy}}
\startdata
\multicolumn{7}{c}{Relations at Fixed Overdensity in Proxy and Mass} \\
\hline 
$T^{cut}_X/8$ keV  & $r_{500}^{WL}$ & $r_{500}^{WL}$ & all & $1.97\pm0.89$ & $1.04 \pm 0.06$ & $0.46\pm0.23$  \\
$T^{cut}_X/8$ keV & $r_{500}^{X}$ & $r_{500}^{X}$ & all &  $1.42\pm0.19$ & $0.96\pm0.02$ & $0.17\pm0.08$  \\
\\
$L^{cut}_X E(z)^{-1} $  & $r_{500}^{WL}$ & $r_{500}^{WL}$ & all & $0.45\pm0.10$ & $0.93\pm0.03$ & $0.35\pm0.07$ \\
$L^{cut}_X E(z)^{-1} $  & $r_{500}^{X}$ & $r_{500}^{X}$ &  all & $0.50\pm0.06$ & $0.91\pm0.02$ & $0.26\pm0.05$ \\
\\
$M_\mr{Gas} E(z)$ & $r_{500}^{WL}$ & $r_{500}^{WL}$ & all & $1.04\pm0.10$ & $0.90\pm0.02$ & $0.15\pm0.06$ \\
&&&{$K_0<70$ keV cm$^2$} & $0.91\pm0.20$ & $0.89\pm0.03$ & $<0.1$ \\
&&&{$K_0>70$ keV cm$^2$} & $1.09\pm0.13$ & $0.90\pm0.02$ & $0.18\pm0.09$ \\
&&&{$D_\mr{BCG} < $ 0.01 Mpc} & $0.93\pm0.13$ & $0.89\pm0.02$ & $<0.06$ \\
&&&{$D_\mr{BCG} > $0.01 Mpc} & $1.13\pm0.18$ & $0.90\pm0.03$ & $0.22\pm0.15$ \\
\\
$Y_X E(z)^{0.6}$ & $r_{500}^{WL}$ & $r_{500}^{WL}$ & all & $0.56\pm0.07$ & $0.45\pm0.07$ & $0.22\pm0.05$ \\
&&& {$K_0<70$ keV cm$^2$}& $0.44\pm0.14$ & $0.53\pm0.11$ & $0.24\pm0.18$ \\
&&& {$K_0>70$ keV cm$^2$}& $0.62\pm0.10$ & $0.41\pm0.09$ & $0.21\pm0.09$ \\
&&& {$D_\mr{BCG} < $ 0.01} & $0.48\pm0.09$ & $0.52\pm0.08$ & $0.17\pm0.11$ \\
&&& {$D_\mr{BCG} > $0.01} & $0.65\pm0.14$ & $0.36\pm0.13$ & $0.27\pm0.17$ \\
\\
\multicolumn{7}{c}{Relations at Other Radii} \\
\hline
$T^{cut}_X$/8 keV (keV) & 1 Mpc & 1 Mpc & all & $1.10\pm0.57$ & $0.80\pm0.02$ & $0.15\pm0.11$ \\
$L^{cut}_X$ &  "& " & all & $0.23\pm0.06$ & $0.79\pm0.02$ & $0.19\pm0.04$ \\
$M_{Gas} $ & " & " & all & $0.83\pm0.14$ & $0.90\pm0.03$ & $0.16\pm0.10$ \\
$Y_X $ & " & " & all & $0.40\pm0.06$ & $0.48\pm0.05$ & $0.12\pm0.04$ \\
$T^{cut}_X$/8 keV & " & $r_{500}^{WL}$ & all & $3.04\pm1.38$ & $1.03\pm0.08$ & $0.46\pm0.31$ \\
$L^{cut}_X $ & " & $r_{500}^{WL}$& all &  $0.50\pm0.13$ & $0.96\pm0.03$ & $0.38\pm0.07$ \\
$M_{Gas}$ & " & $r_{500}^{WL}$& all & $1.73\pm0.59$ & $1.20\pm0.13$ & $0.39\pm0.18$ \\
$Y_X$ & " & $r_{500}^{WL}$ & all & $0.80\pm0.15$ & $0.35\pm0.11$ & $0.28\pm0.14$ \\
\enddata
\tablecomments{All proxies are fit against $M_{WL} E(z)$ at an
  aperture of $r_{500}^{WL}$ or $M_{WL}$ at an aperture of 1 Mpc. All
  masses are in units of $10^{14} M_\odot$. The core-cut X-ray
  luminosity is in units of $10^{45}$ erg s$^{-1}$, and $Y_X$ is in
  units of $10^{14} M_\odot$ keV.  }
\end{deluxetable*}

\begin{deluxetable*}{rlccc}
\tablecaption{X-ray to Weak Lensing Mass Ratios \label{tbl:mhmlsummary}}
\tablehead{Contrast & Sample & $M_X/M_L$ & \colhead{Fractional Scatter} \\
&&&\colhead{in $M_X$ at fixed $M_L$}}
\startdata
$r_{2500}^\mr{WL}$& All & $0.92\pm0.05$ & $0.19\pm0.05$ \\
& $K_0 < 70$ keV cm$^{2}$ & $1.11\pm0.10$ & $<10\%$ \\
& $K_0 > 70$ keV cm$^{2}$ & $0.85\pm0.05$ & $0.19\pm0.06$ \\
& $D_\mr{BCG} < 0.01$ Mpc & $1.04\pm0.07$ & $<0.15$ \\
& $D_\mr{BCG} > 0.01$ Mpc & $0.81\pm0.07$ & $0.24\pm0.07$ \\
\\
$r_{1000}^\mr{WL}$& All & $0.89\pm0.05$ & $0.20\pm0.05$ \\
& $K_0 < 70$ keV cm$^{2}$ & $1.08\pm0.09$ & $<9\%$ \\
& $K_0 > 70$ keV cm$^{2}$ &$0.83\pm0.06$ & $0.20\pm0.06$ \\
& $D_\mr{BCG} < 0.01$ Mpc & $0.97\pm0.07$ & $0.13\pm0.10$ \\
& $D_\mr{BCG} > 0.01$ Mpc & $0.84\pm0.06$ & $0.22\pm0.07$ \\
\\
$r_{500}^\mr{WL}$ & All & $0.88\pm0.05$ & $0.21\pm0.06$ \\
& $K_0 < 70$ keV cm$^{2}$ & $0.97\pm0.10$ & $0.17\pm0.13$ \\
& $K_0 > 70$ keV cm$^{2}$ & $0.83\pm0.07$ & $0.22\pm0.07$ \\
& $D_\mr{BCG} < 0.01$ Mpc & $0.85\pm0.09$ & $0.22\pm0.11$ \\
& $D_\mr{BCG} > 0.01$ Mpc & $0.89\pm0.07$ & $0.20\pm0.08$ \\
\end{deluxetable*}

\section{Conclusion}
\label{sec:conclusion}

We examine archival X-ray data on a sample of 50 clusters of galaxies;
most of the clusters have \emph{Chandra} data, while roughly half have
\emph{XMM-Newton} data of good quality. All clusters have CFHT weak
gravitational lensing data from either the CFH12k or the Megacam
instruments.  

In attempting to combine \emph{Chandra} and \emph{XMM-Newton} data to
maximize both effective area and spatial resolution, we confirm
previously reported systematic calibration differences between the two
observatories. Using multiple calibration releases, we find a
$15\%$ systematic difference in hydrostatic masses between
\emph{Chandra} and \emph{XMM-Newton}. Reassuringly, there is no
intrinsic scatter between the masses for the two observatories,
indicating that the issue is merely a matter of overall gain
calibration and not a more serious spatially dependent issue. We
develop an effective area correction that revises \emph{Chandra}
masses downward into agreement with \emph{XMM-Newton} masses. This
correction is only valid for high temperature ($\gtrsim 5$ keV)
clusters such as ours; at lower temperatures, the two observatories
are more consistent due to the abundant prominence of X-ray lines.

Using the $L_X-T_X$ relation, we find that our sample is consistent with
being randomly drawn from the same parent population as samples with
well understood selection functions, such as HIFLUGS
\citep{Reiprich02} and MACS \citep{Ebeling10}.

We examine several measures of substructure, including central
entropy, BCG to X-ray peak offset, centroid shift variance, and power
ratios. There is a significant correlation among all the substructure
measures. The most strikingly correlated quantities are the BCG to
X-ray peak offset (in Mpc) and the central entropy measured at a
radius of 20 kpc. The hint of bimodality in the joint 2D distribution
of the BCG offset and central entropy indicates a complex connection
between the thermal and dynamical relaxation times of galaxy clusters.

Gas mass is by far the most robust predictor of weak lensing mass,
with $<10\%$ scatter for cool-core clusters and $14\% \pm 6\%$ scatter
for the sample overall. It is followed by the X-ray pseudo-pressure,
$Y_X$, which has $22\% \pm 5$ intrinsic scatter for both cool core
clusters and the sample overall. The mass-temperature relationship has
even higher scatter, $43\% \pm 21\%$ for the sample
overall. All scaling relations have slopes that are consistent with
the expected self-similar value. \hly{We also find that
core-excised X-ray luminosity is somewhat better than temperature at predicting
weak lensing mass, yielding $28\%\pm18\%$
intrinsic scatter for relaxed systems}.

By comparing hydrostatic and weak gravitational lensing masses, we
extend our earlier detection \citep{Mahdavi08} of \hly{non-hydrostatic
  gas, with associated deviations from} hydrostatic equilibrium, in
X-ray clusters of galaxies. We are able to quantify the hydrostatic
mass underestimate separately for cool-core and non-cool-core
clusters. We find that cool-core clusters exhibit \hly{little or no
  difference} between their weak lensing and X-ray masses; the
hydrostatic mass underestimate is consistent with 0\% at both
$r_{2500}^{WL}$ and at $r_{500}^{WL}$. Non cool-core clusters, on the
other hand, have fairly consistent, $\approx 20\% \pm 10\%$,
underestimates \hly{between the same radii}.  \hly{This is broadly
  consistent with N-body gasdynamical simulations of unthermalized
  gas}.
  
Except for the non-core-cut $L_X$-$T_X$ relation, we do not find a
significant correlation between the \emph{residuals} in a given
scaling relation and any of our four substructure measures (central
entropy, BCG offset, centroid shift variance or P3/P0 power ratio). We
interpret this result as indicating that it is not possible to reduce
the intrinsic scatter in a scaling relation (other than the $L_X-T_X$
relation) by applying corrections based on substructure measures to
individual clusters. In essence, clusters of galaxies have
``forgotten'' the sources of their departures from
self-similarity. \hlb{This lack of correlation suggests that we may
  have accounted for most if not all the parameters that could affect
  the cluster selection function for cosmological surveys, and that
  few if any ``hidden'' parameters remain.}

\hly{However, we do find a partial trend with cluster ellipticity: cool-core
  clusters have consistent X-ray and weak lensing masses at
  $r_{2500}^{WL}$; whereas non-cool-core clusters have
  increasing 
  $M_X(<r_{2500}^{WL})/M_L(<r_{2500}^{WL})$ with
  BCG ellipticity at 30 kpc. Clusters with low
  ellipticity BCGs are the most likely to have mismatched X-ray and weak
  lensing masses, while clusters with higher ellipticity are more likely
  to have concordant X-ray and weak lensing masses. We leave it to
  future studies to determine which combination of X-ray
  non-hydrostatic bias and lensing projection bias is contributing to
  this trend.}

We emphasize that the X-ray peak to BCG location offset is perhaps the
most efficient among our inspected substructure measures. Selecting
clusters based on low BCG offset is sufficient to guarantee scatter
consistent with zero in the gas mass-lensing mass relation, at least
for a sample as large or larger than ours.

In summary, we find that cool-core clusters with $K_0< 70$ keV
cm$^{2}$ or BCG offset \hlo{$<0.01$ Mpc} are extremely well-behaved
and regular systems with respect to their X-ray and lensing
properties. \hlo{However, it should be noted that the two cuts do not
  select the same subsamples, because low BCG offset is indicative of
  the dynamical equilibrium, whereas low central entropy is a result
  of thermal equilibrium. While there are clusters that are in both
  thermal and dynamical equilibrium, the overlap is not perfect.}

Clusters with $K_0 > 70$ keV cm$^{2}$ show some intriguing
properties---such as tightly correlated \hlo{P3/P0 power ratios} and
BCG offsets, \hlo{a linear correlation between $M_X/M_L$ and
  ellipticity}, and consistently low X-ray to weak lensing mass
ratios---but larger samples and more careful theoretical studies are
required before we can learn how to use these relations to gain
greater physical insight into their evolution.

\acknowledgments{ The authors would like to acknowledge productive
  discussions with Steve Allen, Hans B\"ohringer, Dick Bond,
  Maru\u{s}a Brada\u{c}, Megan Donahue, Stefano Ettori, Gus Evrard,
  Fabio Gastaldello, Andrey Kravtsov, Dan Marrone, Daisuke Nagai,
  Trevor Ponman, Graham Smith, David Spergel, and Mark Voit.  The
  anonymous referee made comments which improved the paper. AM
  \hly{and TJ} were supported by NASA through Chandra award
  No. AR0-11016A, issued by the Chandra X-ray Observatory Center,
  which is operated by the Smithsonian Astrophysical Observatory for
  and on behalf of NASA under contract NAS8-03060. AM was also
  supported \hly{through NASA ADAP grant 11-ADAP11-0270. AB would also
    like to acknowledge research funding from NSERC Canada through its
    Discovery Grant program as well as support provided by
    J. Criswick}. \hlb{HH acknowledges support from the Netherlands
    organisation for Scientific Research (NWO) through VIDI grant
    639.042.814; HH and CB acknowledge support from Marie Curie IRG
    Grant 230924.} AM and AB acknowledge an especially productive time
  at the Kavli Institute for Theoretical Physics, where this research
  was supported in part by the National Science Foundation under Grant
  No. NSF PHY11-25915.  } \vspace{0.2in}

\vspace{0.1in}

\input{ms.bbl}
\end{document}

%% file: defs.tex
\usepackage{ifthen}
\newcommand{\mr}[1]{\mathrm{#1}}

\ifthenelse{\isundefined{\nonaas}}{}{

\bibpunct{(}{)}{;}{a}{}{,}

  \renewenvironment{thebibliography}[1]{%
    \begin{oldthebibliography}{#1}%
      \setlength{\parskip}{0ex}%
      \parindent 0ex%
      \setlength{\itemsep}{0ex}%
  }%
  {%
    \end{oldthebibliography}%
  }

\def\aj{{AJ}}			% Astronomical Journal
\def\araa{{ARA\&A}}		% Annual Review of Astron and Astrophys
\def\apj{{ApJ}}			% Astrophysical Journal
\def\apjl{{ApJ}}		% Astrophysical Journal, Letters
\def\apjs{{ApJS}}		% Astrophysical Journal, Supplement
\def\aap{{A\&A}}		% Astronomy and Astrophysics
\def\mnras{{MNRAS}}		% Monthly Notices of the RAS
}

\usepackage{ifpdf}
\usepackage{graphicx}

%% file: table1.tex
   \begin{deluxetable*}{lrrrrrrrrr}
   \tablecaption{Basic Properties of the Sample \label{tbl:sample}}
   \tablehead{
   \colhead{Cluster} & \colhead{RA} & \colhead{DEC} & 
\colhead{$z$} & \colhead{{\it Chandra} } & \colhead{Exposure} & \colhead{{\it XMM-Newton} }
 & \colhead{Exposure} & \colhead{$L_\mr{X,all,bol,500}$} & \colhead{$T_\mr{all,500}$} \\
   \colhead{Name} & \colhead{J2000} & \colhead{J2000}  & & \colhead{ObsID} & \colhead{s} & \colhead{ObsID }
 & \colhead{s} & \colhead{keV} & \colhead{$10^{45}$ erg s$^{-1}$}}
   \startdata
3C295 & 14:11:20.52 & +52:12:09.9 & 0.464 &2254 & 87914&\nodata&\nodata & $1.16\pm0.02$ & $5.7\pm0.3$ \\
 Abell0068 & 00:37:06.65 & +09:09:24.0 & 0.255 &3250 & 9986&0084230201 & 14068 & $1.43\pm0.02$ & $6.3\pm0.4$ \\
 Abell0115N & 00:55:50.37 & +26:24:36.6 & 0.197 &3233 & 49719&0203220101 & 21393 & $1.11\pm0.01$ & $5.3\pm0.2$ \\
 Abell0115S & 00:56:00.17 & +26:20:29.5 & 0.197 &3233 & 49719&0203220101 & 21309 & $1.17\pm0.01$ & $5.4\pm0.3$ \\
 Abell0209 & 01:31:53.42 & -13:36:46.3 & 0.206 &3579 & 9986&0084230301 & 11219 & $1.77\pm0.02$ & $7.0\pm0.4$ \\
 Abell0222 & 01:37:34.25 & -12:59:30.8 & 0.207 &4967 & 45078&0502020201 & 23178 & $0.50\pm0.01$ & $4.1\pm0.3$ \\
 Abell0223S & 01:37:56.06 & -12:49:12.8 & 0.207 &4967 & 45078&0502020201 & 23206 & $0.62\pm0.01$ & $6.3\pm0.4$ \\
 Abell0267 & 01:52:42.38 & +01:00:48.0 & 0.231 &3580 & 19624&0084230401 & 10421 & $1.15\pm0.03$ & $6.5\pm0.5$ \\
 Abell0370 & 02:39:53.18 & -01:34:34.9 & 0.375 &515 & 68532&\nodata&\nodata & $1.55\pm0.06$ & $7.2\pm0.6$ \\
 Abell0383 & 02:48:03.33 & -03:31:45.1 & 0.187 &2320 & 19285&0084230501 & 20237 & $0.96\pm0.01$ & $3.7\pm0.1$ \\
 Abell0520 & 04:54:10.10 & +02:55:18.3 & 0.199 &4215 & 66274&0201510101 & 21915 & $1.58\pm0.02$ & $7.3\pm0.2$ \\
 Abell0521 & 04:54:06.30 & -10:13:16.9 & 0.253 &901 & 38626&\nodata&\nodata & $1.37\pm0.02$ & $5.9\pm0.3$ \\
 Abell0586 & 07:32:20.16 & +31:37:56.6 & 0.171 &530 & 10043&\nodata&\nodata & $1.19\pm0.02$ & $5.1\pm0.4$ \\
 Abell0611 & 08:00:56.96 & +36:03:22.0 & 0.288 &3194 & 36114&\nodata&\nodata & $1.61\pm0.05$ & $8.7\pm0.6$ \\
 Abell0697 & 08:42:57.29 & +36:21:56.2 & 0.282 &4217 & 19516&\nodata&\nodata & $3.15\pm0.07$ & $10.3\pm0.7$ \\
 Abell0851 & 09:43:00.39 & +46:59:20.4 & 0.407 &\nodata&\nodata&0106460101 & 15731 & $0.93\pm0.03$ & $6.1\pm0.4$ \\
 Abell0959 & 10:17:35.61 & +59:33:53.4 & 0.286 &\nodata&\nodata&0406630201 & 4134 & $0.58\pm0.03$ & $6.0\pm1.7$ \\
 Abell0963 & 10:17:03.63 & +39:02:48.3 & 0.206 &903 & 36289&0084230701 & 17234 & $1.57\pm0.01$ & $6.2\pm0.2$ \\
 Abell1689 & 13:11:29.52 & -01:20:29.8 & 0.183 &6930 & 76144&0093030101 & 24457 & $4.48\pm0.02$ & $10.9\pm0.2$ \\
 Abell1758E & 13:32:46.43 & +50:32:25.9 & 0.279 &2213 & 55220&\nodata&\nodata & $2.33\pm0.06$ & $8.9\pm0.6$ \\
 Abell1758W & 13:32:38.70 & +50:33:23.0 & 0.279 &2213 & 55220&\nodata&\nodata & $1.46\pm0.06$ & $8.7\pm1.0$ \\
 Abell1763 & 13:35:18.16 & +40:59:57.7 & 0.223 &3591 & 19595&0084230901 & 8852 & $1.91\pm0.03$ & $6.3\pm0.2$ \\
 Abell1835 & 14:01:01.90 & +02:52:42.7 & 0.253 &6880 & 117918&0098010101 & 16021 & $4.51\pm0.02$ & $7.1\pm0.1$ \\
 Abell1914 & 14:26:02.80 & +37:49:27.3 & 0.171 &3593 & 18865&0112230201 & 17025 & $2.97\pm0.02$ & $9.7\pm0.2$ \\
 Abell1942 & 14:38:21.90 & +03:40:12.9 & 0.224 &3290 & 55716&\nodata&\nodata & $0.33\pm0.01$ & $4.0\pm0.3$ \\
 Abell2104 & 15:40:08.09 & -03:18:16.5 & 0.153 &895 & 49199&\nodata&\nodata & $2.32\pm0.04$ & $5.8\pm0.3$ \\
 Abell2111 & 15:39:41.74 & +34:25:01.9 & 0.229 &544 & 10299&\nodata&\nodata & $0.91\pm0.03$ & $5.6\pm0.7$ \\
 Abell2163 & 16:15:46.05 & -06:09:02.6 & 0.203 &1653 & 71148&\nodata&\nodata & $6.45\pm0.10$ & $11.5\pm0.7$ \\
 Abell2204 & 16:32:46.92 & +05:34:32.4 & 0.152 &7940 & 77141&0306490201 & 13093 & $3.99\pm0.01$ & $7.6\pm0.1$ \\
 Abell2218 & 16:35:50.89 & +66:12:36.9 & 0.176 &1666 & 30693&0112980101 & 13111 & $1.28\pm0.02$ & $6.6\pm0.2$ \\
 Abell2219 & 16:40:20.20 & +46:42:35.3 & 0.226 &896 & 42295&\nodata&\nodata & $6.45\pm0.08$ & $8.3\pm0.3$ \\
 Abell2259 & 17:20:07.75 & +27:40:14.7 & 0.164 &3245 & 9986&\nodata&\nodata & $0.77\pm0.02$ & $5.0\pm0.3$ \\
 Abell2261 & 17:22:27.12 & +32:07:58.9 & 0.224 &5007 & 24316&\nodata&\nodata & $2.59\pm0.03$ & $5.6\pm0.3$ \\
 Abell2390 & 21:53:36.82 & +17:41:44.7 & 0.228 &4193 & 93782&0111270101 & 8100 & $5.99\pm0.03$ & $10.1\pm0.3$ \\
 Abell2537 & 23:08:22.23 & -02:11:30.3 & 0.295 &4962 & 36193&0205330501 & 6267 & $1.37\pm0.03$ & $6.9\pm0.7$ \\
 CL0024.0+1652 & 00:26:35.94 & +17:09:46.2 & 0.390 &929 & 39417&\nodata&\nodata & $0.34\pm0.02$ & $4.5\pm0.8$ \\
 MACSJ0717.5+3745 & 07:17:31.39 & +37:45:24.8 & 0.548 &4200 & 58912&\nodata&\nodata & $5.55\pm0.12$ & $11.6\pm0.7$ \\
 MACSJ0913.7+4056 & 09:13:45.49 & +40:56:28.7 & 0.442 &10445 & 76159&\nodata&\nodata & $1.99\pm0.04$ & $6.0\pm0.2$ \\
 MS0015.9+1609 & 00:18:33.74 & +16:26:09.0 & 0.541 &520 & 67410&0111000101 & 22477 & $3.73\pm0.10$ & $9.5\pm0.6$ \\
 MS0440.5+0204 & 04:43:09.99 & +02:10:19.3 & 0.190 &4196 & 22262&\nodata&\nodata & $0.33\pm0.01$ & $3.3\pm0.3$ \\
 MS0451.6-0305 & 04:54:11.24 & -03:00:57.3 & 0.550 &902 & 43420&\nodata&\nodata & $3.25\pm0.12$ & $9.2\pm0.8$ \\
 MS0906.5+1110 & 09:09:12.73 & +10:58:28.4 & 0.174 &924 & 29752&\nodata&\nodata & $1.10\pm0.02$ & $5.6\pm0.3$ \\
 MS1008.1-1224 & 10:10:32.52 & -12:39:53.1 & 0.301 &926 & 25222&\nodata&\nodata & $0.77\pm0.02$ & $5.9\pm0.5$ \\
 MS1231.3+1542 & 12:33:55.01 & +15:26:02.3 & 0.233 &\nodata&\nodata&0404120101 & 26520 & $0.29\pm0.01$ & $4.5\pm0.2$ \\
 MS1358.1+6245 & 13:59:50.56 & +62:31:05.3 & 0.328 &516 & 50989&\nodata&\nodata & $1.19\pm0.03$ & $6.2\pm0.5$ \\
 MS1455.0+2232 & 14:57:15.05 & +22:20:33.2 & 0.258 &4192 & 91626&0108670201 & 22571 & $1.84\pm0.01$ & $4.4\pm0.1$ \\
 MS1512.4+3647 & 15:14:22.47 & +36:36:20.9 & 0.372 &800 & 36400&\nodata&\nodata & $0.48\pm0.02$ & $3.1\pm0.2$ \\
 MS1621.5+2640 & 16:23:35.05 & +26:34:22.1 & 0.426 &546 & 30062&\nodata&\nodata & $0.89\pm0.04$ & $6.9\pm0.8$ \\
 RXJ1347.5-1145 & 13:47:30.59 & -11:45:09.8 & 0.451 &3592 & 57458&0112960101 & 21712 & $10.96\pm0.18$ & $12.1\pm0.4$ \\
 RXJ1524.6+0957 & 15:24:38.85 & +09:57:41.8 & 0.520 &1664 & 49849&\nodata&\nodata & $0.41\pm0.03$ & $4.7\pm1.2$ \\
 \end{deluxetable*}

%% file: table2.tex
\begin{deluxetable*}{lrrrrr@{$\ \pm\ $}rrrrr}
\tablecaption{Mass and Substructure Properties at $r_{500}$ \label{tbl:data}}
\tablehead{
 \colhead{Cluster} & \colhead{$r_{500}^{\mr{WL}}$} & \colhead{$M_\mr{WL}$} & \colhead{$M_\mr{Gas}$} & 
\colhead{$M_\mr{hydro}$} &
\multicolumn{2}{c}{$K_0$} &
\colhead{$D_\mr{BCG}$} &
\colhead{$w_X$} &
\colhead{$P3/P0$} \\
& \colhead{Mpc} & 
\colhead{$10^{14} M_\odot$} & \colhead{$10^{14} M_\odot$} & \colhead{$10^{14} M_\odot$} & \multicolumn{2}{c}{keV cm$^2$} & kpc & $10^{3} \times r_{500}^{\mr{WL}}$  & $\times 10^{-7}$
}
\startdata
3C295 &  $1.06\pm0.06$ & $5.7\pm1.2$ & $0.62\pm0.03$ & $3.9\pm1.0$ & $12.8$&$2.4$ & $12$ & $5.7\pm1.4$ & $0.25\pm0.20$ \\
 Abell0068 &  $1.16\pm0.09$ & $5.9\pm1.6$ & $0.77\pm0.01$ & $5.1\pm1.0$ & $214.2$&$29.5$ & $15$ & $14.0\pm2.1$ & $0.74\pm0.60$ \\
 Abell0115N &  $1.03\pm0.12$ & $3.9\pm1.5$ & $0.60\pm0.01$ & $4.1\pm0.2$ & $30.0$&$2.4$ & $10$ & $59.6\pm0.8$ & $3.31\pm0.82$ \\
 Abell0115S &  $1.14\pm0.07$ & $5.3\pm1.2$ & $0.80\pm0.01$ & $4.2\pm0.3$ & $192.8$&$48.5$ & $143$ & \nodata & \nodata \\
Abell0209 &  $1.24\pm0.07$ & $6.8\pm1.4$ & $1.02\pm0.02$ & $5.6\pm1.1$ & $152.7$&$20.9$ & $16$ & $7.1\pm1.1$ & $1.47\pm1.10$ \\
 Abell0222 &  $1.16\pm0.07$ & $5.7\pm1.3$ & $0.61\pm0.01$ & $2.4\pm0.6$ & $220.2$&$32.2$ & $10$ & $38.4\pm3.2$ & $1.21\pm0.96$ \\
 Abell0223S &  $1.24\pm0.10$ & $6.9\pm2.0$ & $0.68\pm0.04$ & $3.3\pm1.6$ & $133.5$&$20.1$ & $8$ & $36.9\pm2.4$ & $2.26\pm1.62$ \\
 Abell0267 &  $1.13\pm0.09$ & $5.3\pm1.5$ & $0.63\pm0.01$ & $5.7\pm0.6$ & $160.8$&$20.6$ & $77$ & $22.0\pm1.7$ & $0.27\pm0.22$ \\
 Abell0370 &  $1.43\pm0.06$ & $12.8\pm2.0$ & $1.00\pm0.05$ & $8.6\pm6.0$ & $500.1$&$159.8$ & $23$ & $15.8\pm1.3$ & $0.62\pm0.49$ \\
 Abell0383 &  $1.04\pm0.13$ & $4.0\pm1.8$ & $0.39\pm0.01$ & $4.6\pm0.6$ & $21.3$&$1.0$ & $<3 $ & $3.1\pm0.6$ & $0.36\pm0.26$ \\
 Abell0520 &  $1.16\pm0.07$ & $5.6\pm1.3$ & $0.85\pm0.01$ & $7.3\pm0.3$ & $590.1$&$39.4$ & $341$ & $100.7\pm0.7$ & $4.66\pm0.97$ \\
 Abell0521 &  $1.19\pm0.08$ & $6.3\pm1.6$ & $1.06\pm0.02$ & $5.0\pm1.3$ & $75.6$&$18.1$ & $33$ & $58.5\pm1.7$ & $8.59\pm3.41$ \\
 Abell0586 &  $1.18\pm0.09$ & $5.6\pm1.6$ & $0.65\pm0.08$ & $3.9\pm0.6$ & $140.1$&$23.5$ & $11$ & $7.3\pm1.6$ & $0.59\pm0.49$ \\
 Abell0611 &  $1.13\pm0.06$ & $5.7\pm1.3$ & $0.66\pm0.05$ & $6.0\pm0.9$ & $57.0$&$9.0$ & $4$ & $8.0\pm0.8$ & $0.68\pm0.42$ \\
 Abell0697 &  $1.35\pm0.05$ & $9.7\pm1.3$ & $1.56\pm0.03$ & $10.9\pm1.5$ & $240.0$&$45.4$ & $20$ & $9.1\pm1.1$ & $0.20\pm0.16$ \\
 Abell0851 &  $1.32\pm0.09$ & $10.5\pm2.5$ & $0.97\pm0.02$ & $7.4\pm2.3$ & $479.7$&$79.7$ & $278$ & $30.5\pm3.4$ & $13.15\pm7.51$ \\
 Abell0959 &  $1.26\pm0.07$ & $7.8\pm1.7$ & $0.75\pm0.03$ & $5.6\pm0.5$ & $203.8$&$23.7$ & $36$ & $42.6\pm4.5$ & $7.70\pm6.55$ \\
 Abell0963 &  $1.00\pm0.10$ & $3.7\pm1.3$ & $0.57\pm0.01$ & $4.7\pm0.5$ & $63.1$&$5.9$ & $6$ & $5.5\pm0.5$ & $0.12\pm0.10$ \\
 Abell1689 &  $1.57\pm0.09$ & $13.7\pm2.7$ & $1.27\pm0.01$ & $9.7\pm0.6$ & $72.5$&$5.5$ & $5$ & $4.1\pm0.3$ & $0.08\pm0.04$ \\
 Abell1758E &  $1.37\pm0.08$ & $10.1\pm2.3$ & $1.23\pm0.04$ & $9.4\pm0.6$ & $227.4$&$28.5$ & $25$ & $117.9\pm1.2$ & $8.42\pm1.62$ \\
 Abell1758W &  $1.37\pm0.06$ & $10.0\pm1.4$ & $0.93\pm0.07$ & $11.5\pm1.6$ & $194.5$&$19.9$ & $25$ & $117.9\pm1.2$ & $8.42\pm1.62$ \\
 Abell1763 &  $1.40\pm0.10$ & $10.1\pm2.5$ & $1.34\pm0.01$ & $3.9\pm0.7$ & $419.5$&$54.0$ & $7$ & $22.9\pm1.2$ & $0.97\pm0.64$ \\
 Abell1835 &  $1.30\pm0.05$ & $8.4\pm1.3$ & $1.21\pm0.01$ & $9.9\pm0.7$ & $19.7$&$0.4$ & $6$ & $3.9\pm0.2$ & $< 0.1$ \\
Abell1914 &  $1.18\pm0.05$ & $5.6\pm1.0$ & $0.99\pm0.00$ & $9.2\pm0.9$ & $128.7$&$9.5$ & $86$ & $27.8\pm0.5$ & $2.39\pm0.40$ \\
 Abell1942 &  $1.05\pm0.06$ & $4.3\pm1.0$ & $0.44\pm0.01$ & $2.7\pm0.6$ & $230.6$&$72.2$ & $4$ & $9.3\pm1.4$ & $1.57\pm1.21$ \\
 Abell2104 &  $1.22\pm0.08$ & $6.1\pm1.6$ & $0.68\pm0.14$ & $5.8\pm0.8$ & $201.7$&$44.2$ & $8$ & \nodata & \nodata \\
Abell2111 &  $1.07\pm0.10$ & $4.5\pm1.5$ & $0.74\pm0.07$ & $7.3\pm2.5$ & $203.8$&$55.6$ & $129$ & $33.0\pm2.8$ & $3.22\pm2.43$ \\
 Abell2163 &  $1.38\pm0.11$ & $9.5\pm2.5$ & $2.33\pm0.03$ & $12.0\pm1.2$ & $336.0$&$18.0$ & $160$ & $35.3\pm0.4$ & $3.76\pm0.37$ \\
 Abell2204 &  $1.34\pm0.07$ & $8.1\pm1.6$ & $1.16\pm0.01$ & $8.7\pm0.6$ & $17.3$&$0.3$ & $<3 $ & $4.8\pm0.3$ & $< 0.1$ \\
Abell2218 &  $1.14\pm0.08$ & $5.1\pm1.4$ & $0.72\pm0.01$ & $4.3\pm0.6$ & $317.9$&$44.9$ & $60$ & $18.9\pm1.0$ & $1.28\pm0.53$ \\
 Abell2219 &  $1.35\pm0.07$ & $9.1\pm1.9$ & $1.65\pm0.03$ & $7.1\pm0.9$ & $243.2$&$33.3$ & $8$ & \nodata & \nodata \\
Abell2259 &  $1.05\pm0.09$ & $4.0\pm1.2$ & $0.50\pm0.04$ & $4.1\pm0.9$ & $134.7$&$30.1$ & $78$ & $24.1\pm1.7$ & $1.18\pm0.95$ \\
 Abell2261 &  $1.52\pm0.05$ & $12.9\pm1.6$ & $1.46\pm0.13$ & $6.6\pm1.0$ & $60.0$&$9.0$ & $<2 $ & $14.3\pm1.0$ & $0.39\pm0.21$ \\
 Abell2390 &  $1.33\pm0.06$ & $8.6\pm1.5$ & $1.48\pm0.01$ & $11.0\pm0.9$ & $31.6$&$1.1$ & $4$ & $11.1\pm0.9$ & $1.24\pm0.17$ \\
 Abell2537 &  $1.22\pm0.05$ & $7.2\pm1.1$ & $0.86\pm0.06$ & $5.9\pm0.9$ & $91.8$&$21.7$ & $17$ & $8.4\pm1.3$ & $0.99\pm0.74$ \\
 CL0024.0+1652 &  $1.30\pm0.10$ & $9.8\pm2.7$ & $0.45\pm0.08$ & $3.1\pm4.7$ & $61.2$&$15.9$ & $24$ & $73.5\pm11.5$ & $6.46\pm5.23$ \\
 MACSJ0717.5+3745 &  $1.46\pm0.07$ & $16.6\pm3.4$ & $2.35\pm0.03$ & $12.3\pm1.9$ & $396.3$&$80.0$ & $224$ & $23.9\pm0.9$ & $23.09\pm3.24$ \\
 MACSJ0913.7+4056 &  $0.95\pm0.07$ & $4.0\pm1.3$ & $0.53\pm0.02$ & $4.8\pm0.7$ & $17.0$&$1.1$ & $4$ & $4.3\pm1.0$ & $0.40\pm0.19$ \\
 MS0015.9+1609 &  $1.60\pm0.06$ & $21.9\pm3.2$ & $2.01\pm0.06$ & $13.4\pm1.9$ & $171.0$&$20.0$ & $41$ & $8.6\pm1.1$ & $0.58\pm0.38$ \\
 MS0440.5+0204 &  $0.85\pm0.06$ & $2.2\pm0.7$ & $0.24\pm0.05$ & $2.8\pm0.5$ & $30.1$&$5.7$ & $<3 $ & $19.6\pm4.2$ & $1.38\pm1.16$ \\
 MS0451.6-0305 &  $0.95\pm0.10$ & $4.5\pm1.7$ & $1.03\pm0.02$ & $7.8\pm1.0$ & $235.3$&$43.3$ & $28$ & $11.9\pm1.1$ & $1.44\pm0.82$ \\
 MS0906.5+1110 &  $1.36\pm0.09$ & $8.7\pm1.9$ & $0.87\pm0.03$ & $3.5\pm0.5$ & $148.9$&$29.0$ & $3$ & $17.1\pm1.2$ & $0.20\pm0.14$ \\
 MS1008.1-1224 &  $1.06\pm0.05$ & $4.8\pm0.9$ & $0.58\pm0.04$ & $7.3\pm3.1$ & $97.9$&$24.7$ & $10$ & $55.8\pm2.2$ & $4.17\pm2.33$ \\
 MS1231.3+1542 &  $0.54\pm0.11$ & $0.6\pm0.4$ & $0.14\pm0.00$ & $1.4\pm0.1$ & $131.5$&$16.5$ & $72$ & $6.9\pm1.4$ & $5.08\pm3.82$ \\
 MS1358.1+6245 &  $1.12\pm0.09$ & $5.9\pm1.6$ & $0.67\pm0.07$ & $7.6\pm0.9$ & $39.3$&$3.9$ & $4$ & $8.6\pm1.2$ & $0.34\pm0.29$ \\
 MS1455.0+2232 &  $1.04\pm0.05$ & $4.2\pm0.8$ & $0.56\pm0.01$ & $3.1\pm0.2$ & $23.6$&$0.8$ & $3$ & $4.9\pm0.2$ & $0.13\pm0.06$ \\
 MS1512.4+3647 &  $0.85\pm0.18$ & $2.6\pm1.8$ & $0.34\pm0.03$ & $2.1\pm0.7$ & $26.4$&$8.1$ & $6$ & $6.7\pm1.3$ & $1.30\pm1.09$ \\
 MS1621.5+2640 &  $1.19\pm0.07$ & $7.7\pm1.8$ & $0.83\pm0.03$ & $5.4\pm0.8$ & $182.1$&$37.7$ & $41$ & $19.0\pm4.3$ & $7.47\pm5.43$ \\
 RXJ1347.5-1145 &  $1.25\pm0.12$ & $9.3\pm2.9$ & $1.63\pm0.01$ & $13.1\pm1.8$ & $29.7$&$2.1$ & $<4 $ & $12.6\pm1.4$ & $1.30\pm0.41$ \\
 RXJ1524.6+0957 &  $0.87\pm0.12$ & $3.4\pm1.8$ & $0.41\pm0.04$ & $2.7\pm0.4$ & $123.9$&$42.3$ & $22$ & $63.2\pm5.6$ & $22.92\pm15.12$ \\
 \tablecomments{All quantities are measured at $r_{500}^{WL}$, except
  for P3/P0 power ratio, which is measured at
  $r_{2500}^{WL}$, \hlb{and $D_\mr{BCG}$, which is in Mpc}. \hly{$M_\mr{X}$ is the X-ray hydrostatic mass, $K_0$
  is the entropy at 20 kpc, $w_\mr{BCG}$ is the X-ray peak to BCG
  offset, $w_X$ is the centroid shift.} \hlb{Lensing masses are from 
  \protect\cite{Hoekstra12}.}}
\end{deluxetable*}